%% file: paper.tex
\documentclass[%
 reprint,
 nofootinbib,
 amsfonts,amsmath,amssymb,
 aps,
 prd,
]{revtex4-2}

\usepackage{xifthen}
\newcommand{\journal}{arxiv}

\ifthenelse{\equal{\journal}{PRDx}}%
{%
\newcommand{\colorNP}{black}%
\newcommand{\colorW}{black}%
\newcommand{\colorK}{black}%
\newcommand{\colorC}{black}%
}%
{%
\newcommand{\colorNP}{blue}%
\newcommand{\colorW}{orange}%
\newcommand{\colorK}{red}%
\newcommand{\colorC}{olive}%
}%

\usepackage[dvipsnames]{xcolor}
\usepackage[%
  colorlinks=true,
  urlcolor=blue,
  linkcolor=blue,
  citecolor=blue
]{hyperref}

\usepackage{graphicx}

\usepackage{tikz}

\newcommand{\figinput}[1]{\includegraphics{#1.pdf}}%
\newcommand{\tikzinput}[1]{\includegraphics{#1.pdf}}%

\usepackage{lmodern}

\usepackage{dcolumn}

\usepackage{orcidlink}

\usepackage{booktabs}
\usepackage{tabularx}

\usepackage{printlen}

\usepackage{accents}

\usepackage{upgreek}
\usepackage[scr=boondoxo]{mathalpha}
\DeclareMathAlphabet{\mathsfit}{T1}{\sfdefault}{\mddefault}{\sldefault}

\usepackage{amsthm}
\newtheorem*{remark}{Remark}

\newtheorem{definition}{Definition}
\newcommand{\name}[1]{{\tt{#1}}}

\newcommand{\HH}{\mathcal{H}}
\newcommand{\NN}{\mathcal{N}}
\renewcommand{\SS}{\mathcal{S}}
\newcommand{\DD}{\mathcal{D}}
\newcommand{\WW}{\mathcal{W}}
\newcommand{\VV}{\mathcal{V}}
\newcommand{\KK}{\mathrm{K}}

\def\pd{\partial}
\def\d{\mathrm{d}}

\newcommand{\szero}{\SS_0}
\def\surfkappa{\kappa_{(\NPl)}}
\def\Lie{\mathcal{L}}
\def\const{\mathrm{constant}}
\def\horeq{\mathbin{\dot{=}}}
\newcommand{\abs}[1]{\left|#1\right|}

\newcommand{\overbar}[1]{\mkern 2.0mu\overline{\mkern-2.0mu#1\mkern-0.7mu}\mkern 0.7mu}
\newcommand{\cconj}[1]{\overbar{#1}}

\let\oldepsilon=\epsilon
\def\epsilon{\varepsilon}

\newcommand{\negphantom}[1]{\ifmmode\settowidth{\dimen0}{$#1$}\else\settowidth{\dimen0}{#1}\fi\hspace*{-\dimen0}}

\newcommand{\Wt}    {{\color{\colorW}{t}}}
\newcommand{\Wr}    {{\color{\colorW}{r}}}
\newcommand{\Wtheta}{{\color{\colorW}{\theta}}}
\newcommand{\Wphi}  {{\color{\colorW}{\phi}}}
\newcommand{\Wv}    {{\color{\colorW}{v}}} 
\newcommand{\Wrho}  {{\color{\colorW}{\varrho}}}
\newcommand{\Wz}    {{\color{\colorW}{z}}}
\newcommand{\WEx}   {{\color{\colorW}{\hat{x}}}}
\newcommand{\WEz}   {{\color{\colorW}{\hat{z}}}}

\newcommand{\pdWr}{{,\Wr}}
\newcommand{\pdWtheta}{{,\Wtheta}}
\newcommand{\pdWrWr}{{,\Wr\Wr}}
\newcommand{\pdWrWtheta}{{,\Wr\Wtheta}}
\newcommand{\pdWthetaWtheta}{{,\Wtheta\Wtheta}}
\newcommand{\pdWrho}{{,\Wrho}}
\newcommand{\pdWz}{{,\Wz}}
\newcommand{\pdWrhoWrho}{{,\Wrho\Wrho}}
\newcommand{\pdWzWz}{{,\Wz\Wz}}

\newcommand{\Kv}    {{\color{\colorK}{u}}}
\newcommand{\Kr}    {{\color{\colorK}{s}}}
\newcommand{\Ktheta}{{\color{\colorK}{\vartheta}}}
\newcommand{\Kphi}  {{\color{\colorK}{\varphi}}}
\newcommand{\Kzeta} {{\color{\colorK}{\varsigma}}}
\newcommand{\KEx}   {{\color{\colorK}{\hat{X}}}}
\newcommand{\KEz}   {{\color{\colorK}{\hat{Z}}}}
\newcommand{\Kx}    {{\color{\colorK}{\mathsfit{x}}}}

\newcommand{\rE}{\Kr_\text{E}}
\newcommand{\vE}{\Kv_\text{E}}
\newcommand{\KzetaE}{\Kzeta_{\kern 0.08em \text{E}}} 
\newcommand{\KthetaE}{\Ktheta_\text{E}}
\newcommand{\KphiE}{\Kphi_\text{E}}

\newcommand{\OSxi}      {{\color{black}{\hat{\xi}}}}
\newcommand{\OSzeta}    {{\color{black}{\hat{\zeta}}}}

\newcommand{\Salpha}    {{\color{\colorNP}{\alpha}}}
\newcommand{\Sbeta}     {{\color{\colorNP}{\beta}}}
\newcommand{\Smu}       {{\color{\colorNP}{\mu}}}
\newcommand{\Snu}       {{\color{\colorNP}{\nu}}}
\newcommand{\Srho}      {{\color{\colorNP}{\rho}}}
\newcommand{\Ssigma}    {{\color{\colorNP}{\sigma}}}
\newcommand{\Skappa}    {{\color{\colorNP}{\kappa}}}
\newcommand{\Stau}      {{\color{\colorNP}{\tau}}}
\newcommand{\Sgamma}    {{\color{\colorNP}{\gamma}}}
\newcommand{\Slambda}   {{\color{\colorNP}{\lambda}}}
\newcommand{\Sepsilon}  {{\color{\colorNP}{\epsilon}}}
\newcommand{\Spi}       {{\color{\colorNP}{\pi}}}
\newcommand{\Sa}        {{\color{\colorNP}{a}}}

\newcommand{\NPl}       {{\color{\colorNP}{\ell\kern 0.02em}}}
\newcommand{\NPn}       {{\color{\colorNP}{n}}}
\newcommand{\NPm}       {{\color{\colorNP}{m}}}

\newcommand{\NPPsi}     {{\color{\colorNP}{\kern -0.10em \mathit{\Psi} \kern -0.08em}}}
\newcommand{\NPPhi}     {{\color{\colorNP}{\kern -0.08em \mathit{\Phi} \kern -0.04em}}}

\newcommand{\NPLambda}  {{\color{\colorNP}{\kern -0.08em \mathit{\Lambda}}}}

\newcommand{\NPD}       {{\color{\colorNP}{\mathrm{D}}}}
\newcommand{\NPDelta}   {{\color{\colorNP}{\Delta}}}
\newcommand{\NPdelta}   {{\color{\colorNP}{\updelta}}}

\newcommand{\Wlambda}   {\zeta}
\newcommand{\Wnu}       {\psi}
\newcommand{\Schf}      {{\color{\colorC}{f}}}
\newcommand{\Womega}    {\omega}

\newcommand{\TU}{U}
\newcommand{\TX}{X}
\newcommand{\TOmega}{\mathit{\Omega}}
\newcommand{\Txi}{\xi}

\newcommand{\area}{A}
\newcommand{\massBH}    {{\color{\colorC}{M}}} 
\newcommand{\massDisk}  {{\color{\colorC}{\mathcal{M}}}} 
\newcommand{\diskR}     {{\color{\colorC}{b}}}
\newcommand{\distDisk}  {{\color{\colorC}{d}}}
\newcommand{\bM}        {{\color{\colorC}{\mathsfit{b}}}}
\newcommand{\EuclidR}   {{\color{\colorC}{\mathcal{R}}}} 
\newcommand{\Cpi}{\kern 0.04em \uppi \kern 0.04em} 
\newcommand{\dss}{{\d\mathscr{s}^2}}
\newcommand{\ChGamma}{{\mathit{\Gamma}}}
\newcommand{\simfun}{{{\mathit{\Sigma}}}}

\DeclareMathOperator{\arccot}{arccot} 
\let\Re\relax
\let\Im\relax
\DeclareMathOperator\Re{Re}
\DeclareMathOperator\Im{Im}
\newcommand{\bigo}{\mathcal{O}}
\def\eu{\mathrm{e}}
\def\ii{\mathrm{i}}
\newcommand\Laplace{\mathop{}\!\mathbin\bigtriangleup}

\usepackage[normalem]{ulem}

\newcommand{\remove}[1]{}
\newcommand{\DIFadd}[1]{#1}
\newcommand{\issueB}[1]{#1}
\newcommand{\issueC}[1]{#1}
\newcommand{\issueD}[1]{#1}
\newcommand{\eqc}{}

\begin{document}

\preprint{APS/123-QED}

\title{
    Initial data for a deformed isolated horizon
}

\author{Ale\v{s} Flandera\,\orcidlink{0000-0002-9288-8910}}
\email{flandera.ales@utf.mff.cuni.cz}
\author{David Kofro\v{n}\,\orcidlink{0000-0002-0278-7009}}
\email{d.kofron@gmail.com}
\author{Tom\'{a}\v{s} Ledvinka\,\orcidlink{0000-0002-6341-2227}}
\email{tomas.ledvinka@mff.cuni.cz}

\affiliation{
    Institute of Theoretical Physics, Faculty of Mathematics and Physics, 
    Charles University, V Hole\v{s}ovi\v{c}k\'ach 2, 180~00 Prague, Czech Republic
}

\date{\today}

\begin{abstract}
    Within the isolated horizon formalism, we investigate a static axisymmetric
    space-time of a black hole influenced by matter in its neighborhood.
    To illustrate the role of ingredients and assumptions in this formalism,
    we first show how, in spherical symmetry, the field equations and gauge
    conditions imply the isolated horizon initial data leading to the
    Schwarzschild space-time. Then, we construct the initial
    data for a static axisymmetric isolated horizon representing a deformed
    black hole. The space-time description in the Bondi-like coordinates is
    then found as a series expansion in the vicinity of the horizon. To graphically
    illustrate this construction, we also find a numerical solution for a black
    hole deformed by a particular analytic model of a thin accretion disk. 
    We also discuss how an accretion disk affects
    the analytical properties of the horizon geometry.
\end{abstract}


\maketitle


\section{Introduction}
    \label{sec:introduction}
    The \emph{isolated horizon}~(IH) formalism~\cite{Ashtekar-2004LR} has proven to
    be a powerful framework for the description of black holes in equilibrium. 
    These black holes can have arbitrary time-independent intrinsic geometry, 
    the space-time does not have to be asymptotically flat, and outside the 
    horizon it may be dynamical due to outgoing gravitational and 
    electromagnetic radiation.
    One of the key features of the isolated horizon formalism is the
    applicability of the standard laws of black hole 
    thermodynamics~\cite{Ashtekar-2000b}.
    Another important characteristic is the
    quasi-locality of the description.
    Thanks to it, unlike event horizons, isolated horizons do not require knowledge of 
    the entire space-time and avoid having central quantities defined both at 
    infinity (angular momentum, mass) and at the horizon (surface gravity, 
    angular velocity). This quasi-locality was essential in the computation of 
    entropy in quantum gravity \cite{Ashtekar-1998}. 
    The more general scenario, where in-falling matter is present, has also been
    explored within the more universal framework of dynamical horizons,
    \cite{Hayward-1994, Ashtekar-2002b}.

    Both the mechanics of isolated horizons, \cite{Ashtekar-1999, Ashtekar-2000a, 
    Ashtekar-2000b, Ashtekar-2000c, Ashtekar-2001}, and their
    intrinsic geometry, \cite{Ashtekar-2002,
    Lewandowski-2002, Lewandowski-2006, Adamo-2009}, have been extensively studied.
    In~\cite{Lewandowski-2000}, general solutions of the Einstein field equations
    admitting isolated horizons were studied. Interestingly, while the intrinsic
    geometry can coincide with the Kerr--Newman black hole, the neighborhood
    can be distorted. The geometry around an isolated horizon has been 
    investigated in~\cite{Krishnan-2012}, where the isolated horizon was specified
    as the characteristic initial value formulation. A general review of such
    a formulation can be
    found in~\cite{Stewart-1993}. For more information, see the references in%
    ~\cite{Krishnan-2012}. However, the construction in~\cite{Krishnan-2012} did
    not provide any clues on how to choose the initial data. This issue was 
    partially addressed in~\cite{Scholtz-2017}, where the Kinnersley tetrad was 
    transformed to obtain the initial data for Kerr space-time.
    Nevertheless, the obtained tetrad was not in an explicit form. 
    This has been recently revised in~\cite{Kofron-2024}.

    Moreover, while the Kerr example is physically interesting, the solution of%
    ~\cite{Scholtz-2017} 
    was only obtained thanks to the prior knowledge of the solution and
    its properties. Hence, we shall start by following the construction of a 
    \emph{tetrad} describing an isolated horizon as shown by
    \cite{Krishnan-2012} (performed using the \emph{Newman--Penrose} (NP) formalism
    \cite{Newman-1962}) and finding the initial data for
    spherical symmetry in the isolated horizon formalism. Although this is a
    simpler (and astrophysically less important) example, no prior knowledge
    of the resulting metric is used, the properties of the isolated horizons
    are demonstrated in action, and the result shall serve as a starting point
    for the next section.

    Isolated horizons are also suitable for describing models of astrophysical 
    black holes, as they can describe deformations of the horizon geometry caused
    by static distribution of matter without compromising the horizon isolation.
    The metric of a black hole and surrounding matter have been studied, e.g. in 
    \cite{Poisson-2005, Poisson-2010} for the Schwarzschild and Kerr black holes.
    In this paper, we investigate a Schwarzschild black hole deformed by
    the presence of matter in its neighborhood (such as an ``accretion'' disk)
    in the isolated horizon formalism using the Weyl superimposed metric from
    \cite{Frolov-2003}. A disk derived in \cite{Kofron-2023} 
    is later used as a particular example.
    We shall see that, while the Weyl coordinates are usually easier to use due 
    to their simplicity, since they are singular at the horizon, they are not
    always well suited. The coordinate transformation to Bondi-like coordinates 
    is not obvious. Our work might thus be useful whenever checking the 
    on-horizon limit of a certain process is required or desirable.
    Importantly, the results are not based on the perturbation approach;
    the Weyl metric allows to consider strong-field deformations of the
    black hole horizon. Very recently, first-order tidal deformations of
    isolated horizons were studied in~\cite{Metidieri-2024}.

    In the next section, we briefly introduce the isolated horizons and the
    coordinates and tetrad of~\cite{Krishnan-2012}. We review the initial
    value problem and an axial isolated horizon as introduced by Ashtekar.
    Then, in Sec.~\ref{sec:spherical_symmetry}, we discuss spherical symmetry
    in the isolated horizon formalism and construct the initial data and
    tetrad describing it. This is then used in Sec.%
    ~\ref{sec:schwarzschild_deformaiton} as a starting point for deformation
    of a non-rotating black hole in the isolated horizon formalism. 
    The set of initial data for isolated horizons is large and not thoroughly 
    studied. We select the subspace corresponding to an arbitrarily strong 
    deformation of the Schwarzschild black hole horizon and construct both 
    initial data and the near-horizon series expansion of the solution in
    Bondi-like coordinates.
    While the deformation is considered analytically (as a series)
    and generally, it is complemented in Sec.~\ref{sec:particular_solution}
    with one particular example produced numerically. Finally, in Sec.%
    ~\ref{sec:static_solution_geometry}, we discuss some important properties 
    such as the convergence of the solution. The description of the numerical
    solution, its precision, and comparison are, together with the review of the
    NP formalism and the particular solution of the disk, postponed to the
    appendices.

    A geometrized unit system is used, where $c = 1$ and $G = 1$.
    Moreover, the space-time metric $g_{ab}$ has the signature $({+}{-}{-}{-})$ 
    and the Riemann tensor is defined by $2\nabla_{[c}\nabla_{d]}\,X^a = 
    - R^a{}_{bcd} X^b$.

\section{Isolated horizons, coordinates and tetrad}
    \label{sec:coordinates_and_tetrad}

    A Bondi-like coordinate system and adapted Newman--Penrose  null tetrad
    for weakly isolated horizons have been introduced in \cite{Krishnan-2012}
    and the details of the construction were spelled out in detail in%
    ~\cite{Gurlebeck-2018}. Here we do not repeat the construction and only
    review the results.%
    \footnote{
    Since we shall work with multiple coordinates
    and formalisms, and their usual naming conventions are oftentimes
    incompatible, we shall use non-standard names for some of the variables.
    This includes e.g.\ $\Wv$ and $\Kv$ not being advanced and retarded null
    coordinates but two different time coordinates while $\Wt$ is yet another
    one.
    \ifthenelse{\equal{\journal}{PRD}}
    {}
    {To help the reader navigate the paper, the Bondi-like coordinates are
    highlighted in {\color{red}{red}} color while Weyl coordinates (and closely
    related coordinates) are colored {\color{orange}{orange}}. Moreover, the NP
    quantities such as the tetrad vectors, derivatives and scalars are marked in
    {\color{blue}{blue}}.  Finally, {\color{olive}{green}} is used for physical
    constants.}
    }
    For a brief review of the Newman--Penrose formalism, see Appendix%
    ~\ref{app:np_formalism}.
    The coordinates will be denoted by $\Kx^\mu = (\Kv, \Kr, \Ktheta, \Kphi)$,
    $\mu=0,1,2,3$ where, on the horizon $\HH$, $\Kv$ is the parameter along
    the null generators of $\HH$ and the submanifolds $\Kv=\const$ are null
    hypersurfaces intersecting $\HH$ (given by $\Kr=0$) at topological
    2-spheres with coordinates $\Kx^I$, $I=2,3$. In particular, the sphere given
    by $\Kr=0,\Kv=0$ will be denoted by $\SS_0$.

    Appropriate NP tetrad $e_\mu{}^a$ adapted to these coordinates consists of 
    four null vectors named $\NPl^a$, $\NPn^a$ and $\NPm^a$ and its complex 
    conjugate $\cconj{\NPm}^a$. Their coordinate expressions read as follows:
    \begin{align}
        \begin{aligned}
            \boldsymbol{\NPl} &= \pd_\Kv + \TU\,\pd_\Kr + \TX^I\,\pd_I \,, \\
            \boldsymbol{\NPn} &= -\,\pd_\Kr \,, \\
            \boldsymbol{\NPm} &= \TOmega\,\pd_\Kr + \Txi^I\,\pd_I \,.
        \end{aligned}
        \label{eq:general_tetrad}
    \end{align}
    Since the NP projections of the fields and the spin coefficients depend on
    the geometry itself as well as on the tetrad, we understand all tetrad
    projections with respect to this geometrically motivated tetrad.
    
    Following the standard NP notation, \cite{Newman-1962}, we introduce the 
    directional derivatives associated with the respective null vectors as
    \begin{align}
        \NPD &\equiv \NPl^a \nabla_a \,,&
        \NPDelta &\equiv \NPn^a \nabla_a \,,&
        \NPdelta &\equiv \NPm^a \nabla_a \,.
    \end{align}
    The coefficients appearing in the expressions of the vectors will be 
    referred to as the \emph{metric functions}. Functions $\TU, \TX^I$ and 
    $\TOmega$ vanish on the horizon, i.e.
    \begin{align}
        \TU \horeq \TX^I \horeq \TOmega \horeq 0 \,,
    \end{align}
    where we use symbol $\horeq$ for the equality of two quantities on the 
    horizon $\HH$ but not necessarily elsewhere. Hence, $\boldsymbol{\NPl}\horeq \pd_\Kv$ 
    is tangent to the null geodesics generating $\HH$ but, in general, it is 
    not affinely parameterized, its acceleration being the surface gravity 
    $\surfkappa$ defined by
    \begin{align}
        \NPD \NPl^a &\horeq \surfkappa\,\NPl^a\,.
    \end{align}
    On the horizon, vectors $\NPm^a$ and $\cconj{\NPm}^a$ span the tangent space to
    the 2-spheres $\Kr=0$, $\Kv = \const$. We can choose these vectors (or
    equivalently functions $\Txi^I$) arbitrarily on $\SS_0$ and
    propagate them along $\HH$ using the Lie derivative $\Lie$ by demanding
    \begin{align}
        \Lie_{\boldsymbol{\NPl}} \NPm^a &\horeq 0 \,.
    \end{align}
    The choice of $\NPl^a$ and $\NPm^a$ on $\HH$ uniquely determines vector $\NPn^a$.
    Finally, we propagate the tetrad from the horizon to its neighborhood by
    conditions
    \begin{align}
        \NPDelta \NPl^a = \NPDelta \NPn^a = \NPDelta \NPm^a = 0 \,.
        \label{eq:off_horizon_transport}
    \end{align}

    Moreover, we require $\HH$ to be a non-expanding horizon. That is, $\HH$ is 
    a null hypersurface with vanishing expansion, the Einstein field equations 
    and an energy condition are satisfied on the horizon.
    The non-expanding horizon is then paired with a (class of) null normal(s)
    $\NPl^a$ to create either a \emph{weakly isolated horizon} or (strongly)
    \emph{isolated horizon} in the sense of~\cite{Ashtekar-2002}.
    For a weakly isolated horizon, the intrinsic geometry has to satisfy
    \begin{align}
        [\Lie_{\boldsymbol{\NPl}},\DD_a]\NPl^b\horeq 0 \,, 
        \label{eq:WIH_cond}
    \end{align}
    where the covariant derivative $\DD_a$ intrinsic to the horizon is defined by 
    $Y^a \DD_a X^b\horeq Y^a \nabla_a X^b$ for any $X^a$ and $Y^a$ tangent to 
    $\HH$. Any non-expanding horizon can be made into a weakly isolated horizon
    and condition~\eqref{eq:WIH_cond} already makes the surface gravity 
    $\surfkappa$ constant across the horizon ($0^\text{th}$ law of thermodynamics).

    The stronger condition for an isolated horizon is
    \begin{align}
        [\Lie_{\boldsymbol{\NPl}},\DD_a]X^b\horeq 0 \,, 
        \label{eq:IH_cond}
    \end{align}
    for an arbitrary vector $X^b$ tangent to $\HH$. This astrophysically
    more relevant choice is in fact a restriction on the geometry of the 
    space-time (in contrast with the weakly isolated horizons) and
    makes all geometrical quantities time independent ($\Kv$ independent) on 
    $\HH$. It can be shown that Eq.\ (\ref{eq:IH_cond}) is equivalent to 
    $\NPl^a$ satisfying the Killing equations up to the second order in $\Kr$. 

    Conditions imposed on the tetrad enumerated above translate into the
    properties of the NP spin coefficients as follows:
    \begin{align}
        \begin{aligned}
            \Smu &= \cconj{\Smu} \,, \\
            \Stau &= \Sgamma = \Snu = 0 \,,\\
            \Spi &= \Salpha + \cconj{\Sbeta} \,, \\
            \Sepsilon & \horeq \cconj{\Sepsilon} = \frac{\surfkappa}{2} \,, 
                \quad \surfkappa \horeq \const \,,\\
            \Srho &\horeq \Ssigma \horeq \Skappa \horeq 0 \,.
        \end{aligned}
        \label{eq:spin_coeffs}
    \end{align}
    \remove{%
    The fact that $\Smu$ is real implies that congruence $\NPn^a$ is twist-free; 
    vanishing of spin coefficients $\Srho, \Ssigma$ and $\Skappa$ on the horizon 
    expresses the fact that $\HH$ is both non-expanding, non-twisting and 
    shear-free and $\NPl^a$ is geodesic.}\\
    It is also convenient to introduce quantity
    \begin{equation}
        \Sa = \Salpha - \cconj{\Sbeta} 
        = \NPm^a\, \cconj{\NPdelta}\cconj{\NPm}_a \,.
    \end{equation}
    On the horizon, it encodes the induced 2-dimensional connection on $\SS_0$. 

    \issueB{While the meaning of the spin coefficients can be read off the so-called
    transport equations, see Eqs.~\eqref{eq:transport_equations}, some of the
    coefficients also have a direct interpretation as optical scalars for
    null congruences --- expansion, shear, and twist. For a review
    see~\cite{Chandrasekhar-1983} or an
    alternative description in \cite{Wald-1984}. There are two
    congruences, one along each of the vectors $\NPl^a$ and $\NPn^a$. 
    For vector $\NPl^a$, $\Srho$ encodes expansion and twist (in real and
    imaginary parts, respectively), and $\Ssigma$ represents shear.
    Since $\Srho$ and $\Ssigma$ vanish on the horizon, all three optical scalars
    are zero there (this is already required by the existence of a non-expanding
    horizon, \cite{Ashtekar-2004}). Similarly, for the vector $\NPn^a$, $\Smu$
    gives expansion and twist, while $\Slambda$ is the shear. Vanishing twist
    along $\NPn^a$ implies that $\Smu$ is real. The congruence along $\NPn^a$
    is geodesic ($\Snu = 0$) and affinely parametrized ($\Sgamma = 0$).
    Moreover, $\Skappa \horeq 0$ means that $\NPl^a$ is geodesic on $\HH$.
    The vector $\NPl^a$ is parallelly transported along $\NPn^a$ ($\Stau = 0$).}

    \issueB{The interpretation of the Weyl scalars, which are projections of the
    Weyl tensor, is far less straightforward. These quantities are
    tetrad-dependent. 
    While a common interpretation is that $\NPPsi_0$ represents ingoing and
    $\NPPsi_4$ outgoing gravitational waves at infinity, this is true only in
    the linearized regime of asymptotically flat space-times with a well-adapted
    tetrad and coordinates, \cite{Misner-2017, Teukolsky-1973, Brito-2015}.
    Additionally, an appropriate fall-off at radial infinity ($\sim r^{-1}$)
    is expected for the gravitational radiation. An analogous interpretation of
    $\NPPsi_0$ as gravitational radiation falling down the black hole horizon
    (in linearized gravity) is provided in \cite{Hawking-1972, Brito-2015}.
    In full nonlinear general relativity, the interpretation of $\NPPsi_4$ as
    gravitational radiation is clear only in type N space-times. Another interpretation of
    the Weyl scalars is done by investigating the geodesic deviation
    equation. In our case, all of the Weyl scalars can be non-vanishing, but this
    does not necessarily indicate the presence of gravitational radiation, as
    our exact solutions will be manifestly \emph{static}. Recall that this can
    occur even in the pure Schwarzschild solution if one does not choose the tetrad
    adapted to the principal null directions of the Weyl tensor (for the
    transformations of Weyl scalars see e.g.~\cite{Stewart-1993}).}

    \subsection{Initial value problem}
        \label{sec:geometric_description}
        Space-time near an isolated horizon can be reconstructed as the 
        solution of characteristic initial value problem with the initial data 
        given on $\HH$ itself and another null hypersurface intersecting $\HH$,
        \cite{Krishnan-2012}. In the Bondi-like coordinates introduced in 
        Sec.~\ref{sec:coordinates_and_tetrad}, such hypersurfaces $\NN_\Kv$ are 
        conveniently 
        represented by equation $\Kv = \const$. Then the 2-spheres foliating the 
        horizon are simply $\SS_\Kv = \HH \cap \NN_\Kv$. In our setting, we will 
        formulate the initial value problem on $\HH$ and $\NN_0$ which intersect
        at $\SS_0$.

        As discussed in \cite{Krishnan-2012}, the free initial data consist of 
        the following quantities.\footnote{We omit the electromagnetic field 
        in this paper.}
        \begin{itemize}
            \item Free data on $\SS_0$:
                \begin{itemize}
                    \item constant surface gravity $\surfkappa = \Sepsilon
                        +\cconj{\Sepsilon}$,
                    \item spin coefficients $\Spi, \Sa, \Slambda, \Smu$,
                    \item metric functions $\Txi^I$,
                \end{itemize}
            \item Free data on $\NN_0$:
                \begin{itemize}
                    \item Weyl component $\NPPsi_4$;
                \end{itemize}
        \end{itemize}
        \begin{figure}
            \centering
            \tikzinput{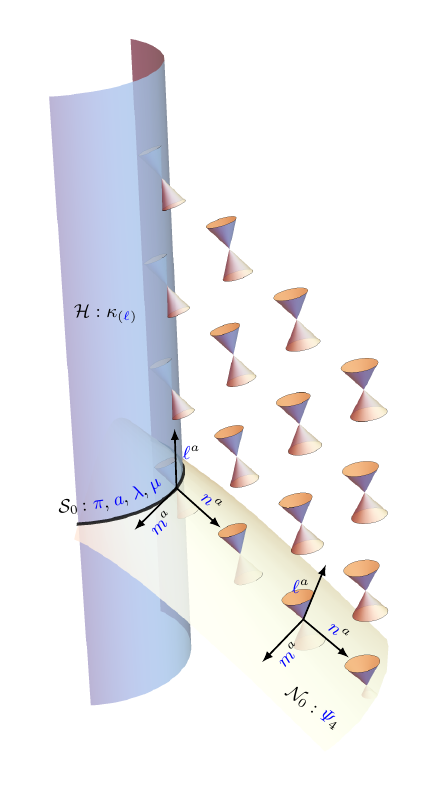}
            \caption[The neighborhood of the horizon and initial value 
            surfaces.]{The neighborhood of the horizon and initial value 
            surfaces. The scalars enlisted after the colons have to be 
            introduced on the corresponding hypersurface. Moreover, we
            need to know the complete tetrad on the horizon.}
            \label{img:spacetime_structure}
        \end{figure}
        The situation is schematically visualized in Fig.~\ref{img:spacetime_structure}.

        Properties of a vacuum weakly isolated horizon enforce relations 
        \eqref{eq:spin_coeffs} and, in addition,
        \begin{align}
            \NPPsi_0 \horeq \NPPsi_1 \horeq 0 \,.
        \end{align}
        Data on $\SS_0$ can be propagated along the entire horizon as follows. 
        The Weyl scalar $\NPPsi_3$ on $\HH$ can be computed from the Ricci 
        identity \eqref{app:np:RI:deltalambda} and scalar $\NPPsi_4$ is propagated 
        via Bianchi identity \eqref{np:BI:DPsi4}.
        On $\SS_0$, one can use $\Sa$ and $\Spi$ to calculate the Weyl scalar 
        $\NPPsi_2$ by taking the real and imaginary parts of the Ricci identity 
        (\ref{app:np:RI:deltaalpha}),
        \begin{subequations}
        \begin{align}
            \Re\! \NPPsi_2 &\horeq\Sa \cconj{\Sa} - \frac{1}{2}\left(\NPdelta \Sa
            + \cconj{\NPdelta}\cconj{\Sa}\right) , \\
            \Im\! \NPPsi_2 &\horeq - \Im \left(\NPdelta \Spi - \Sa\Spi\right) .
            \label{eq:im_psi_2}
        \end{align}
        \end{subequations}
        The Bianchi identity \eqref{np:BI:DPsi2} shows that $\NPPsi_2$ is 
        $\Kv$ independent on $\HH$,
        \begin{align}
            \NPD\NPPsi_2 &\horeq 0 \,.
        \end{align}
        Ricci identities \eqref{app:np:RI:Dalpha} and \eqref{app:np:RI:Dbeta} imply
        \begin{align}
            \NPD\Spi &\horeq \NPD \Sa \horeq 0 \,,
        \end{align}
        while the evolution of $\Slambda$ and $\Smu$ along $\HH$ is given by 
        Ricci identities \eqref{app:np:RI:Dlambda} and \eqref{app:np:RI:Dmu}:
        \begin{align}
            \begin{aligned}
                \NPD\Slambda &\horeq 
                    \cconj{\NPdelta} \Spi - \cconj{\Sa}\Spi 
                    + \Spi^2 - \surfkappa\,\Slambda \,, \\
                \NPD\Smu &\horeq 
                    \NPdelta \Spi - \Sa\Spi 
                    + |\Spi|^2 + \NPPsi_2 - \surfkappa\,\Smu \,.
            \end{aligned}
            \label{eq:Dlambdamu}
        \end{align}
        Hence, the geometry of $\HH$ is fully determined by the initial data on 
        $\SS_0$. What remains is to specify scalar $\NPPsi_4$
        on $\NN_0$ which is free, unconstrained function of $\Kr, \Ktheta, \Kphi$. 
        Notice that for weakly isolated horizons, the NP scalars $\Smu, \Slambda, 
        \NPPsi_3$ and $\NPPsi_4$ in general depend on time $\Kv$.

        After a short discussion of weakly isolated horizons in Sec.%
        ~\ref{sec:spherical_symmetry}, we will focus 
        on isolated horizons (rather than weakly isolated 
        ones) with axial symmetry, which imposes additional constraints. 
        Consequently, all NP scalars are independent of coordinates $\Kv$ and 
        $\Kphi$, in particular
        \begin{align}
            \NPD\Smu &\horeq \NPD\Slambda 
            \horeq \NPD\NPPsi_3 
            \horeq \NPD\NPPsi_4 \horeq 0 \,.
            \label{eq:time_independence_strong_isolation}
        \end{align}
        Therefore, Eqs.
        \eqref{eq:Dlambdamu} simplify to
        \begin{align}
            \begin{aligned}
                \cconj{\NPdelta} \Spi - \cconj{\Sa}\Spi 
                + \Spi^2 &\horeq \surfkappa\,\Slambda \,, 
                \\
                \NPdelta \Spi - \Sa\Spi 
                + |\Spi|^2+\NPPsi_2 &\horeq \surfkappa\,\Smu \,, 
            \end{aligned}
        \end{align}
        and they further constrain 
        $\Spi$ and $\NPPsi_2$ provided that 
        $\Slambda$ and $\Smu$ are free functions on $\SS_0$.

    \subsection{Axial symmetry}
        \label{sec:axial_symmetry_by_ashtekar}
        Let us now introduce an axially symmetric (weakly) isolated horizon as
        defined in~\cite{Ashtekar-2004}.
        \begin{definition}[Axially symmetric isolated horizon]
            \label{def:axial_ih}
            An isolated horizon is axially symmetric if any 2-sphere
            $\mathcal{S}_\Kv$, defined by $\Kr = 0$, $\Kv = \const$,
            possesses a space-like Killing vector with closed orbits whose
            norm vanishes exactly at two points called the north and south poles.
        \end{definition}
        Note that the axial symmetry is only defined for the isolated horizon
        itself, not the space-time admitting the horizon. 
        Then, as shown in~\cite{Ashtekar-2004}, one can introduce the
        coordinates $\Kx^I = (\Kzeta, \Kphi)$ with $\Kzeta \in [-1, 1]$ and 
        $\Kphi \in [0, 2\Cpi)$. In these coordinates
        the metric on $\mathcal{S}_\Kv$ is of the form
        \begin{equation}
            \boldsymbol{q} = - \EuclidR^2 \left(h(\Kzeta)^{-1}\, \d \Kzeta^2 
            + h(\Kzeta) \, \d \Kphi^2\right) . 
            \label{eq:metric_axi_ih}
        \end{equation}
        Here, $\EuclidR$ is the Euclidean radius defined by 
        $\area = 4 \Cpi\EuclidR^2$ and $\area$ is the geometrically 
        defined area of the 2-sphere $\mathcal{S}_\Kv$.
        The positive function $h(\Kzeta)$ is arbitrary, up to satisfying the
        conditions
        \begin{align}
            \begin{aligned}
                h(\pm 1) &= 0 \,, \\
                h'(\pm 1) &= \mp 2 \,, 
            \end{aligned}
            \label{eq:axial_h_cond}
        \end{align}
        which ensure the regularity of the axis.

        From the assumption that the horizon is foliated by 2-spheres of the
        form \eqref{eq:metric_axi_ih}, we can naturally choose the vector
        $\NPm^a$ to be
        \begin{equation}
            \boldsymbol{\NPm} \horeq \frac{1}{\sqrt{2}\,\EuclidR} \left(\sqrt{h}\,
            \partial_\Kzeta + \frac{\ii}{\sqrt{h}}\, \partial_\Kphi\right) .
            \label{eq:axial_m_vector}
        \end{equation}
        \begin{remark}
            All other possible choices of vector $\NPm^a$ can also be
            obtained by the spin transformation introduced in Sec.%
            ~\ref{app:np:lorentz_transformation} and, hence, give no new
            results.
        \end{remark}
        The form of the vector $\NPm^a$ gives us two more scalars from the set 
        of the initial data:
        \begin{align}
                \Txi^2 &\horeq \frac{1}{\EuclidR} \sqrt{\frac{h}{2}} \,, &
                \Txi^3 &\horeq \frac{\ii}{\EuclidR \, \sqrt{2 h}} \,.
        \end{align}
        This is already enough to provide us with the induced 2-dimensional connection 
        on $\mathcal{S}_0$ which is given by $\Sa$
        and can be computed to be
        \begin{equation}
            \Sa = \NPm^a\, \cconj{\NPdelta} \cconj{\NPm}_a \horeq - \frac{h'}{2 \sqrt{2h}\, \EuclidR} \,. 
            \label{eq:a_hor}
        \end{equation}
        Notice that in the axially symmetric scenario, the coefficient $\Sa$ is,
        real.

\section{Spherically symmetric initial data}
    \label{sec:spherical_symmetry}
    Obviously, the Schwarzschild black hole represents an isolated horizon.
    If we compare the Schwarzschild metric in the ingoing Eddington--Finkelstein
    coordinates with Def.~\ref{def:axial_ih} of axially symmetric
    isolated horizon, it is simple to reckon the appropriate Krishnan tetrad
    and, hence, the initial data, in the form of~\cite{Krishnan-2012}.
    Not surprisingly, many quantities are identically zero. The Schwarzschild 
    IH initial data can also be obtained as the non-rotating limit of the Kerr 
    initial data~\cite{Scholtz-2017}. In both cases, the data for the 
    IH are only available by using previous knowledge of the full solution in 
    the metric form.

    In this section, we show how these quantities arise from the assumptions of
    the IH formalism under spherical symmetry. We sequentially reduce a large 
    set of partial differential equations into a single differential equation
    and show that the isolated 
    horizons assumption then selects its single time-independent solution. 
    Because IH formalism already assumes certain time independence of the
    geometry, rather than providing another proof of Birkhoff's theorem,
    this section illustrates how, in spherical symmetry, various ingredients
    of IH formalism join to provide the expected result.

    Spherical symmetry is given by the existence of 
    three angular Killing vectors that meet the commutation relations
    \newcommand{\Ki}{K^{(i)}}
    \newcommand{\Kj}{K^{(j)}}
    \newcommand{\Kk}{K^{(k)}}
    \newcommand{\cKi}{\cconj{K}^{(i)}}
    \newcommand{\cKj}{\cconj{K}^{(j)}}
    \newcommand{\cKk}{\cconj{K}^{(k)}}
    \begin{equation}
        \left[\Ki, \Kj\right] = \oldepsilon_{ijk} \Kk \,,
        \label{eq:Killing_commutation}
    \end{equation}
    where $i,j,k$ denote the three different Killing vectors and
    $\oldepsilon_{ijk}$ is the Levi-Civita permutation symbol, \cite{Carroll-2019}. 
    Let $K_a$ be a Killing (co-)vector of a space-time. We can expand it into
    the Newman--Penrose tetrad as
    \begin{equation}
        K_a = K_0\,\NPl_a + K_1\,\NPn_a + K_2\,\NPm_a + \cconj{K}_2\,\cconj{\NPm}_a
        \,. 
    \end{equation}
    See Appendix~\ref{app:np:killing_eq} for a review of Killing vectors in the
    Newman--Penrose formalism.
    
    We shall assume a general set of Killing vectors without specifying
    any particular form. However, we can set both $K_0 \horeq K_1 \horeq 0$.
    The coefficient $K_0$ determines the projection in the $\NPl^a$
    direction.  Consider that the vector $\NPl^a$ is a Killing vector
    with respect to the induced metric on the horizon (see 
    e.g.~\cite{Krishnan-2012}) and
    that any linear combination of Killing vectors is also a Killing vector.
    Therefore, we can, without loss of generality, choose the coefficient $K_0$
    to vanish on the horizon. To show that also $K_1 \horeq 0$, consider that
    the horizon itself is a sphere. The Killing vectors of spherical symmetry
    must therefore lie within the horizon itself and, hence, can have no
    radial part. But this radial part can, on the horizon, be given only by
    the vector $\NPn^a$. Thus, $K_1$ also vanishes on the horizon:
    \begin{align}
        \begin{aligned}
            0 &= K^a \nabla_a \Kr = K_0\, \NPD \Kr + K_1\, \NPDelta \Kr + K_2\, \NPdelta \Kr
                + \cconj{K}_2\, \cconj{\NPdelta} \Kr \\
              &\qquad{}= K_0\, \TU - K_1 + K_2\, \TOmega
                + \cconj{K}_2\, \cconj{\TOmega} \horeq - K_1 \,.
        \end{aligned}
    \end{align}

    Let us now consider the Killing equations%
    ~\eqref{app:np:eq:Killing:Killing_equations_projected},
    but also plug in the conditions imposed by the Krishnan tetrad $\Sgamma = \Stau
    = \Snu = 0$, and general isolated horizons $\Smu = \cconj{\Smu}$ and
    $\Spi = \Salpha + \cconj{\Sbeta}$ into them. We can extract the following
    information:
    \begin{subequations}
        \begin{align}
            \NPDelta K_0 &= 0 \,, \\
            \NPDelta K_1 &= - (\Sepsilon + \cconj{\Sepsilon}) K_0 -\NPD K_0
            - \cconj{\Spi} K_2 
            - \Spi \cconj{K}_2 \,, \\
            \NPDelta K_2 &= \Spi K_0 + \cconj{\NPdelta} K_0
            + \Smu K_2 + \Slambda \cconj{K}_2 \,,
        \end{align}
        \label{eq:radial_Killing_conditions}
    \end{subequations}
    and
    \begin{subequations}
        \begin{align}
            \NPD K_1 &= (\Sepsilon + \cconj{\Sepsilon}) K_1
            + \Skappa K_2 + \cconj{\Skappa} \cconj{K}_2 \,, \label{eq:DK1}\\
            \begin{split}
                \NPD K_2 &= \cconj{\Skappa} K_0
                - 2 \Spi K_1 + \cconj{\NPdelta} K_1 \\
                &\qquad{}- (\Sepsilon - \cconj{\Sepsilon} + \Srho) K_2 
                - \cconj{\Ssigma} \cconj{K}_2 \,,
            \end{split}
        \end{align}
        \label{eq:time_Killing_conditions}
    \end{subequations}
    and
    \begin{subequations}
        \begin{align}
            \NPdelta \cconj{K}_2 &= \Ssigma K_0 - \cconj{\Slambda} K_1 - 
            (\cconj{\Salpha} - \Sbeta) \cconj{K}_2 \,, \\
            \begin{split}
                \NPdelta K_2 + \cconj{\NPdelta}\cconj{K}_2 &= (\Srho + \cconj{\Srho})
                K_0 - 2 \Smu K_1 \\
                &\qquad{}+ (\cconj{\Salpha} - \Sbeta) K_2 
                + (\Salpha - \cconj{\Sbeta}) \cconj{K}_2 \,.
            \end{split}
        \end{align}
        \label{eq:angular_Killing_conditions}
    \end{subequations}
    The first set of equations tells us about the radial evolution of the 
    coefficients, and we immediately see that, in fact, $K_0 = 0$. 
    We can also check that Eq.~\eqref{eq:DK1}
    is in accordance with $K_1 \horeq 0$ because $\Skappa \horeq 0$.
    In the last set, which sheds light on the angular evolution, there is
    neither an equation for $K_0$ nor for $K_1$, but
    since both $K_0$ and $K_1$ vanish on the horizon, we have all the equations
    we need. However, the equation for $\NPdelta K_2$ is missing,
    we shall see this term in the resulting equations.

    We can evaluate the commutation relation~\eqref{eq:Killing_commutation}
    for the individual components by plugging in the Killing
    equations. For the Killing vector $\Kk$ we get:
    \begin{subequations}
        \begin{align}
            \Kk_0 &= 0 \,,
            \label{eq:K0_commu}
            \\[1.5ex]
            \begin{split}
                \Kk_1 &= \cconj{\Srho} \Kj_2 \cKi_2 - \Srho \Kj_2 \cKi_2
                + \Srho \Ki_2 \cKj_2 \\
                &\qquad{}- \cconj{\Srho} \Ki_2 \cKj_2 
                + \Ki_2 \NPdelta \Kj_1 - \Kj_2 \NPdelta \Ki_1 \\
                &\qquad{}+ \cKi_2 \cconj{\NPdelta} \Kj_1 - \cKj_2 \cconj{\NPdelta} \Ki_1 \,,
                \label{eq:K1_commu}
            \end{split}
            \\[1.5ex]
            \begin{split}
                \Kk_2 &= \Slambda \left(\Ki_1 \cKj_2 - \Kj_1 \cKi_2\right) \\
                &\qquad{}+ \Ki_2 \NPdelta \Kj_2 - \Kj_2 \NPdelta \Ki_2 \,.
                \label{eq:K2_commu}
            \end{split}
        \end{align}
    \end{subequations}
    We shall soon make use of these relations.

    \subsection{Data on the horizon}
        \label{sec:spherical_symmetry_on_H}
        The integrability conditions for the existence of a Killing vector%
        ~\eqref{eq:integrability_condition}   
        imply certain restrictions on the initial data. Because%
        ~\eqref{eq:integrability_condition} has 40 independent projections 
        (16 complex + 8 real equations), we illustrate the derivation in detail
        only for one particular projection,
        $\NPl^a \cconj{\NPm}^b \NPn^c \left(\nabla_c \nabla_a K_b
        - R_{abcd}\,K^d \right)$: 
        \begin{equation}
            \NPDelta \NPD K_2 + \Spi\, \NPDelta K_1 + K_2 \left(\NPDelta \Sepsilon
            - \NPDelta \cconj{\Sepsilon} - \cconj{\NPPsi}_2 \right) \horeq 0 \,. 
        \end{equation}
        First, we use the commutation relations~\eqref{app:np:eq:commutators}
        to make $\NPDelta$ the innermost operator,
        \begin{align}
            \begin{aligned}
                &\NPD \NPDelta K_2 + 2 \Sepsilon\, \NPDelta K_2 + \Spi\, \NPDelta K_1
                - \Spi\, \NPdelta K_2 - \cconj{\Spi}\, \cconj{\NPdelta} K_2 \\
                &\qquad{}+ K_2 \left(\NPDelta \Sepsilon - \NPDelta \cconj{\Sepsilon}
                - \cconj{\NPPsi}_2 \right) \horeq 0 \,.  
            \end{aligned}
        \end{align}
        We proceed with the substitution of the terms for radial evolution 
        using~\eqref{eq:radial_Killing_conditions} and \eqref{app:np:RI},
        \begin{align}
            \begin{aligned}
                &\NPD\!\left(\Slambda \cconj{K}_2 + \Smu K_2\right) + 2\Sepsilon \left(\Slambda
                \cconj{K}_2 + \Smu K_2\right) \\
                &\qquad{}- \Spi \left(\cconj{\Spi} K_2
                + \Spi \cconj{K}_2\right) - \Spi\, \NPdelta K_2
                - \cconj{\Spi}\, \cconj{\NPdelta} K_2 \\
                &\qquad{}+ K_2 \left(\Spi \Sa
                - \cconj{\Spi} \Sa - \NPPsi_2\right) \horeq 0 \,.
            \end{aligned}
        \end{align}
        In a similar manner, we replace also the operator $\NPD$,
        \begin{equation}
            K_2\, \NPdelta \Spi + \cconj{K}_2\, \cconj{\NPdelta} \Spi
            - \Spi\, \NPdelta K_2 - \cconj{\Spi}\, \cconj{\NPdelta} K_2
            - K_2 \cconj{\Spi} \Sa + \cconj{K}_2 \Spi \Sa \horeq 0 \,.
        \end{equation}
        The last set of Killing equations~\eqref{eq:angular_Killing_conditions}
        simplifies the equation even further:
        \begin{equation}
            \Sa \cconj{K}_2 \Spi - \Spi\, \NPdelta K_2 + K_2\, \NPdelta
            \Spi + \cconj{K}_2\, \cconj{\NPdelta} \Spi \horeq 0 \,.
        \end{equation}

        Following the same process for the rest of the projections, we find
        that most of them are trivially satisfied or are dependent on
        others, usually they are a complex conjugate of another projection.
        Let us show the final set of independent equations:
        \begin{subequations}
            \begin{align}
                &&\NPl^a \cconj{\NPm}^b \NPn^c &\mapsto \nonumber\\
                &&0 &\horeq \Sa \cconj{K}_2 \Spi - \Spi\, \NPdelta K_2 + K_2\, \NPdelta
                \Spi + \cconj{K}_2\, \cconj{\NPdelta} \Spi \,,
                \label{eq:pi_projection} \\[1.5ex]
                &&\NPn^a \cconj{\NPm}^b \NPm^c &\mapsto \nonumber\\
                &&0 &\horeq K_2\, \NPdelta \Smu + \cconj{K}_2\, \cconj{\NPdelta} \Smu \,,
                \label{eq:mu_projection} \\[1.5ex]
                &&\NPn^a \cconj{\NPm}^b \cconj{\NPm}^c &\mapsto \nonumber\\
                &&0 &\horeq 2 \Sa \cconj{K}_2 \Slambda - 2 \Slambda \NPdelta K_2 + K_2 \NPdelta
                \Slambda + \cconj{K}_2 \cconj{\NPdelta} \Slambda \,,
                \label{eq:lambda_projection} \\[1.5ex]
                &&\NPm^a \NPm^b \cconj{\NPm}^c &\mapsto \nonumber\\
                &&&\negphantom{0}\begin{aligned}
                    0 &\horeq \NPdelta \NPdelta K_2 - \Sa \NPdelta K_2 - \left(K_2 + \cconj{K}_2\right)
                    \NPdelta \Sa \\
                    &\qquad{}- \cconj{K}_2 \cconj{\NPdelta} \Sa + 2 \Sa^2 \cconj{K}_2 \,,
                \end{aligned}
                \label{eq:a_projection}
            \end{align}
        \end{subequations}
        where in the last equation, we have employed $\cconj{\Sa} \horeq \Sa$.

        Let us now discuss what the above equations tell us about the
        scalars. All of them must be satisfied for particular values
        of $\Spi, \Smu, \Slambda$ and $\Sa$ and particular operators
        $\NPdelta$ and $\cconj{\NPdelta}$ but three different values of
        $K_2$, which moreover have to obey conditions given by the
        commutation relation~\eqref{eq:Killing_commutation}.

        Let us start with Eq.~\eqref{eq:mu_projection}.
        As we have mentioned, we have a set of three equations, hence, let us
        rewrite the equation as
        \begin{subequations}
            \begin{align}
                \Ki_2 \NPdelta \Smu + \cKi_2 \cconj{\NPdelta} \Smu &\horeq 0 \,, \\
                \Kj_2 \NPdelta \Smu + \cKj_2 \cconj{\NPdelta} \Smu &\horeq 0 \,, \\
                \Kk_2 \NPdelta \Smu + \cKk_2 \cconj{\NPdelta} \Smu &\horeq 0 \,. 
            \end{align}
        \end{subequations}
        Into the last one, we plug the expression from the commutation relations%
        ~\eqref{eq:K2_commu} expressed on the horizon:
        \begin{align}
            \begin{split}
                \Kk_2 &\horeq \Ki_2 \NPdelta \Kj_2 - \Kj_2 \NPdelta \Ki_2 \,,
            \end{split}
            \\
            \begin{split}
                \cKk_2 &\horeq \cKj_2 \NPdelta \Ki_2 - \cKi_2 \NPdelta \Kj_2 \\
                &\qquad{}+ \cconj{\Sa} \left( \Kj_2 \cKi_2 - \Ki_2 \cKj_2 \right) ,
            \end{split}
        \end{align}
        and from the two equations for $i$ and $j$
        we replace $\cKi_2 \cconj{\NPdelta} \Smu$ with $- \Ki_2 \NPdelta \Smu$,
        and the same for $j$. Altogether, we get
        \begin{equation}
            \NPdelta \Smu \left(\Ki_2 \NPdelta \Kj_2 - \Kj_2 \NPdelta \Ki_2 \right) \horeq 0 \,.
        \end{equation}
        If the term in parentheses is zero, then $\Kk_2 \horeq 0$. Compare
        the term with the commutator~\eqref{eq:K2_commu} on the horizon, and
        since $K_0 \horeq K_1 \horeq 0$, the vector would be trivial on the
        horizon, which is not admissible. Hence, the only solution of the
        condition \remove{is}
        \begin{equation}
            \NPdelta \Smu \horeq 0 \,
            \label{eq:delta_mu_on_H_spherical}
        \end{equation}
        \issueB{implies that expansion along $\NPn^a$ is constant
        on $\SS_\Kv$.}

        Because both equations~\eqref{eq:pi_projection}
        and~\eqref{eq:lambda_projection} are very similar (up to the factor
        $2$), we only show how to tackle the first. 
        An obvious solution of this equation is to
        set $\Spi \horeq 0$. More importantly, let us show that it is, in fact,
        the only possible solution of the equation. Let us suppose that
        $\Spi \neq 0$ on the horizon.
        Then we can find from~\eqref{eq:pi_projection} that
        \begin{equation}
            \NPdelta K_2 \horeq \frac{K_2 \NPdelta \Spi + \cconj{K}_2 \cconj{\NPdelta} \Spi
            + \cconj{K}_2 \Spi \Sa}{\Spi} \,.
        \end{equation}
        This can be substituted into the condition on $K_2$ from the commutation
        relation~\eqref{eq:K2_commu} evaluated on the horizon, which reads
        \begin{equation}
            \Kk_2 \horeq \Ki_2 \NPdelta \Kj_2 - \Kj_2 \NPdelta \Ki_2 \,.
        \end{equation}
        It gives us:
        \begin{equation}
            \Kk_2 \horeq \frac{\Sa \Spi + \cconj{\NPdelta} \Spi}{\Spi}
            \left( \Ki_2 \cKj_2 - \Kj_2 \cKi_2 \right) ,
        \end{equation}
        which must be valid for each of the three cyclic combinations of
        the values of $i, j, k$. Moreover, the term
        $\frac{\Sa \Spi + \cconj{\NPdelta} \Spi}{\Spi}$ remains the same. The
        only solution is then $\Kk_2 \horeq 0$ for each value of $k$, but since
        we already have $K_0 \horeq K_1 \horeq 0$ the Killing vectors
        would be trivial. Thus, the only possibility is the already mentioned
        solution of Eq.~\eqref{eq:pi_projection}:
        \begin{equation}
            \Spi \horeq 0 \,.
            \label{eq:pi_on_H_spherical}
        \end{equation}
        \issueB{Thus, according to Eq.~\eqref{eq:im_psi_2}, $\Im\NPPsi_2 \horeq 0$,
        this indicates no angular momentum of the horizon, which is given as a
        weighted integral of $\Im\NPPsi_2$, \cite{Ashtekar-2004}.
        Similarly, Eq.~\eqref{eq:lambda_projection} implies}
        \remove{In an analogous way, the equation ~\eqref{eq:lambda_projection} implies}
        \begin{equation}
            \Slambda \horeq 0 \,,
            \label{eq:lambda_on_H_spherical}
        \end{equation}
        \issueB{meaning that, analogously to~\eqref{eq:delta_mu_on_H_spherical},
        the ingoing transversal shear vanishes.}
        We cannot find similarly clear implications of Eq.~\eqref{eq:a_projection}.

    \subsection{Data outside the horizon}
        Unlike other initial data quantities, the curvature scalar $\NPPsi_4$
        must also be given outside the horizon. With an adapted tetrad, in
        spherical symmetry, we expect $\NPPsi_4 = 0$. To show this, we start
        with an assumption of spherical symmetries of the Riemann and the Weyl
        tensors
        \begin{equation}
            \Lie_{\Ki} R_{abcd} = \Lie_{\Ki} W_{abcd} = 0 \,.
        \end{equation}
        We can then proceed in a similar way to
        Sec.~\ref{sec:spherical_symmetry_on_H}. In the end, we obtain only five
        independent equations, which can be nicely written as
        \begin{alignat}{4}
                 &{}&           {}&\KK(\NPPsi_4) +  2&\WW&\NPPsi_4& &{}= 0 \,, \label{eq:Weyl_pro_4} \\
             \VV &\NPPsi_4&{} + {}&\KK(\NPPsi_3) + {}&\WW&\NPPsi_3& &{}= 0 \,, \\
            2\VV &\NPPsi_3&{} + {}&\KK(\NPPsi_2)   {}&   &        & &{}= 0 \,, \label{eq:Weyl_pro_2} \\
            3\VV &\NPPsi_2&{} + {}&\KK(\NPPsi_1) - {}&\WW&\NPPsi_1& &{}= 0 \,, \\
            4\VV &\NPPsi_1&{} + {}&\KK(\NPPsi_0) -  2&\WW&\NPPsi_0& &{}= 0 \,, 
        \end{alignat}
        where we introduced scalars $\VV$ and $\WW$ and operator $\KK$ as follows:
        \begin{align}
            \VV &= - \left(\cconj{\Salpha} + \Sbeta\right) K_1 + \NPdelta K_1
            + \left(\Srho - \cconj{\Srho}\right) \cconj{K}_2 \,, \\
            \WW &= - \Smu K_1 - \NPdelta K_2
            + \left(\Salpha - \cconj{\Sbeta}\right) \cconj{K}_2 \,, \\
            \KK(\NPPsi_j) &= \left(K_1 \NPDelta + K_2 \NPdelta
            + \cconj{K}_2 \cconj{\NPdelta}\right) \NPPsi_j \,. 
        \end{align}
        Equation~\eqref{eq:Weyl_pro_4} does not involve any other $\NPPsi_j$
        except $\NPPsi_4$. Let us write the equation down explicitly on the
        horizon:
        \begin{equation}
            2 \Sa \cconj{K}_2 \NPPsi_4 - 2 \NPPsi_4\, \NPdelta K_2 + K_2\, \NPdelta \NPPsi_4
            + \cconj{K}_2\, \cconj{\NPdelta} \NPPsi_4 \horeq 0 \,. \label{eq:Psi4_on_H}
        \end{equation}
        Comparing Eq.~\eqref{eq:Psi4_on_H} to~\eqref{eq:lambda_projection}
        we can see that they are the same except for the change of $\Slambda$
        to $\NPPsi_4$.  Hence, the solution is also the same:
        \begin{equation}
            \NPPsi_4 \horeq 0 \,. 
        \end{equation}

        Let us now have a at look the same equation, \eqref{eq:Weyl_pro_4}, but
        outside the horizon. Written explicitly, it is
        \begin{align}
            \begin{aligned}
                &2 \Sa \cconj{K}_2 \NPPsi_4 - 2 \NPPsi_4\, \NPdelta K_2 + K_2\, \NPdelta \NPPsi_4
                + \cconj{K}_2\, \cconj{\NPdelta} \NPPsi_4 \\
                &\qquad{}+ K_1\, \NPDelta \NPPsi_4
                - 2 K_1 \Smu \NPPsi_4 = 0 \,. 
            \end{aligned}
        \end{align}
        If $K_1 = 0$, the equation reduces to~\eqref{eq:Psi4_on_H}. And
        because the commutator~\eqref{eq:K2_commu} has for $K_1 = 0$ the
        same form as on the horizon everywhere, the only solution can be
        that $\NPPsi_4$
        is zero. On the other hand, if $K_1$ is non-zero, we can express
        $\NPDelta \NPPsi_4$:
        \begin{align}
            \begin{aligned}
                \NPDelta \NPPsi_4 = - \frac{1}{K_1}&\Big(\!\left(2 \Sa \cconj{K}_2 
                    - 2 \NPdelta K_2 - 2 K_1 \Smu\right) \NPPsi_4 \\
                    &\qquad{}+ K_2 \NPdelta \NPPsi_4
                + \cconj{K}_2 \cconj{\NPdelta} \NPPsi_4\Big) .
            \end{aligned}
        \end{align}
        If $\NPPsi_4$ vanishes on a radial hypersurface, its radial derivative
        is zero as well. Since we already know that $\NPPsi_4 \horeq 0$,
        the only possible solution is $\NPPsi_4 = 0$. This equality is valid
        everywhere, but it would be enough to know $\NPPsi_4$ on the
        transversal hypersurface $\mathcal{N}_0$.

    \subsection{Coordinates on the horizon}
        So far, we have only imposed condition~\eqref{eq:axial_h_cond} 
        onto the function $h(\Kzeta)$, it can otherwise be arbitrary
        to satisfy axial symmetry. Let us show that this is not the case in
        spherical symmetry.

        Since $\Spi \horeq 0$, we can rewrite the Ricci 
        equation~\eqref{app:np:RI:deltaalpha} into
        \begin{equation}
            \NPPsi_2 \horeq \Sa \cconj{\Sa} - \frac{1}{2}\left(\NPdelta \Sa
            + \cconj{\NPdelta}\cconj{\Sa}\right) ,
        \end{equation}
        and use the value of $\Sa$ on the horizon~\eqref{eq:a_hor} to get
        \begin{equation}
            \NPPsi_2 \horeq \frac{h''(\Kzeta)}{4 \EuclidR^2} \,.
        \end{equation}
        However, if we evaluate Eq.~\eqref{eq:Weyl_pro_2} on the
        horizon:
        \begin{equation}
            K_2\, \NPdelta \NPPsi_2 + \cconj{K}_2\, \cconj{\NPdelta} \NPPsi_2 \horeq 0 \,,
        \end{equation}
        we find that because $\NPPsi_2$ is independent of $\Kphi$ due to the
        axial symmetry, $\NPPsi_2$ is also independent of $\Kzeta$ on the
        horizon.
        Hence, $h''(\Kzeta) \horeq \const$. Owing to \eqref{eq:axial_h_cond},
        there is only one such function $h(\Kzeta)$ and we get
        \begin{align}
            h(\Kzeta) &\horeq 1 - \Kzeta^2 \,, \label{eq:h_on_H} \\
            \NPPsi_2 &\horeq - \frac{1}{2 \EuclidR^2} \,. \label{eq:Psi2_on_H}
        \end{align}
        
        In order to obtain a result that matches the more standard form, we
        shall additionally consider the trivial transformation
        $\Kzeta = \cos\Ktheta$.

    \subsection{Initial data for spherical symmetry}
        Let us summarize what we have found. Out of the set of the sought
        initial data, only $\Sepsilon$ and $\Smu$ are
        non-zero. While $\Sepsilon$ is a constant on the horizon (let
        $\Sepsilon_0$ be its value), $\Smu$ is only needed on the sphere
        $\szero$, let us denote its value there by $\Smu_{\szero}$, and similarly
        it is constant there.

        In \eqref{eq:Psi2_on_H}, we have already found that spherical symmetry
        yields $\NPPsi_2$ as a time-independent quantity. Similarly, all the other
        initial data are constant on the entire horizon under spherical
        symmetry. We are left to find $\Smu$, which is given by Ricci
        equation~\eqref{app:np:RI:Dmu} which (under all conditions involved) is
        \begin{equation}
            \NPD \Smu \horeq - 2 \Sepsilon_0 \Smu + \NPPsi_2 \,.
            \label{eq:time_evolution_mu}
        \end{equation}
        We have already mentioned that under strong isolation, conditions
        \eqref{eq:time_independence_strong_isolation} are satisfied.
        Here, it would mean $\NPD\Smu\horeq0$.
        To distinguish implications of the spherical symmetry from those of
        horizon isolation (and to include a weakly isolated horizon), we
        postpone enforcing the strong isolation until the general solution is
        found. Moreover, we shall show how the condition $\NPD\Smu \horeq 0$
        manifests. We thus treat Eq.~\eqref{eq:time_evolution_mu}
        as a differential equation for $\Smu(\Kv)$ and using \eqref{eq:Psi2_on_H}
        we find that
        \begin{align}
            \Smu \horeq \frac{\eu^{-2\Kv\Sepsilon_0} \left(1 
            + 4 \EuclidR^2 \Sepsilon_0 \Smu_{\szero}\right)-1}{4 \EuclidR^2 \Sepsilon_0} \,.
            \label{eq:mu_sol}
        \end{align}
        \issueB{This provides the admissible values of the expansion along
        $\NPn^a$.}

    \subsection{The Schwarzschild solution}
        \label{sec:schwarzschild_solution}
        Obviously, the spherically symmetric initial data we just found lead to
        the well-known Schwarzschild solution. Thus, let us only briefly
        outline how the equations separate into easily solvable ones.
        
        Since $\NPPsi_4 = 0$, the Ricci equations
        paired with the initial data~\eqref{eq:lambda_on_H_spherical}
        and~\eqref{eq:mu_sol} give us $\Slambda = 0$ and 
        \begin{equation}
            \Smu = \frac{\eu^{-2\Kv\Sepsilon_0} \left(1 
            + 4 \EuclidR^2 \Sepsilon_0 \Smu_{\szero}\right)-1}
            {\simfun} \,,
        \end{equation}
        where we have denoted
        \begin{equation}
            \simfun = 4 \EuclidR^2 \Sepsilon_0 - \Kr \big(\eu^{-2\Kv\Sepsilon_0}
                \left(1 + 4 \EuclidR^2 \Sepsilon_0 \Smu_{\szero}\right)-1\big)
            \,. 
        \end{equation}
        We find the on-horizon values of $\NPPsi_2$ and $\NPPsi_3$ from Ricci
        equations and the off-horizon using the Bianchi ones arriving finally at
        \begin{align}
            \NPPsi_2 &= \frac{-32\EuclidR^4\,\Sepsilon_0{}^3}{\simfun^3} \,, \\
            \NPPsi_3 &= 0 \,,
        \end{align}
        using also that $\Spi = 0$ similarly to $\Slambda$. For $\Sepsilon$, we
        have
        \begin{equation}
            \Sepsilon = \frac{\Sepsilon_0\left(\left(\simfun - \Kr\right)^2
                - \eu^{-2\Kv \Sepsilon_0} 
            \left(1 + 4\EuclidR^2\Sepsilon_0\Smu_{\szero}\right) \Kr^2\right)}
                {\simfun^2} \,.
        \end{equation}
        Note that for $\Kr = 0$ this expression simplifies to 
        $\Sepsilon \horeq \Sepsilon_0$ as it should, and the surface gravity
        is constant.

        The metric functions are then computed from%
        ~\eqref{app:np:eq:frame_equations_radial} to be
        \begin{align}
            \begin{aligned}
                \TX^I &= 0 \,, \\
                \TOmega &= 0 \,, \\
                \TU &= \frac{2\Kr\Sepsilon_0\left(\simfun - \Kr\right)}{\simfun} \,, \\
                \Txi^2 &= \frac{2\sqrt{2}\, \EuclidR \Sepsilon_0}{\simfun} \,, \\
                \Txi^3 &= \frac{2\sqrt{2}\, \ii\, \EuclidR \Sepsilon_0}{\simfun\, \sin\Ktheta} \,. \\
            \end{aligned}
            \label{eq:time_dependent_Schw_metric_func}
        \end{align}

        This is everything needed to write down the metric:
        \begin{align}
            \begin{aligned}
                \dss &=
                \frac{4 \left(\simfun - \Kr\right)\Kr \Sepsilon_0}
                {\simfun} \d \Kv^2
                - 2 \d \Kv \, \d \Kr \\
                &\qquad{}- \frac{\simfun^2}{16 \EuclidR^2\, \Sepsilon_0{}^2}
                    \Big(\d \Ktheta^2 + \sin^2\Ktheta\, \d \Kphi^2\Big)
                .
            \end{aligned}
            \label{eq:general_spherical_metric}
        \end{align}

        Seemingly, we have arrived at a whole class of metrics parameterized
        by $\Sepsilon_0$ and $\Smu_{\szero}$ which is unexpected.
        Let us show that it is, in fact, not the case. There exists a coordinate 
        transformation
        \begin{align}
            \begin{aligned}
                \rE &= \frac{\simfun}
                {4 \EuclidR\, \Sepsilon_0} \,, \\
                \vE &= 2 \EuclidR \ln\!\left|1 - \eu^{2\Kv\Sepsilon_0} 
                    + 4 \EuclidR^2\, \Sepsilon_0 \Smu_{\SS_0} \right| .
            \end{aligned}
            \label{eq:transformation_v_time_dependent}
        \end{align}
        which converts the metric~\eqref{eq:general_spherical_metric}
        to the Schwarzschild metric in the Eddington--Finkelstein
        coordinates, which is
        \begin{align}
            \begin{aligned}
                \dss &= \left(1 - \frac{\EuclidR}{\rE}\right)\d \vE{}^2 
                - 2\, \d \vE\, \d \rE \\
                     &\qquad{}-\rE^2 \left(\d \KthetaE{}^2 
                      + \sin^2\KthetaE\, \d \KphiE{}^2\right) . 
            \end{aligned}
            \label{eq:schwarzschild_metric}
        \end{align}

        Hence, we have found that by keeping $\Spi$, $\Slambda$ and $\NPPsi_4$
        zero and $\Sa$ and $\Txi^I$ in the form given by Def.%
        ~\ref{def:axial_ih} any pair of constants $\Sepsilon_0$ and
        $\Smu_{\szero}$ lead to the Schwarzschild solution.
        However, there is one particular choice that leads directly to the
        Schwarzschild metric~\eqref{eq:schwarzschild_metric} without the
        transformation~\eqref{eq:transformation_v_time_dependent}.%
        \footnote{Up to an offset of the coordinate $\rE$.}
        The choice is
        \begin{align}
            \Sepsilon_0 &= \frac{1}{4\EuclidR} \,, &
            \Smu_{\szero} &= - \frac{1}{\EuclidR} \,.
            \label{eq:simplest_choice_of_data}
        \end{align}

        Recall that we obtained 
        \eqref{eq:time_dependent_Schw_metric_func} and 
        \eqref{eq:general_spherical_metric} 
        by solving the Bianchi, Ricci, and frame equations using the imposed
        initial conditions
        \begin{align}
            \begin{aligned}
                \begin{aligned}
                \Spi &\doteq 0 \,, &
                \Slambda &\doteq 0 \,, &
                \NPdelta \Smu &\doteq 0 \,, &
                \end{aligned}
                \\
                \begin{aligned}
                \Sepsilon &\doteq \Sepsilon_0 \,, &
                \Smu|_{\Kv=0} &\doteq \Smu_{\szero} \,,
                \end{aligned}
            \end{aligned}
        \end{align}
        and
        \begin{align}
            \NPPsi_0 &\doteq 0 \,, &
            \NPPsi_1 &\doteq 0 \,, &
            \NPPsi_4 &= 0 \,,
        \end{align}
        together with the condition defining the \emph{weakly isolated horizons}
        \begin{align}
            [\Lie_{\boldsymbol{\NPl}},\DD_a]\, \NPl^b &\doteq 0 \,.
        \end{align}

        If we want to impose (strongly) \emph{isolated horizons}, 
        we need to strengthen the condition to, recalling~\eqref{eq:IH_cond},
        \begin{align}
            [\Lie_{\boldsymbol{\NPl}},\DD_a] &\doteq 0 \,,
        \end{align}
        which makes the connection time independent on the entire horizon,
        and together with \eqref{eq:time_dependent_Schw_metric_func} and 
        \eqref{eq:general_spherical_metric},
        leads to 
        \begin{align}
            1 + 4 \EuclidR^2 \Sepsilon_0 \Smu_{\szero} & = 0 \,.
            \label{eq:strong_isolation_condition}
        \end{align}
        Owing to its direct relation to the surface gravity~\eqref{eq:spin_coeffs},
        we may consider $\Sepsilon_0$ as a free parameter and then determine
        $\Smu_{\szero}$ from \eqref{eq:strong_isolation_condition}.
        To arrive at the metric~\eqref{eq:schwarzschild_metric}, the particular
        value of $\Sepsilon_0$ given in~\eqref{eq:simplest_choice_of_data} has
        to be chosen.  The transformation~\eqref{eq:transformation_v_time_dependent}
        shows that, in the case of spherical symmetry, all weakly isolated 
        horizons are equivalent to the strongly isolated one.
        
        Finally, let us present the simplest form of the correct Krishnan tetrad:
        \begin{align}
            \begin{aligned}
                \boldsymbol{\NPl} &= \pd_\Kv + \frac{\Kr}
                {2\left(2\massBH + \Kr\right)} \pd_\Kr \,,\\
                \boldsymbol{\NPn} &= - \pd_\Kr \,,\\
                \boldsymbol{\NPm} &= \frac{1}{\sqrt{2}
                \left(2\massBH + \Kr\right)} \pd_\Ktheta + 
                \frac{\ii}{\sqrt{2}
                \left(2\massBH + \Kr\right)\sin\Ktheta} \pd_\Kphi \,,
            \end{aligned}
        \end{align}
        where we identified $\EuclidR = 2\massBH$.

        As mentioned in Sec.~\ref{sec:geometric_description}, to specify a given 
        IH space-time, we need initial data in the form of functions $\Sepsilon$,
        $\Spi$, $\Sa$, $\Slambda$, $\Smu$ and $\Txi^I$ on the horizon and 
        $\NPPsi_4$ on $\NN$.
        In this section, we showed that the Krishnan gauge, together with the 
        assumption of spherical symmetry everywhere, leads all these quantities
        to be constant on the horizon, except for $\Smu$, which is constant only 
        in the case of (strongly) isolated horizons. Owing to the spherical symmetry,
        $\Slambda$, $\Spi$ and $\NPPsi_4$ vanish (everywhere).
        The 2-dimensional connection $\Sa$ is given by Eqs.%
        ~\eqref{eq:a_hor} and \eqref{eq:h_on_H}, and $\Smu$ is,
        in the case of strong isolation, restricted by%
        ~\eqref{eq:simplest_choice_of_data}.
        The remaining quantity $\Sepsilon$, connected to the surface gravity,
        can be chosen on the horizon and represents gauge freedom.
        Thus, for a spherically symmetric (strongly) isolated horizon of a
        black hole with circumference $2\Cpi \EuclidR$, all initial data are
        uniquely determined by~$\EuclidR$ up to the choice of $\Sepsilon$.

\section{Axially symmetric non-rotating isolated horizon}
    \label{sec:schwarzschild_deformaiton}
    As mentioned in the Introduction, astrophysical black holes are often 
    surrounded by an approximately axisymmetric distribution of matter.
    For static space-times, 
    the Einstein equations allow for simple superposition of a static 
    axisymmetric and reflection-symmetric (with respect to the equatorial
    plane) vacuum gravitational field with the 
    Schwarzschild black hole, \cite{Semerak-1999}.

    In this section, we will use an a priori known solution (in Weyl form)
    and provide a general procedure on how to cast it in the IH formulation
    (although not fully explicitly). Because the Weyl metric is not regular
    at the horizon, we will have to pay attention to the horizon where
    complications arise in both the computation of the series expansion and
    the numerical solution.

    First of all, we set up the proper null tetrad on the horizon.
    This will be done with the help of a coordinate transformation regularizing
    the horizon.
    Then, solving (either as a series expansion or numerically) the geodesic
    equation with the initial conditions given by (a) the position on the
    horizon and (b) the vector $\NPn^a$ localized
    there, we obtain a Bondi-like coordinate system
    with the affine parameter along the geodesics being the new radial coordinate.
    Having the null geodesic emanating from the horizon, we can parallelly
    propagate the remaining tetrad vectors along these geodesics.
    
    In this way, we obtain the complete tetrad in the vicinity of the horizon
    and we can evaluate NP projections of the Weyl tensor as well as the spin
    coefficients. Here we have to circumvent the problems that arise from the
    fact that we are actually working in the Weyl coordinates, yet we need to
    calculate the derivatives with respect to the IH coordinates.

    \subsection{Weyl superimposed metric}
        Any stationary axisymmetric vacuum metric can be written in the
        Weyl--Lewis--Papapetrou coordinates $(\Wt, \Wrho, \Wz, \Wphi)$, cf. 
        \cite{Semerak-2002}, as
        \footnote{Note that we use the mostly
        positive metric signature $({+}{-}{-}{-})$ and hence the metric has opposite
        signs with respect to the most usual form.}
        \begin{align}
            \begin{aligned}
                \dss &= \eu^{2\Wnu} \d \Wt^2 
                - B^2 \Wrho^2 \eu^{-2\Wnu}\left(\d \Wphi - \Womega\, \d \Wt\right)^2 \\
                &\qquad{}- \eu^{2\Wlambda - 2\Wnu}\left(\d \Wrho^2 + \d \Wz^2\right) ,
            \end{aligned}
        \end{align}
        where the four functions $B$, $\Womega$, $\Wnu$ and $\Wlambda$ depend on
        $\Wrho$ and $\Wz$ only. We are going to discuss only the static solutions
        (since no stationary solutions of a black hole surrounded by a disk are 
        known). Hence, we have $\Womega=0$, and the metric reduces to the Weyl 
        metric. For the function
        $B$ there is a standard choice of $B = 1$ providing the metric in the
        Weyl coordinates
        \begin{equation}    
            \dss = \eu^{2\Wnu} \d \Wt^2 - \Wrho^2 \eu^{-2\Wnu}\, \d \Wphi^2
            - \eu^{2\Wlambda - 2\Wnu}\left(\d \Wrho^2 + \d \Wz^2\right) .
            \label{eq:Weyl_metric}
        \end{equation}
        The potentials $\Wlambda$ and $\Wnu$ must satisfy the Einstein equations
        which, in vacuum regions, reduce to
        \begin{subequations}
            \begin{align}
                \Laplace_3\,\Wnu &\equiv \Wnu_\pdWrhoWrho + \frac{\Wnu_\pdWrho}{\Wrho} + \Wnu_\pdWzWz = 0 \,, 
                \label{eq:einstein_in_weyl_form_laplace}\\
                \Wlambda_\pdWrho &= \Wrho \left({\Wnu_\pdWrho}^2 - {\Wnu_\pdWz}^2\right) ,\\
                \Wlambda_\pdWz &= 2 \Wrho\,\, \Wnu_\pdWrho\, \Wnu_\pdWz \,,
            \end{align}%
            \label{eq:einstein_in_weyl_form}%
        \end{subequations}%
        where the first equation is the axially symmetric Laplace
        equation in an auxiliary flat 3-dimensional space and $\Wnu_\pdWz$
        denotes $\pd_\Wz \Wnu$.

        As discussed in~\cite{Semerak-1999}, the most important solutions of
        equations~\eqref{eq:einstein_in_weyl_form} are line sources (rods) located on
        the $\Wz$ axis, which include the special case of the Schwarzschild
        solution given by 
        \begin{align}
            \Wnu_\text{S} &= \frac{1}{2} \ln\frac{L-\massBH}{L+\massBH} \,,&
            \Wlambda_\text{S} &= \frac{1}{2} \ln\frac{L^2-\massBH^2}{l_\text{p}\,
            l_\text{m}} \,, 
        \end{align}
        where
        \begin{align}
            \begin{aligned}
                L &= \frac{1}{2}\left(l_\text{p} + l_\text{m}\right) , \\
                l_\text{p} &= \sqrt{\Wrho^2 + \left(\Wz + \massBH\right)^2} \,, \\
                l_\text{m} &= \sqrt{\Wrho^2 + \left(\Wz - \massBH\right)^2} \,.
            \end{aligned}
            \label{eq:lplm}
        \end{align}
        It was shown, e.g. in~\cite{Frolov-2003}, that the horizon, having the
        form of a line segment on the $\Wz$ axis, can be expanded into a sphere
        with radius $2\massBH$ (coincident with the horizon surface) using the 
        transformation
        \begin{align}
            \Wrho &= \sqrt{\Wr\left(\Wr - 2\massBH\right)} \sin\Wtheta\,, &
            \Wz &= \left(\Wr - \massBH\right) \cos\Wtheta \,.
            \label{eq:frolov_transformation}
        \end{align}
        from the Weyl metric~\eqref{eq:Weyl_metric}.

        Following~\cite{Frolov-2003} further, we can split the functions $\Wnu$
        and $\Wlambda$ into the Schwarzschild solution and a (not necessarily 
        small) perturbation as
        \begin{align}
            \Wnu &= \Wnu_\text{S} + \Wnu_\text{P}\,, &
            \Wlambda &= \Wlambda_\text{S} + \Wlambda_\text{P} \,.
            \label{eq:schwarzschild_plus_perturbation}
        \end{align}

        Using the transformation~\eqref{eq:frolov_transformation} together with%
        ~\eqref{eq:schwarzschild_plus_perturbation} we can rewrite%
        ~\eqref{eq:Weyl_metric} into the superimposed Weyl metric.
        Moreover, we introduce $\Schf = 1 - \frac{2\massBH}{\Wr}$ and to
        simplify further equations, we drop the index denoting perturbation
        ($\bullet_\text{P}$) and write the metric as:
        \begin{align}
            \begin{aligned}
                \dss &= \Schf \eu^{2\Wnu} \d \Wt^2 
                - \Wr^2 \eu^{-2\Wnu} \sin^2 \Wtheta\, \d \Wphi^2 \\
                &\qquad{}- \eu^{2\Wlambda - 2\Wnu} 
                \left(\Schf^{-1} \d \Wr^2 + \Wr^2 \d \Wtheta^2\right).
            \end{aligned}
            \label{eq:weyl_superimposed}
        \end{align}
        Owing to this notation, henceforth, $\Wnu=\Wlambda=0$ means an
        unperturbed Schwarzschild black hole.

        As discussed in~\cite{Frolov-2003}, since the black hole is deformed, 
        its area is not necessarily $16\Cpi \massBH^2$, but it is rather given by
        \begin{equation}
            \area = 16\Cpi\massBH^2\eu^{-2\Wnu_\text{pole}} \,.
        \end{equation}
        where $\Wnu_\text{pole}$ is the value of $\Wnu$ on the poles (of the
        coordinate sphere $r=2M$ representing the horizon).

        \subsubsection{Regularization of coordinates}
            In the metric~\eqref{eq:weyl_superimposed}, given 
            in~\cite{Frolov-2003}, $g_{\Wr\Wr}$ diverges on the horizon 
            similarly to a pure Schwarzschild solution. This is crucial for
            isolated horizons, and in the pure Schwarzschild case, it can be
            overcome by using the Eddington--Finkelstein coordinates which 
            are given by the transformation (for the ingoing variant)
            \begin{equation}
                \Wt = \Wv - \left(\Wr + 2\massBH \ln \abs{\frac{\Wr}{2\massBH} - 1}\right) .
            \end{equation}
            However, this simple transformation does not work for our more
            general case, since we would not get a regular expression on
            the horizon.

            Nevertheless, thanks to an ansatz $\Wt = \Wv - F(\Wr, \Wtheta)$ 
            we can find a more general version of this transformation:%
            \footnote{There is, of course, also the second (outgoing) variant 
            with $\Wv + F$.}
            \begin{align}
                \begin{aligned}
                    \Wt &= \Wv - F(\Wr, \Wtheta)\,, \\
                    F(\Wr, \Wtheta) &= \left(\int \frac{\eu^{\Wlambda 
                    - 2\Wnu}}{\Schf}\, \d \Wr + H(\Wtheta)\right) . 
                \end{aligned}
                \label{eq:star_transformation}
            \end{align}
            Unfortunately, we cannot calculate the integral (except numerically)
            but in the resulting metric we only find its derivatives
            \begin{align}
                F_\pdWr &= \frac{\eu^{\Wlambda - 2\Wnu}}{\Schf} \,, \\
                F_\pdWtheta &\horeq H_\pdWtheta \,. 
            \end{align}
            The resulting metric reads
            \begin{align}
                \begin{aligned}
                    \dss &= \eu^{2\Wnu} \Schf \d \Wv^2 - 2 \eu^\Wlambda \d \Wv\, \d \Wr
                        - 2 \eu^{2\Wnu} \Schf F_\pdWtheta\, \d \Wv\, \d \Wtheta \\
                        &\qquad{}+ 2 \eu^\Wlambda F_\pdWtheta\, \d \Wr\, \d \Wtheta 
                        - \eu^{-2\Wnu} \Wr^2 \sin^2\Wtheta\, \d \Wphi^2 \\
                        &\qquad{}+ \left(\eu^{2\Wnu} \Schf {F_\pdWtheta}^2 
                        - \eu^{2\Wlambda - 2\Wnu} \Wr^2\right) \d \Wtheta^2 
                         \,. 
                \end{aligned}
                \label{eq:weyl_star_metric}
            \end{align}

            \issueC{The transformation~\eqref{eq:star_transformation}
            has been chosen to eliminate the (diverging) term $g_{\Wr\Wr}$, 
            since the Krishnan tetrad, \cite{Krishnan-2012}, does not allow
            non-zero $g_{\Wr\Wr}$. While the unknown function $F(\Wr,\Wtheta)$
            shall not play a crucial role in our computation,
            for other gauges, one may find an explicit transformation,
            see e.g.~\cite{Fairhurst-2001}.}

            The derivatives of the
            functions $\Wlambda$ and $\Wnu$ have to satisfy the Einstein
            equations. For the superimposed metric~\eqref{eq:weyl_superimposed}
            the equations~\eqref{eq:einstein_in_weyl_form} take the form
            \begin{widetext}
                \begin{subequations}
                    \begin{align}
                        0 &= 
                            \Wr\left(\Wr - 2\massBH\right) \Wnu_\pdWrWr 
                            + \Wnu_\pdWthetaWtheta 
                            - 2\left(\massBH - \Wr\right)\Wnu_\pdWr 
                            + \cot\Wtheta\, \Wnu_\pdWtheta
                            \,, \label{eq:lambda_nu_einstein_eq:laplace} \\[1.5ex]%
                        \Wlambda_\pdWr &= \frac{\left(\massBH - \Wr\right)}{\massBH^2 
                            + \Schf \Wr^2 \csc^2\Wtheta} \left({\Wnu_\pdWtheta}^2 
                            + \left(\Schf - 1\right) \Wr \Wnu_\pdWr 
                            - \Schf \Wr^2 {\Wnu_\pdWr}^2 + 2\Wnu_\pdWtheta 
                            \left(\massBH + \Schf \Wr^2 \Wnu_\pdWr\right) 
                            \frac{\cot\Wtheta}{\massBH - \Wr}\right) ,
                            \label{eq:lambda_nu_einstein_eq:lambda_r} \\[1.5ex]%
                        \Wlambda_\pdWtheta &= \frac{\Schf \Wr^2 \cot\Wtheta}
                            {\massBH^2 + \Schf \Wr^2 \csc^2\Wtheta} 
                            \left({\Wnu_\pdWtheta}^2 + \left(\Schf - 1\right) \Wr \Wnu_\pdWr
                            - \Schf \Wr^2 {\Wnu_\pdWr}^2 - 2\Wnu_\pdWtheta 
                            \left(\massBH + \Schf \Wr^2 \Wnu_\pdWr\right) 
                            \frac{\left(\massBH - \Wr\right)\tan\Wtheta}{\Schf \Wr^2}\right),
                            \label{eq:lambda_nu_einstein_eq:lambda_theta}
                    \end{align}
                    \label{eq:lambda_nu_einstein_eq}
                \end{subequations}
            \end{widetext}
            where we have used $R_{\Wt\Wt}$, $R^\Wr_\Wr - R^\Wtheta_\Wtheta$
            and $R_{\Wr\Wtheta}$ components of the Ricci tensor. On the horizon,
            we have explicitly
            \begin{align}
                \begin{aligned}
                    \Wnu_\pdWr &\horeq -\frac{\Wnu_\pdWthetaWtheta 
                        + \cot\Wtheta\,\Wnu_\pdWtheta}{2\massBH} \,, 
                        \\
                    \Wlambda_\pdWr &\horeq - \frac{\left({\Wnu_\pdWtheta}^2 
                        - \cot \Wtheta\,\Wnu_\pdWtheta
                        + \Wnu_\pdWthetaWtheta\right)}{\massBH} \,, 
                        \\
                    \Wlambda_\pdWtheta &\horeq 2\Wnu_\pdWtheta \,. 
                \end{aligned}
                \label{eq:lambda_nu_einstein_eq_hor}
            \end{align}
            Were we to compute the same expressions for the transformed 
            metric~\eqref{eq:weyl_star_metric},
            we would arrive at more complex expressions that would simplify
            to the same results (recall that we transformed only the coordinate
            $\Wv$ and consider that $\Wlambda$ and $\Wnu$ depend only on $\Wr$ 
            and $\Wtheta$). The missing derivative $\Wnu_\pdWtheta$ is free data.

    \subsection{Tetrad on the horizon}
        \label{sec:tetrad_on_the_horizon}
        We want to find such a tetrad on the horizon that satisfies all the
        necessary conditions of isolated horizons. Let us start with a
        completely general tetrad where each component depends on the
        coordinates $\Wr$ and $\Wtheta$.
        We have omitted the dependence on the coordinates $\Wv$ and $\Wphi$,
        since the problem we solve is static and axially symmetric. 
        Next, we explore all the different conditions and find the 
        resulting conditions on the tetrad.

        Since both vectors $\NPl^a$ and $\NPm^a$ have to be \textbf{tangent} to the
        horizon, they cannot have any radial component, hence:
        \begin{align}
            \NPl^\Wr &\horeq 0 \,, \\
            \NPm^\Wr &\horeq 0 \label{eq:axial_m_radial_on_h} \,.
        \end{align}

        All tetrad vectors have to be \textbf{null}. From the first contraction $\NPl_a
        \NPl^a$ we get:
        \begin{equation}
            -4 \eu^{- 2\Wnu} \massBH^2 \left(\eu^{2\Wlambda} (\NPl^\Wtheta)^2 
            + (\NPl^\Wphi)^2 \sin^2\Wtheta\right) \horeq 0 \,.
        \end{equation}
        We can deduce that
        \begin{equation}
            \NPl^\Wphi \horeq \ii\, \eu^{\Wlambda} \frac{\NPl^\Wtheta}{\sin\Wtheta} \,.
        \end{equation}

        Analogically for $\NPm^a$ from $\NPm_a \NPm^a = 0$ follows
        \begin{equation}
            \NPm^\Wphi \horeq \ii\, \eu^{\Wlambda} \frac{\NPm^\Wtheta}{\sin\Wtheta} \,.
            \label{eq:axial_m_parts_on_h}
        \end{equation}

        For the vector $\NPn^a$ we get a more complicated expression:
        \begin{align}
            \begin{aligned}
                &- 2 \eu^{\Wlambda} \NPn^\Wr \left(\NPn^\Wv - H_\pdWtheta\,
                \NPn^\Wtheta\right) \\
                &\qquad{}- 4 \eu^{- 2\Wnu} \massBH^2 \left(\eu^{2\Wlambda}
                (\NPn^\Wtheta)^2 + (\NPn^\Wphi)^2 \sin^2\Wtheta\right) \horeq 0 \,.
            \end{aligned}
        \end{align}
        In this case, we can find the $\Wv$ component ($\NPn^a$ is the only 
        vector with a radial component on the horizon)\remove{.}\DIFadd{:}
        \begin{equation}
            \begin{aligned}
                \NPn^\Wv &\horeq - \frac{2\eu^{-\Wlambda-2\Wnu} 
                    \massBH^2\left(\eu^{2\Wlambda} (\NPn^\Wtheta)^2
                + (\NPn^\Wphi)^2 \sin^2\Wtheta\right)}{\NPn^\Wr} \\
                         &\qquad{}+ H_\pdWtheta\, \NPn^\Wtheta  \,.
            \end{aligned}
        \end{equation}

        Expressing the \textbf{normalization condition} $\NPl_a \NPn^a = 1$ in
        coordinates, we get
        \begin{align}
            \begin{aligned}
                &\eu^\Wlambda \left(-\NPl^\Wv \NPn^\Wr + \NPl^\Wtheta
                \left(H_\pdWtheta\, \NPn^\Wr \right.\right.\\
                &\qquad\left.\left.{}- 4 \eu^{-2\Wnu}
                \massBH^2 \left(\eu^\Wlambda \NPn^\Wtheta 
                + \ii \NPn^\Wphi \sin \Wtheta\right)\right)\right) \horeq 1 \,,
            \end{aligned}
        \end{align}
        from which we can solve for $\NPl^\Wtheta$:
        \begin{equation}
            \NPl^\Wtheta \horeq \frac{\eu^{-\Wlambda} \left(1 + \eu^{\Wlambda}\, \NPl^\Wv \NPn^\Wr\right)}
            {H_\pdWtheta\, \NPn^\Wr - 4 \eu^{-2\Wnu} \massBH^2 \left(\eu^\Wlambda\, \NPn^\Wtheta
            + \ii\, \NPn^\Wphi \sin \Wtheta\right)} \,.
        \end{equation}
        The second normalization condition $\NPm_a \cconj{\NPm}^a = -1$ yields
        \begin{equation}
            -8\eu^{2\Wlambda - 2\Wnu} \massBH^2 \cconj{\NPm}^\Wtheta \NPm^\Wtheta 
            \horeq -1 \,.
        \end{equation}
        We know that the vector $\NPm^a$ is complex, but in analogy with
        Schwarzschild and Kerr solutions, let us suppose that $\NPm^\Wphi$
        is complex while $\NPm^\Wtheta$ is real. This is without loss of
        generality, since we can use spin rotation to set one of the
        components to be real. 
        Then, we can find the solution for $\NPm^\Wtheta$ to be
        \begin{equation}
            \NPm^\Wtheta \horeq \cconj{\NPm}^\Wtheta \horeq 
            \frac{\eu^{-\Wlambda + \Wnu}}{2\sqrt{2}\massBH} \,.
            \label{eq:axial_m_theta_on_h}
        \end{equation}

        All the other \textbf{contractions} between the vectors are supposed to be zero.
        From $\NPl_a \cconj{\NPm}^a = 0$, we get
        \begin{equation}
            - \frac{2\sqrt{2}\, \eu^{\Wnu} \massBH \left(1 + \eu^{\Wlambda}\, 
            \NPl^\Wv \NPn^\Wr\right)}{\eu^{2\Wnu}\,
            H_\pdWtheta\,\NPn^\Wr - 4 \massBH^2 \left(\eu^\Wlambda\,\NPn^\Wtheta 
            + \ii\, \NPn^\Wphi \sin\Wtheta\right)} \horeq 0 \,,
        \end{equation}
        and we can deduce
        \begin{equation}
            \NPn^\Wr \horeq - \frac{\eu^{-\Wlambda}}{\NPl^\Wv} \,.
        \end{equation}

        Similarly, from $\NPn_a \NPm^a = 0$, we find
        \begin{equation}
            \begin{aligned}
                &-\frac{\eu^{-\Wlambda+\Wnu}\,H_\pdWtheta}{2\sqrt{2}\,\NPl^\Wv}
                + \frac{\NPm^\Wv}{\NPl^\Wv} \\
                &\qquad{}- \sqrt{2}\eu^{-\Wnu}\massBH
                \left(\eu^\Wlambda\, \NPm^\Wtheta + \ii\, \NPn^\Wphi \sin\Wtheta\right) 
                \horeq 0 \,,
            \end{aligned}
        \end{equation}
        and
        \begin{align}
            \begin{aligned}
                \NPn^\Wtheta &\horeq \frac{\eu^{-2\Wlambda}}{4\massBH^2 \NPl^\Wv} \Big(-\eu^{2\Wnu}\,
            H_\pdWtheta + 2 \eu^\Wlambda \massBH \big(\sqrt{2}\, \eu^\Wnu \NPm^\Wv \\ 
                &\qquad{}- 2 \ii \,\massBH \NPl^\Wv \NPn^\Wphi \sin\Wtheta \big)\!\Big) .
            \end{aligned}
        \end{align}

        The last contraction is $\NPn_a \cconj{\NPm}^a = 0$ and yields
        \begin{equation}
            \frac{1}{\NPl^\Wv}\left(\cconj{\NPm}^\Wv - \NPm^\Wv 
            + 2 \ii \sqrt{2} \eu^{-\Wnu} \massBH \NPl^\Wv \NPn^\Wphi\sin\Wtheta\right) \horeq 0 \,.
        \end{equation}
        It follows that
        \begin{equation}
            \NPn^\Wphi \horeq \frac{\ii\, \eu^\Wnu \left(\cconj{\NPm}^\Wv 
            - \NPm^\Wv\right)}{2\sqrt{2}\, \NPl^\Wv \sin\Wtheta} \,.
            \label{eq:n_phi_component}
        \end{equation}

        We can check that $\NPm^a$ given by Eqs.~\eqref{eq:axial_m_radial_on_h},
        \eqref{eq:axial_m_parts_on_h} and \eqref{eq:axial_m_theta_on_h}
        satisfies the condition of being Lie dragged,
        $\Lie_{\boldsymbol{\NPl}} \NPm^a \horeq 0$, thanks stationarity and
        independence of $\NPl^a$ on $\Wtheta$.

        We can use a similar approach as for Lie dragging to compute the
        \textbf{spin coefficients}. While some of them do not contain any
        radial derivatives (on the horizon), others do and cannot be expressed
        without the knowledge of the full solution. Let us start by
        dividing the coefficients into these two groups. The result is quite
        simple: the coefficients $\Stau$, $\Sgamma$, and $\Snu$ cannot be
        computed while all the other ones can.

        First, we are interested in the coefficients $\Srho$ and $\Ssigma$ 
        which are connected to the expansion, shear, and twist of the vector 
        $\NPl^a$ and, hence, have to be zero. Moreover, $\Skappa$ also needs
        to vanish on the horizon. Using the proposed tetrad, we find that 
        they are indeed zero.

        Similarly, the spin $\Smu$ has to be real, and it can be shown that 
        it indeed is (we do not present it here due to its length).

        We also have to satisfy the condition
        \begin{equation}
            \Spi = \Salpha + \cconj{\Sbeta} \,.
        \end{equation}
        If we use our tetrad candidate, we find
        \begin{equation}
            \Spi - \left(\Salpha + \cconj{\Sbeta}\right) \horeq 
            \frac{\eu^{-\Wlambda + 2\Wnu} \left(\cconj{\NPm}^\Wv 
            - \NPm^\Wv\right)}{8\massBH} = 0 \,.
        \end{equation}
        We can thus deduce that $\NPm^\Wv$ component is real. This also has
        other consequences, most importantly~\eqref{eq:n_phi_component}
        gives that $\NPn^\Wphi = 0$.

        Next, let us compute the spin coefficient $\Sepsilon$, which also gives
        the \textbf{surface gravity} by
        $\surfkappa = \Sepsilon + \cconj{\Sepsilon}$. We find that
        \begin{equation}
            \Sepsilon \horeq \frac{\eu^{-\Wlambda + 2\Wnu}\, \NPl^\Wv}{8\massBH} \,.
            \label{eq:condition_on_lv}
        \end{equation}
        This has to match the correct surface gravity corresponding
        to the metric~\eqref{eq:weyl_superimposed}. Although there is no
        universal definition for surface gravity, if the horizon is a Killing
        horizon, which is our case, the surface gravity is given by
        \begin{equation}
            K^a \nabla_a K^b = \surfkappa K^b \,, 
        \end{equation}
        \remove{where $K^a=\pd_\Wv$ is the Killing vector that generates the horizon.}
        \issueD{where $K^a$ is the Killing vector that generates the horizon.
        This still does not uniquely determine the surface gravity
        because of the possibility of rescaling of the Killing vector. However, in
        a static asymptotically flat space-time, it is common to require the
        magnitude of the Killing vector to be unit at infinity, \cite{Fairhurst-2001}.
        Unlike for a general isolated horizon, the
        coordinates~\eqref{eq:weyl_superimposed} also cover the infinity
        where the potentials $\Wlambda$ and $\Wnu$ must go to zero. This
        suggests the choice $K^a=\pd_\Wv$.
        Moreover, the metric~\eqref{eq:weyl_superimposed} includes the
        Schwarzschild solution as a special case and it is
        natural to choose the surface gravity in such a way that Schwarzschild
        case has the expected surface gravity. This again supports fixing
        $K^a=\pd_\Wv$.
        }

        For the metric~\eqref{eq:weyl_superimposed}, the resulting surface
        gravity is
        \begin{equation}
            \surfkappa = \frac{\eu^{-\Wlambda + 2\Wnu}}{4\massBH} \,. 
            \label{eq:surface_gravity}
        \end{equation}
        Comparing~\eqref{eq:condition_on_lv} (recall that
        $\surfkappa = 2 \Sepsilon$) and~\eqref{eq:surface_gravity} 
        we should choose $\NPl^\Wv = 1$.
        \issueD{Were we to choose a different surface gravity (scaled by a 
        constant) we would arrive at a rescaled $\NPl^a$.
        This is consistent with the surface gravity of an isolated
        horizon depending on the choice of the normal $\NPl^a$.}

        This implies that $\Sepsilon$ is real and due to%
        ~\eqref{eq:lambda_nu_einstein_eq_hor}
        we also get that the surface gravity is constant on the horizon
        \begin{equation}
            \surfkappa = \frac{\eu^{2\Wnu_\text{pole}}}{4\massBH} \,. 
            \label{eq:surface_gravity_2}
        \end{equation}

        The choice of $\NPm^\Wv$ is equivalent to rotating the tetrad about
        the vector $\NPl^a$, given by~\eqref{app:np:eq:rotation_l}.
        In consistency with~\cite{Krishnan-2012}, we choose the vectors
        $\NPm^a$ (and $\cconj{\NPm}^a$) to be tangent to the spheres
        $\mathcal{S}_\Wv$ and therefore, $\NPm^\Wv \horeq 0$.

        In the Bondi-like coordinates of~\cite{Krishnan-2012}, the coordinates
        are propagated off the horizon by
        \begin{equation}
            \NPDelta \Kv = \NPDelta \Kx^I = 0 \,.
        \end{equation}
        To have our coordinates identical to the
        Krishnan ones on the horizon, we have to choose $H_\pdWtheta = 0$.

        At this point, we have already obtained the correct tetrad on the horizon
        (i.e., it fulfills all geometrical properties imposed by the IH framework),
        but it is still not manifestly in the form given by
        Eq.~\eqref{eq:general_tetrad}, since the coordinates ($\Wv$, $\Wr$,
        $\Wtheta$, $\Wphi$) of the Weyl metric~\eqref{eq:weyl_star_metric} 
        are not the
        natural coordinates of IH formalism ($\Kv$, $\Kr$, $\Ktheta$, $\Kphi$).
        It would be difficult to find the coordinate transformation
        $\Wr=\Wr(\Kr,\Ktheta)$, $\Wtheta=\Wtheta(\Kr,\Ktheta)$
        since we would have to find the inverse function of an integral
        depending on $\Wlambda$. This could, in theory, be done for one
        particular $\Wlambda$ (although they tend to be very complex).
        We shall bypass this issue in Sec.~\ref{sec:singular_ode}
        by constructing $\NPn^a$ from a solution to the geodesic equation
        $\NPDelta \NPn^a = 0$ and taking its affine parameter as the correct
        radial coordinate $\Kr$ while also obtaining the coordinate $\Ktheta$
        assigned to each geodesic ray.

        For a \textbf{stationary and axially symmetric horizon},
        the surface gravity must also satisfy
        \begin{equation}
            \cconj{\NPdelta} \Spi - \cconj{\Sa}\Spi
            + \Spi^2 \horeq \surfkappa \Slambda \,.
            \label{eq:stataxi_pi}
        \end{equation}
        This condition from~\cite{Lewandowski-2003} was in our context
        of non-extremal horizons discussed in~\cite{Guerlebeck-2017}.
        It is a consequence of the integrability condition assuring the
        existence of a Killing vector
        \begin{equation}
            \nabla_c \nabla_a K_b = R_{abcd} K^d \,.
        \end{equation}
        When written in the NP formalism for the natural choice of
        $K^a = \NPl^a$ and restricted to the horizon, it gives%
        ~\eqref{eq:stataxi_pi}.
        For our tetrad candidate, Eq.~\eqref{eq:stataxi_pi} is already
        satisfied, which can be checked by rewriting the equation in coordinates
        (using partial derivatives of $\Wlambda$ and $\Wnu$) and
        employing Einstein equations~\eqref{eq:lambda_nu_einstein_eq_hor}.

    \subsection{Tetrad propagated from the horizon}
        \label{sec:singular_ode}
        So far, we have discussed the situation on the horizon itself.
        To get information about the outside, we use Eq.%
        ~\eqref{eq:off_horizon_transport}, which is an affinely parametrized
        geodesic equation.
        The other vectors, and all other quantities, will be transported
        from the horizon along the geodesic congruence in the direction
        of $\NPn^a$.

        In Sec.~\ref{sec:tetrad_on_the_horizon}, we used 
        transformation~\eqref{eq:star_transformation} to regularize the
        horizon and to find the tetrad. However, since the function
        $F(\Wr, \Wtheta)$ would be an unnecessary quantity outside the horizon,
        we revert to the coordinates of~\eqref{eq:weyl_superimposed},
        despite their singularity on the horizon. 
        In addition to causing some differential equations to become singular
        on the horizon and leading to diverging components of the tetrad, it
        also has more subtle consequences, which we will discuss later in this
        section.

        Let us assume that the tetrad is in the form of a series
        in the affine parameter, given by the Krishnan radial coordinate
        $\Kr$, around the horizon ($\Kr = 0$)
        \begin{align}
                \NPl^a &= \NPl_0^a + \NPl_1^a\,\Kr + \frac{1}{2}\, \NPl_2^a\,\Kr^2 
                    + \dots \,, \\
                \NPm^a &= \NPm_0^a + \NPm_1^a\,\Kr + \frac{1}{2}\, \NPm_2^a\,\Kr^2 
                    + \dots \,. 
        \end{align}
        For the vector $\NPn^a$, we also need to include a term proportional to
        the negative power of the affine parameter:
        \begin{align}
            \NPn^a = \NPn_{-1}^a\,\frac{1}{\Kr} + \NPn_0^a + \NPn_1^a\,\Kr 
            + \frac{1}{2}\, \NPn_2^a\,\Kr^2 + \dots \,.
        \end{align}
        This term stems from the singularity and can be seen in the transformation 
        between the coordinates $\Wt$ and $\Wv$ which includes $1/\Schf$.
        Note that for the series expansion, we work in the Weyl coordinates and
        $\NPl^a = (\NPl^\Wt, \NPl^\Wr, \NPl^\Wtheta, \NPl^\Wphi)$, rather than
        $a$ being an abstract index.

        To find the coefficients of the series, we will use the equation for
        parallel transport along the vector $\NPn^b$:
            \begin{equation}
                \NPDelta \eta^a = \NPn^b \nabla_b \eta^a
                = - \frac{\d \eta^a}{\d \Kr} 
                + \ChGamma^a{}_{bc}\,\NPn^b \eta^c = 0 \,, 
                \label{eq:geodesics:parallel_transport}
            \end{equation}
        and the contractions between the vectors of the tetrad, the only 
            non-zero being
            \begin{align}
                \NPl^a \NPn_a &= 1\,, & \NPm^a \cconj{\NPm}_a &= -1 \,.
                \label{eq:geodesics:contractions}
            \end{align}

        Similarly to the tetrad itself, the coordinates, obtained by solving
        \footnote{The minus sign here denotes that while we have outgoing 
            coordinates (to give $\Wr$ a physical sense), we have, on the 
            other hand, ingoing vector $\NPn^a$ (to be consistent 
            with~\cite{Krishnan-2012}).}
        \begin{equation}
            \NPn^a = -\frac{\d \Kx^a}{\d \Kr} \,
        \end{equation}
        are also expanded:
        \begin{subequations}
            \begin{align}
                \Wt &= \Wt_0 + \Wt_1 \Kr + \dots \,, \\
                \Wr &= 2\massBH + \Wr_1 \Kr + \dots \,, \\
                \Wtheta &= \Ktheta + \Wtheta_1 \Kr + \dots \,, \\
                \Wphi &= \Wphi_0 + \Wphi_1 \Kr + \dots \,,
            \end{align}
        \end{subequations}
        Coefficients $\Wr_1$ and $\Wtheta_1$ correspond to (minus) the zeroth order 
        of the vector $\NPn^a$ and cannot be found from the equations%
        ~(\ref{eq:geodesics:parallel_transport}--\ref{eq:geodesics:contractions}).
        However, they are given by the on-horizon values of $\NPn^\Wr$ and
        $\NPn^\Wtheta$, hence,
        \begin{align}
            \begin{aligned}
                \Wr_1 &= \eu^{-\Wlambda} \,, \\
                \Wtheta_1 &= 0 \,.
            \end{aligned}
        \end{align}

        We shall start with the \textbf{vector} $\boldsymbol{\NPn^a}$ since we
        are transporting other vectors along this one. If we want to get the
        vector up to the order $j$, we can do it by solving the orders
        $-2,\dots, j-1$ of the geodesic equation and orders
        $0, \dots, j$ of the normalization condition.
        We find that up to the first order, the vector is
        \begin{subequations}
            \begin{align}
                \begin{split}
                    \NPn^\Wt &= 2 \eu^{\Wlambda - 2\Wnu} \massBH \Kr^{-1} 
                    +  \eu^{-2\Wnu} \left(1 + \massBH \Wlambda_\pdWr
                    - 4 \massBH \Wnu_\pdWr\right) \\
                        &\qquad{}
                        - \frac{\eu^{-\Wlambda - 2\Wnu}}{6} \Big(12 \Wnu_\pdWr
                        + 4 \Wnu_\pdWtheta \left(\Wlambda_\pdWrWtheta - 2\Wnu_\pdWrWtheta\right) 
                        \\
                        &\qquad{}
                        + \massBH \left(\Wlambda_\pdWr{}^2 - 24\Wnu_\pdWr{}^2
                        - 2\Wlambda_\pdWrWr + 12\Wnu_\pdWrWr\right)\!\Big) \Kr \\
                        &\qquad{}+ \bigo(\Kr^2) \,, 
                \end{split} \\
                \NPn^\Wr &= -\eu^{-\Wlambda} 
                    + \eu^{-2\Wlambda} \Wlambda_\pdWr \,\Kr 
                    + \bigo(\Kr^2) \,, \\
                \NPn^\Wtheta &= - \frac{\eu^{-2\Wlambda}\left(
                \Wlambda_\pdWrWtheta - 2\Wnu_\pdWrWtheta\right)}{2\massBH} \Kr 
                    + \bigo(\Kr^2) \,, \\
                \NPn^\Wphi &= 0 \,. 
            \end{align}
        \end{subequations}
        For the component $\NPn^\Wphi$ the solution would give $\bigo(\Kr^2)$
        but from the space-time symmetries, we know that it vanishes.
        Although we construct the solution outside the horizon, the values of
        $\Wlambda, \Wnu$ and their derivatives appearing in the series
        expansions here are those at the horizon, for example,
        $\Wnu \equiv \Wnu(\Wr=2\massBH,\Wtheta)$.

        For the \textbf{vector} $\boldsymbol{\NPl^a}$ we yet again need to investigate orders
        $-2, \dots, j-1$ of the transport equation and
        $0, \dots, j$ of the normalization condition, but, moreover, we need 
        orders $0, \dots, j$ of the contraction with vector $\NPn^a$. 
        We must also set $\NPl^{\Wphi}_0 = 0$ taking into account the space-time
        symmetries.
        Up to the first order, we find
        \begin{subequations}
            \begin{align}
                \NPl^\Wt &= \frac{1}{2} + \frac{\eu^{-\Wlambda} 
                    \Wnu_\pdWtheta{}^2}{4\massBH} \Kr 
                    + \bigo(\Kr^2) \,, \\
                \NPl^\Wr &= \frac{\eu^{-2\Wlambda + 2\Wnu}}{4\massBH} \Kr 
                    + \bigo(\Kr^2)  \,, \\
                \NPl^\Wtheta &= \frac{\eu^{-2\Wlambda 
                    + 2\Wnu} \Wnu_\pdWtheta}{4\massBH^2} \Kr 
                    + \bigo(\Kr^2) \,, \\
                \NPl^\Wphi &= 0 \,. 
            \end{align}
        \end{subequations}
        The entire component $\NPl^\Wphi$ is zero due to symmetries similarly to
        $\NPn^\Wphi$.

        In the case of the \textbf{vector} $\boldsymbol{\NPm^a}$, we need the orders
        $-2, \dots, j-1$ of the transport equation 
        and $0, \dots, j-1$ of the normalization
        and $j$ of the contraction with $\NPn^a$. Up to the 
        first order, we find
        \begingroup
        \begin{subequations}
        \begin{align}
            \begin{split}
                \NPm^\Wt &=  \frac{\eu^{-\Wnu} \Wnu_\pdWtheta}{\sqrt{2}} 
                    + \frac{\eu^{-\Wlambda-\Wnu}}{4\sqrt{2}\massBH}
                    \Big(\Wnu_\pdWtheta \left(1 - 6 \massBH \Wnu_\pdWr\right) \\
                    &\qquad{}- \massBH \left(\Wlambda_\pdWrWtheta 
                    - 4\Wnu_\pdWrWtheta\right)\!\Big) \Kr 
                    + \bigo(\Kr^2) \,, 
            \end{split}
            \\
            \begin{split}
                \NPm^\Wr &=  - \frac{\eu^{-2\Wlambda + \Wnu} \Wnu_\pdWtheta}
                    {2 \sqrt{2} \massBH} \Kr 
                    + \bigo(\Kr^2) \,, 
            \end{split}
            \\
            \begin{split}
                \NPm^\Wtheta &= \frac{\eu^{-\Wlambda + \Wnu}}{2\sqrt{2} \massBH} 
                    - \frac{\eu^{-\Wlambda + \Wnu}}{4\sqrt{2} \massBH^2} 
                    \left(1 + 2\massBH \Wlambda_\pdWr - 2\massBH \Wnu_\pdWr\right) \Kr \\
                    &\qquad{}+ \bigo(\Kr^2) \,, 
            \end{split}
            \\
            \begin{split}
                \NPm^\Wphi &= \frac{\ii \eu^\Wnu}{2\sqrt{2}\massBH \sin\Ktheta} 
                    - \frac{\ii \eu^{-\Wlambda + \Wnu} 
                    \left(1 - 2\massBH \Wnu_\pdWr\right)}{4 \sqrt{2} \massBH^2 \sin\Ktheta} \Kr \\
                    &\qquad{}+ \bigo(\Kr^2) \,. 
            \end{split}
        \end{align}
        \end{subequations}
        \endgroup

        In Secs.~\ref{sec:coordinates_and_tetrad}--\ref{sec:spherical_symmetry},
        and \ref{sec:tetrad_on_the_horizon}, both $\NPl^\Kv = 1$ and
        $\NPl^\Wv(\Wr=2\massBH) = 1$. Thus, it can be puzzling
        why the presented series has $\NPl^\Wt(\Wr=2\massBH) = 1/2$.
        As we have already mentioned, coordinates ($\Wt$, $\Wr$, $\Wtheta$,
        $\Wphi$) are singular on the horizon, and the inverse of the transformation%
        ~\eqref{eq:star_transformation} changes the tetrad found in Sec.%
        ~\ref{sec:tetrad_on_the_horizon} in such a way that
        while only $\NPl_\Wv \NPn^\Wv=1$ is non-zero in the scalar
        product $\NPl_a \NPn^a$ on the horizon, in the singular
        coordinates of~\eqref{eq:weyl_superimposed}, both $\NPl_\Wt \NPn^\Wt$ and
        $\NPl_\Wr \NPn^\Wr$ approach $1/2$.


        Using the technique from Sec.~\ref{sec:singular_ode} we can also find
        the series expansion of the \textbf{scalars} such as spin coefficients
        and components of the Weyl tensor. However, for the spin coefficients 
        we also need to know the derivatives of the vectors themselves.

        Let us demonstrate the procedure on the evaluation of the spin
        coefficient $\Sepsilon$.
        Since $\Sepsilon$ gives the surface gravity ($\surfkappa = \Sepsilon
        + \cconj{\Sepsilon}$), we can check that after the transformation into
        metric~\eqref{eq:weyl_superimposed} we still get the correct surface
        gravity derived for metric~\eqref{eq:weyl_star_metric}.
        Moreover, it is a nice example of the fact that while the surface
        gravity is a quantity intrinsic to the horizon, it is \emph{not}
        possible to compute $\surfkappa$ using the on-horizon terms of 
        the series only.

        First, we assume a general tetrad whose components are functions 
        of $\Wr$ and $\Wtheta$. Using the definition
        \begin{equation}
            \Sepsilon = \frac{1}{2}\left(\NPn^a \NPl^b \nabla_b \NPl_a
            - \cconj{\NPm}^a \NPl^b \nabla_b \NPm_a\right) 
        \end{equation}
        and the Christoffel symbols $\ChGamma^a{}_{bc}$ corresponding to the
        metric~\eqref{eq:weyl_superimposed} we can find a general expression.
        Then, we replace each
        component of the vectors by its series as presented in
        Sec.~\ref{sec:singular_ode} and perform the series expansion of the
        coordinates. Moreover, as indicated above, we need to find the
        expansion for the derivatives.

        To find the derivatives of the components of the tetrad, let us
        reiterate about the coordinates. Our series expansions of the tetrad are
        in terms of the Krishnan coordinates $\Kr, \Ktheta$ ($\Ktheta$ is
        constant along each of the geodesics):
        \begin{equation}
            \eta^a = \eta^a_0(\Ktheta) + \eta^a_1(\Ktheta) \Kr + \dots \,.
        \end{equation}
        while we need the derivatives in the coordinates $\Wr, \Wtheta$
        given by
        \begin{align}
            \Wr &= \Wr(\Kr, \Ktheta)\,,& \Wtheta &= \Wtheta(\Kr, \Ktheta) \,.
            \label{eq:coor_from_krishnan}
        \end{align}
        We can find the derivatives as
        \begin{align}
            \begin{aligned}
                \frac{\pd \eta^a}{\pd \Wr} &= \frac{\pd \eta^a}{\pd \Kr}\frac{\pd \Kr}{\pd \Wr}
                + \frac{\pd \eta^a}{\pd \Ktheta}\frac{\pd \Ktheta}{\pd \Wr} \,, \\
                \frac{\pd \eta^a}{\pd \Wtheta} &= \frac{\pd \eta^a}{\pd \Kr}\frac{\pd \Kr}{\pd \Wtheta}
                + \frac{\pd \eta^a}{\pd \Ktheta}\frac{\pd \Ktheta}{\pd \Wtheta} \,.
            \end{aligned}
        \end{align}
        Factors of the type ${\pd \Kr}/{\pd \Wr}$ can be found from
        the inverse of the Jacobian from~\eqref{eq:coor_from_krishnan}.

        After formally expanding all the quantities we shall substitute the
        values of the coefficients from Sec.~\ref{sec:singular_ode}. Inspecting
        the zeroth order for the spin coefficient $\Sepsilon$ (for which we need
        $\pd_\Wr \NPl^\Wr$ and $\pd_\Wtheta \NPl^\Wr$) we find
        \begin{align}
            \Sepsilon &= \frac{\eu^{-\Wlambda + 2\Wnu}}{8\massBH} 
            - \frac{\eu^{-2\Wlambda+2\Wnu}\left(1 - 2\Wnu_\pdWtheta{}^2
            + \massBH \left(\Wlambda_\pdWr - 4 \Wnu_\pdWr\right)\right)}{8 \massBH^2} \Kr 
            \nonumber \\
            &\qquad{}+ \bigo(\Kr^2) \,,
        \end{align}
        which yields the correct surface gravity~\eqref{eq:surface_gravity}.

        We can compute all the other spin coefficients in the same way, and
        it is easy enough to check that they do satisfy the conditions
        required by isolated horizons as stated by~\cite{Krishnan-2012},
        e.g.\ we can check that $\Smu$ is real on the horizon, or that 
        $\Stau = \Sgamma = \Snu = 0$.

        As an example of a non-trivial spin coefficient, we can show
        $\Ssigma$, which is moreover connected to the shear of the geodesic
        congruence. We get that
        \begin{align}
            \begin{aligned}
                \Ssigma &= \frac{\eu^{-2\Wlambda + 2\Wnu} \left(
                    \cot\Ktheta\,
                    \Wnu_\pdWtheta - \Wnu_\pdWtheta{}^2 
                    - \Wnu_\pdWthetaWtheta - \massBH \Wlambda_\pdWr\right)}
                    {8\massBH^2} \Kr \\
                    &\qquad{}+ \bigo(\Kr^2)\,.
            \end{aligned}
        \end{align}
        We can see that it is zero on the horizon and, unsurprisingly,
        outside the horizon it generally does not vanish since the only such
        space-time is the pure Schwarzschild.

        The series expansions of all spin coefficients and Weyl scalars can be 
        found in Appendix~\ref{app:scalar_expansions}.
        For isolated horizons, $\NPPsi_4$ must be specified as part of the
        initial data on the null hypersurface $\NN$. The one obtained by the
        envisaged procedure and given approximately in Appendix%
        ~\ref{app:scalar_expansions} is the particular one that evolves into
        a static space-time.

\section{Example: Horizon deformed by a disk}
    \label{sec:particular_solution}
    In this section, as an example to illustrate the obtained results, 
    we consider a black hole surrounded by a finite thin disk. It is
    remarkable that for a Weyl space-time modelling this configuration,
    both $\Wlambda$ and $\Wnu$ can be found in analytic form, \cite{Kofron-2023}.
    While we could start with an arbitrary horizon geometry, as will be
    discussed in Sec.~\ref{sec:static_solution_geometry}, considering%
    ~\cite{Kofron-2023} allowed us to relate the observed behavior to a known
    matter source near the black hole.

    We suppose that the disk mass $\massDisk = \massBH / 2$ and that the
    parameter $\diskR$ defining the inner edge of the disk is $\diskR = 3\massBH$ which 
    thanks to~\eqref{eq:inner_disk_radius} translates to 
    $\Wr = \left(1 + \sqrt{10}\right)\massBH \approx 4.16 \massBH$. 
    Although this choice is quite extreme since the inner edge of the disk is
    under the innermost stable circular orbit, it is well above the
    photon orbit. See Appendix~\ref{app:disk} for an overview of the disk
    and the reasoning for the choice of the radius of its inner edge.

    The following figures are plotted in the coordinates $\WEx$ and $\WEz$
    given by%
    \footnote{Notice the hat that ensures the distinction of $\WEz$ and $\Wz$
    which, while having the same orientation, differ in magnitude.}
    \begin{align}
        \WEx &= \Wr \sin\Wtheta\,, & \WEz &= \Wr \cos\Wtheta \,,
    \end{align}
    and produced using a numerical solution to the problem rather than using
    the analytical series expansions. For a commentary on how it was done, the
    numerical precision and comparison with the analytical expansion
    see Appendix~\ref{app:numerical_solution}.

\begin{figure}[ht!]
        \centering
        \figinput{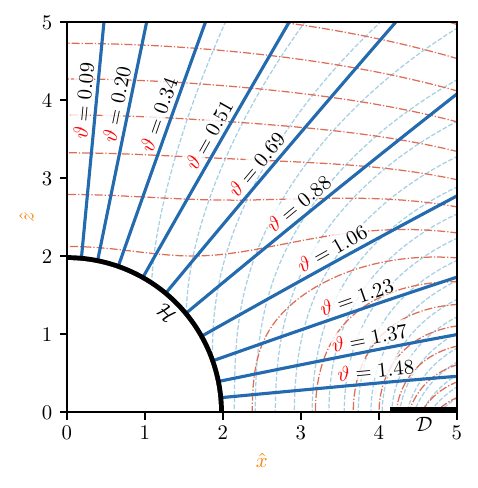}
        \caption[Geodesics lines resolved by the numerical solver.]
        {The solid blue lines represent some of the geodesic lines resolved
        by the numerical solver by starting from the horizon $\HH$ at the
        angle $\Ktheta$ ($\horeq\, \Wtheta$) as shown next to the line.
        Light blue dashed lines represent the potential $\Wlambda$
        and light red dash-dotted lines show $\Wnu$. The thick black line at
        $\WEz=0$ is the disk $\DD$.
        }
        \label{fig:geodesics}
    \end{figure}

\begin{figure}[ht!]
        \centering
        \figinput{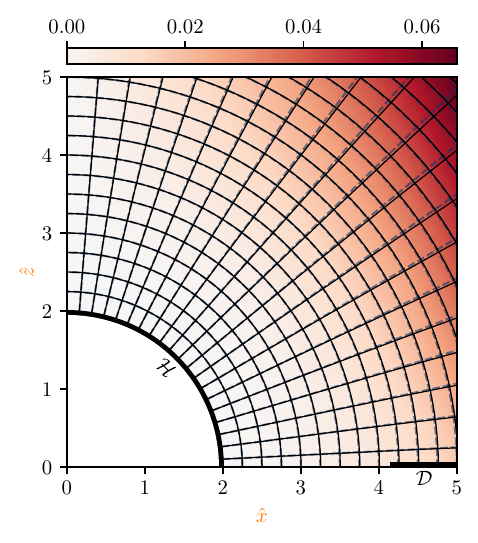}
        \caption[The Krishnan coordinates $\Kr$ and $\Ktheta$ in the region
        between the black hole and the disk]{The Krishnan coordinates $\Kr$ 
        and $\Ktheta$ in the region
        between the black hole and the disk are plotted in solid black lines
        while light blue dashed lines are the Weyl coordinates $\Wr$ and
        $\Wtheta$. The difference between the two coordinate systems, 
        measured by 
        $\sqrt{\left(\WEx - \KEx\right)^2 + \left(\WEz - \KEz\right)^2}$
        , is shown by the saturation of the background.}
        \label{fig:coordinates}
    \end{figure}

    Since the tetrad, which is paramount in the isolated horizon formalism,
    is propagated parallelly along the vector $\NPn^a$, let us start by showing,
    in Fig.~\ref{fig:geodesics}, the corresponding geodesic congruence 
    together with the Weyl potentials $\Wlambda$ and $\Wnu$.
    While it is not obvious from the figure, the geodesics are, of course,
    deformed by the presence of the disk, the effect can be seen in greater detail in Fig.%
    ~\ref{fig:geodesics_at_disk}. Even the difference between the Weyl 
    coordinates $\WEx$ and $\WEz$ and the Krishnan equivalents $\KEx$ and $\KEz$,
    which have a similar definition except for a correction of $2\massBH$ which
    removes the difference of choice of the value at the horizon,
    \begin{align}
        \KEx &= (\Kr + 2\massBH) \sin\Ktheta\,, & \KEz &= (\Kr + 2\massBH) \cos\Ktheta \,,
    \end{align}
    is only evident further away from the black hole, as can be seen in Fig.%
    ~\ref{fig:coordinates}.

    \begin{figure}
        \centering
        \figinput{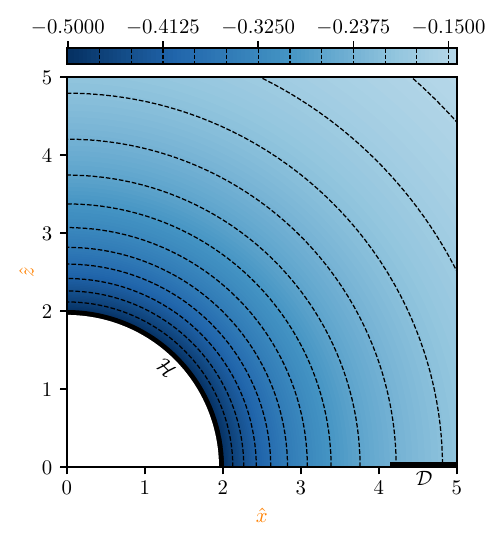}
        \caption[Spin coefficient $\Smu$.]{Spin coefficient $\Smu$ 
        (expansion of $\NPn^a$) in the region between the black hole and 
        the disk.}
        \label{fig:mu_contour}
    \end{figure}
    \begin{figure}
        \centering
        \figinput{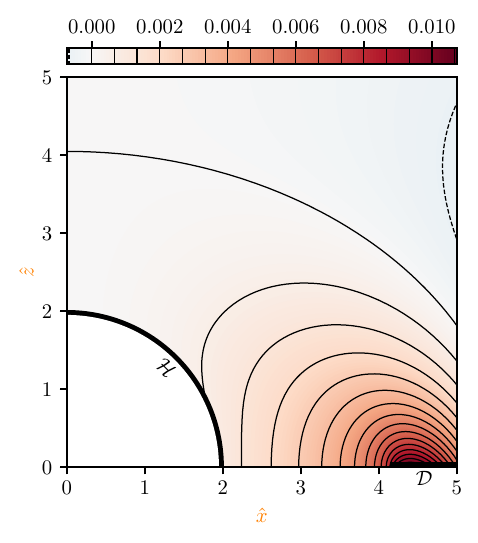}
        \caption[Spin coefficient $\Slambda$.]{Spin coefficient $\Slambda$ 
        (shear of $\NPn^a$) in the region between the black hole and the
        disk.}
        \label{fig:lambda_contour}
    \end{figure}
    \begin{figure}
        \centering
        \figinput{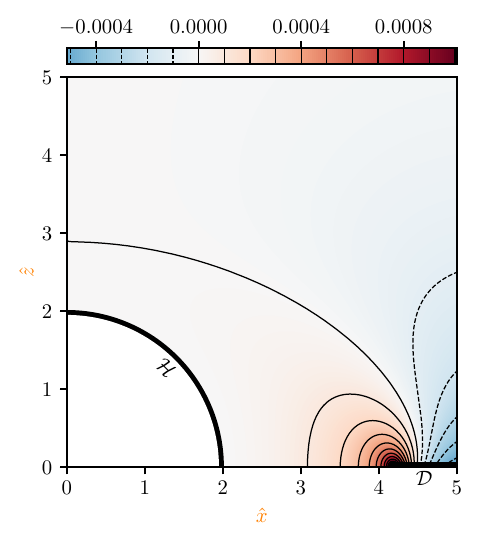}
        \caption[$\NPPsi_0$ in the region between the black hole and the
        disk.]{$\NPPsi_0$ in the region between the black hole and the
        disk.}
        \label{fig:Psi0_contour}
    \end{figure}
    \begin{figure}
        \centering
        \figinput{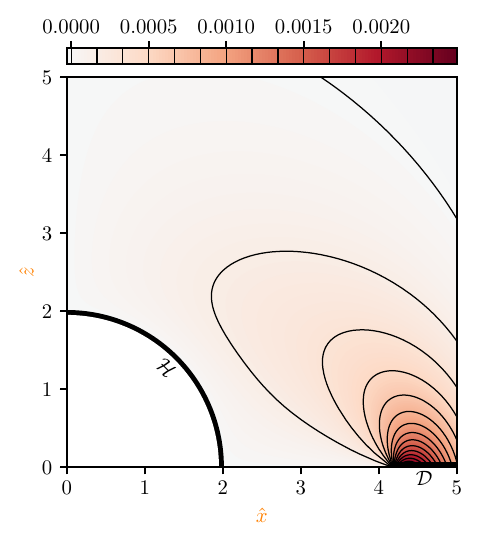}
        \caption[$\NPPsi_1$ in the region between the black hole and the
        disk.]{$\NPPsi_1$ in the region between the black hole and the
        disk.}
        \label{fig:Psi1_contour}
    \end{figure}
    \begin{figure}
        \centering
        \figinput{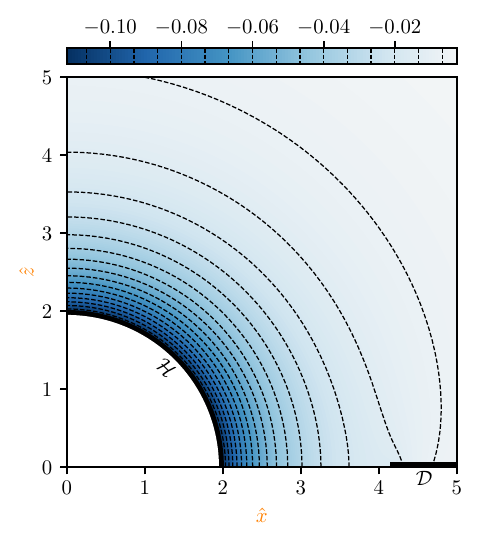}
        \caption[$\NPPsi_2$ in the region between the black hole and the
        disk.]{$\NPPsi_2$ in the region between the black hole and the
        disk.}
        \label{fig:Psi2_contour}
    \end{figure}
    \begin{figure}
        \centering
        \figinput{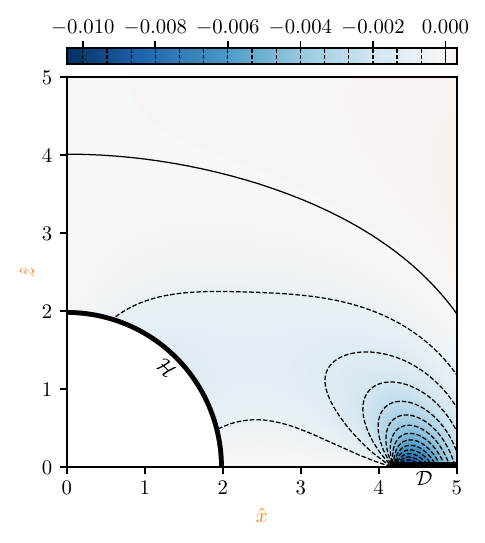}
        \caption[$\NPPsi_3$ in the region between the black hole and the
        disk.]{$\NPPsi_3$ in the region between the black hole and the
        disk.}
        \label{fig:Psi3_contour}
    \end{figure}
    \begin{figure}
        \centering
        \figinput{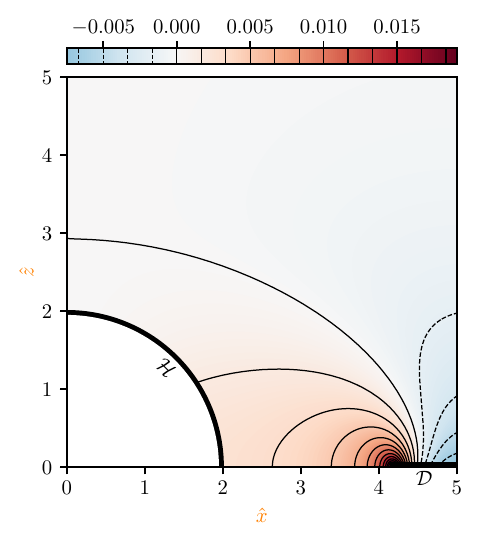}
        \caption[$\NPPsi_4$ in the region between the black hole and the
        disk.]{$\NPPsi_4$ in the region between the black hole and the
        disk.}
        \label{fig:Psi4_contour}
    \end{figure}
    \subsection{NP scalars}
        In the following figures, the saturation of contour shading corresponds
        to the magnitude, while blue regions (with dashed contours)
        have negative values and red ones (with solid contours) positive ones. 
        Zero saturation (white) denotes zero value (or the interior of the black
        hole). However, the (color) scale is not the same for all plots.

        Let us start with the spin coefficients $\Smu$ and $\Slambda$ in
        Figs.~\ref{fig:mu_contour} and \ref{fig:lambda_contour} which
        represent the expansion and shear of the null congruence generated by
        $\NPn^a$. Although the value is different, the profile of $\Smu$ is almost
        the same as in the Schwarzschild case. On the other hand, $\Slambda$
        is no longer vanishing, as with Schwarzschild, and its value grows in the
        vicinity of the inner edge of the disk, where the density of the disk is
        the highest.

        We shall proceed with the Weyl scalars, which can be found in Figs.%
        ~\ref{fig:Psi0_contour} to \ref{fig:Psi4_contour}. 
        The presence of the disk deforms the geometry in such a way that all
        Weyl scalars are non-vanishing, and the magnitude of their deviation is
        highest in the vicinity of the disk. In the pure Schwarzschild, we have
        only $\NPPsi_2=-\massBH/\Kr^3$.

\section{Analytic aspects of horizon geometry}
    \label{sec:static_solution_geometry}
    In the previous section, we used a particular solution%
    ~\eqref{app:eq:disk_potentials} of the Einstein 
    equations and solved the geodesic equations for null rays providing the null 
    tetrad vectors in this given space-time. By appropriate projections and 
    derivatives, we obtained the IH/NP quantities.  We already mentioned the 
    possibility that, because it is much easier to obtain $\Wnu$ as solution 
    of the linear Laplace equation
    than to find $\Wlambda$, we may, for a given $\Wnu$, solve%
    ~\eqref{eq:lambda_nu_einstein_eq:lambda_r} to get 
    $\Wlambda$ along the rays. Owing to%
    ~\eqref{eq:lambda_nu_einstein_eq_hor}, we know $\Wlambda$ on the horizon 
    which is the starting point of each ray.
    
    In the IH formalism, the space-time geometry is determined by
    the geometry of the horizon and a few additional fields. It would thus be
    nice if we could, instead of choosing a given solution $\Wnu$ of the Laplace
    equation \eqref{eq:einstein_in_weyl_form_laplace}, construct $\Wnu$ from the horizon
    geometry.

    Under the assumption that the space-time is static, i.e., there exist Weyl
    coordinates in which the Laplace equation \eqref{eq:einstein_in_weyl_form_laplace} 
    for $\Wnu$ holds. For this field equation, the horizon geometry then becomes
    the boundary condition. With such a construction, we can determine all 
    space-time quantities from the horizon geometry. Indeed, this approach 
    assumes time and axial symmetries of the space-time, which not only lead
    to the Laplace equation for $\Wnu$ but also determine $\NPPsi_4$ everywhere 
    (and thus also on $\NN_0$).

    If the horizon geometry is given by the constant $\EuclidR$ and the function 
    $h(\Kzeta)$ appearing in~\eqref{eq:metric_axi_ih} we may introduce the 
    coordinate $\Wtheta$ so that $\Kzeta=\cos\Wtheta$ and define 
    \begin{equation}
        \tilde \Wnu(\Wtheta)= -\frac{1}{2} \ln\!\left(
        \frac{h(\Kzeta)}{\sin^2\Wtheta}\right) .
    \end{equation}
    Conditions~\eqref{eq:axial_h_cond} fix the value of
    $\tilde \Wnu(\Wtheta=0) = \tilde \Wnu(\Wtheta=\Cpi)=0$.
    
    Then the $\Kv=\const$ section of the (Killing) horizon is a topological 
    sphere with
    \begin{equation}
        \d\mathscr{s}_{\szero}{}^2 = \EuclidR^2 \left( \eu^{2\hat\Wnu}  \d\Wtheta^2
        +\eu^{-2\hat\Wnu} \sin^2\Wtheta\, \d\Wphi^2\right).
    \end{equation}
    Owing to the assumed form of $\Schf=1-\frac{2\massBH}{\Wr}$ we put $\Wr=2\massBH$ at 
    the horizon and from~\eqref{eq:lambda_nu_einstein_eq_hor}, we get
    \begin{align}
        \begin{aligned}
            \eqc
            \Wlambda(\Wr\!=\!2\massBH,\Wtheta) &\eqc= 2\big(\Wnu(\Wr\!=\!2\massBH,\Wtheta) 
            -\Wnu_\text{pole}\big) = \hat \Wnu(\Wtheta) \,, \\
            \massBH &= \tfrac{1}{2}\EuclidR\, \eu^{\Wnu_\text{pole}} \,.
        \end{aligned}
    \end{align}
    The surface gravity is given by~\eqref{eq:surface_gravity_2}.
    Once we have determined $\massBH$ and the function 
    $\Wnu(\Wr=2\massBH,\Wtheta)$ on the horizon, we can write $\Wnu$ both at 
    and outside the horizon as
    \begin{equation}
        \Wnu(\Wr,\Wtheta) = \sum_{l} \Wnu_l\, P_l\!\left(\frac{\Wr}{\massBH}
        - 1\right) P_l(\cos\Wtheta) \,,
        \label{eq:nu_above_horizon}
    \end{equation}
    where $P_l$ are the Legendre polynomials and $l=0,2,4,\dots$. 
    This can be shown from the well-known form of the 
    inner multipole expansion of the regular solution of the Laplace equation in 
    the prolate spheroidal coordinates as well by a direct substitution into%
    ~\eqref{eq:lambda_nu_einstein_eq:laplace}. 
    Because Legendre polynomials satisfy $P_l(1)=1$, the coefficients $\Wnu_l$ 
    are given directly as the multipole expansion of the potential 
    $\Wnu(\Wr=2\massBH,\Wtheta)$ on the horizon
    \begin{equation}
        \Wnu(\Wr=2\massBH,\Wtheta) = \sum_l \Wnu_l P_l(\cos\Wtheta) \,.
    \end{equation}
    
    When we specify $\Wnu$ on the horizon, Eq.~\eqref{eq:nu_above_horizon}
    provides its extension outside the black hole. Obviously, it can hold only
    in the vacuum part of the space-time, and we will now demonstrate the
    imprint of the presence of matter on the analytic properties of the horizon
    geometry, which will be related to 
    the convergence radius of the expansion~\eqref{eq:nu_above_horizon}.
    Owing to its particular form, the values on the symmetry axis are directly
    related to those on the horizon
    \begin{equation}
        \Wnu(\Wr=2\massBH,\Wtheta) = \Wnu(\Wz=\massBH\cos\Wtheta,\Wrho=0) \,,
    \end{equation}
    where $\Wrho, \Wz$ are the cylindrical Weyl coordinates%
    ~\eqref{eq:frolov_transformation}. The convergence
    is obvious for $|\Wz|\le \massBH$, because the transformation%
    ~\eqref{eq:frolov_transformation}
    maps the horizon on the linear interval on the $\Wz$ axis.
    But if we investigate the convergence outside the horizon, this relation
    says that $\Wnu_l$ has to fall off quickly enough so that
    $\sum_l \Wnu_l P_l(\Wz/\massBH)$ converges also for $|\Wz|>\massBH$.
    Thus, because the convergence criteria of $\sum a_l P_l(\cos\Wtheta)$ 
    and $\sum a_l$ are the same (e.g. $\lim_{n\to\infty}|a_l|^{1/l}\le 1$),
    if $\sum_l \Wnu_l P_l(\Wz/\massBH)$ converges for $|\Wz|< \Wz_\text{max}$,
    Eq.~\eqref{eq:nu_above_horizon} converges for $\Wr<\Wz_\text{max}+\massBH$.

    Using $\Kzeta=\cos\Wtheta$ again, if the real convergence radius of the 
    Legendre series for 
    \begin{equation}
        \Wnu_\HH(\Kzeta) = \sum_l \Wnu_l\, P_l(\Kzeta)
    \end{equation}
    is $\Kzeta_\text{max}>1$, we know that~\eqref{eq:nu_above_horizon} will 
    converge for $\Wr<\Wr_\text{max}$, where
    \begin{equation}
        \Wr_\text{max}\equiv \massBH(1+\Kzeta_\text{max})>2\massBH \,,
    \end{equation}
    because $|P_l(\cos\Wtheta)|\le 1$.
    
    For the considered disk perturbation $\Wnu$ we have
    \begin{align}
        \begin{aligned}
            \Wnu_\HH(\Kzeta) &= \frac{2 \massDisk}{3 \Cpi 
            \massBH \Kzeta^5} \Big(5 \bM \Kzeta^3 + 3\bM^3 \Kzeta \\
                             &\qquad{}- 3 \left(\bM^2 + \Kzeta ^2\right)^2 
                         \arctan\frac{\Kzeta}{\bM}\Big) ,
        \end{aligned}
    \end{align}
    where $\bM = \diskR/\massBH$ and the value of $\Kzeta_\text{max}=\sqrt{\bM^2+1}$
    for this function is known for its Chebyshev expansion because the $\arctan$
    function is the well-known showcase in the theory of approximations,
    \cite{Boyd-2001}. 
    The same value of $\Kzeta_\text{max}$ is also valid for
    the Legendre series, which share the same
    Bernstein ellipse that determines the exponential fall-off of the expansion 
    coefficients $\Wnu_l$, \cite{Wang-2012}.
    But because the ratio of Legendre and Chebyshev polynomials has
    \begin{equation}
        \lim_{l\to\infty} \left| \frac{P_l(s)}{T_l(s)} \right|^{1/l} = 1 \,,
    \end{equation}
    for $s>1$, also the series~\eqref{eq:nu_above_horizon}
    will converge for $2\massBH\le \Wr< \Wr_\text{max}$.
    For the potential studied, the value of 
    \begin{equation}
        \Wr_\text{max}=\massBH +\sqrt{\massBH^2+\diskR^2}
        \label{eq:inner_disk_radius}
    \end{equation}
    is equal to the inner radius of the disk.
    
    Let us summarize this finding. The geometry of the horizon influenced by the 
    external matter is specified by a single function $\Wnu_\HH(\Kzeta)$,
    where $\Kzeta=\cos\Wtheta$, and therefore belongs to the interval $[-1,1]$.
    However, the analytic properties of the horizon geometry allow summation up to 
    $\Kzeta_\text{max}>1$, and this value directly determines the region where 
    we can construct the set of space-time quantities starting with $\Wnu$ 
    given by~\eqref{eq:nu_above_horizon}. For the considered exact solution of
    superimposed fields~\eqref{app:eq:disk_potentials} taken from \cite{Kofron-2023}
    we then see that this region of space-time recovered from the horizon 
    geometry (and symmetry assumptions) is $2\massBH\le \Wr<\Wr_\text{max}$
    where $\Wr_\text{max}$ is the inner radius of the disk.

\section{Conclusion}
    The isolated horizons framework allows for a quasi-local description of
    deformed black holes in equilibrium, but the involved equations are quite
    complicated. We have shown how, in spherical symmetry, these equations
    and gauge conditions imply the initial data leading to the Schwarzschild
    space-time. Then we considered axisymmetric initial data and found the
    geometry of a static space-time surrounding the horizon in the form
    of a series expansion of NP quantities in the neighborhood
    of the horizon. Although the solution was found only up to a finite order of 
    expansion, this solution will reflect the changes of the geometry due to 
    the presence of external matter. Moreover, we solved the same problem
    numerically for one particular example
    in the form of a simplified model of an accretion disk (inverted
    Morgan--Morgan disk) and discussed the convergence of both solutions.
    The numerical solution also enabled us to plot the most interesting NP
    scalars.

    Since we followed the same basic procedure, we arrived at similar
    series expansions as in~\cite{Krishnan-2012}. However, while the expressions
    in \cite{Krishnan-2012} represent a general solution, they do not give
    any clues for the right choice of the initial data for any specific
    space-time. The expansions we have found represent one particular
    choice which has been illustrated in the figures. While we have used
    one particular disk for these illustrations, any suitable solution
    could be used. Since the function $\Wlambda$ is not always
    available as an analytical solution, it is convenient that it can be
    obtained from the horizon geometry together with NP quantities.

\begin{acknowledgments}
    We offer sincere gratitude to the late Dr. Martin Scholtz, whose expertise and 
    guidance were instrumental in starting this research. May his legacy live
    through the ideas and insights he shared with us during the early stages of
    this work. His scientific inquiry continues to inspire us.

    Ale\v{s} Flandera acknowledges support from Grant GAUK
    664420 of Charles University.
    David Kofro\v{n} and Tom\'{a}\v{s} Ledvinka acknowledge support from 
    Grant GACR 21-11268S of the Czech Science Foundation.
\end{acknowledgments}

\appendix

\section{Newman--Penrose formalism}
    \label{app:np_formalism}
    We only briefly review the Newman--Penrose formalism with
    emphasis on the relations we shall need. For a more in-depth review
    of the formalism, see \cite{Stewart-1993}.

    The NP null tetrad consists of four null vectors 
    $\NPl^a, \NPn^a, \NPm^a$ and $\cconj{\NPm}^a$ normalized by the conditions
    \begin{align}
        \NPl^a\,\NPn_a &= 1 \,, & \NPm^a\,\cconj{\NPm}_a &= -1 \,, 
    \end{align}
    while all remaining contractions vanish. These vectors constitute the basis
    of the tangent space, so that any tensor can be expanded in terms of
    them. We employ the convention that any real vector $X^a$ has the
    decomposition
    \begin{equation}
        X^a = X_0\,\NPl^a + X_1\,\NPn^a + X_2\,\NPm^a + \cconj{X}_2\,\cconj{\NPm}^a \,.
    \end{equation}
    The metric tensor in terms of the null tetrad reads
    \begin{equation}
        g_{ab} =2\,\NPl_{(a}\,\NPn_{b)} - 2\,\NPm_{(a}\cconj{\NPm}_{b)} \,,
        \label{app:np:eq_expression_for_metric}
    \end{equation}
    where the parentheses denote symmetrization.

    In the NP formalism, the spin coefficients are the Ricci rotation
    coefficients with respect to the null tetrad $\left(\NPl^a,\NPn^a,\NPm^a,\cconj{\NPm}^a\right)$
    and they encode the connection. The twelve independent complex coefficients
    are defined by
    \ifthenelse{\equal{\journal}{PRD}}
    {%
    \addtolength{\tabcolsep}{7pt}
    \begin{align}
        \begin{tabularx}{\linewidth}{*{4}{c}X}
            \midrule
            \midrule
            $\nabla$ & $\NPm^a\,\nabla \NPl_a$ & $\frac{1}{2}\left(\NPn^a\, \nabla \NPl_a 
            - \cconj{\NPm}^a\, \nabla \NPm_a\right)$ & $-\cconj{\NPm}^a\, \nabla \NPn_a$ & \\
            \midrule
            $\NPD$ & $\Skappa$ & $\Sepsilon$ & $\Spi$ & \\
            $\NPDelta$ & $\Stau$ & $\Sgamma$ & $\Snu$ & \\
            $\NPdelta$ & $\Ssigma$ & $\Sbeta$ & $\Smu$ & \\
            $\cconj{\NPdelta}$ & $\Srho$ & $\Salpha$ & $\Slambda$ & \\
            \midrule
            \midrule
        \end{tabularx}
        \nonumber
    \end{align}
    \addtolength{\tabcolsep}{-7pt}
    }%
    {%
    \addtolength{\tabcolsep}{4pt}
    \begin{align}
        \begin{tabular}{*{4}{c}}
            \toprule
            $\nabla$ & $\NPm^a\,\nabla \NPl_a$ & $\frac{1}{2}\left(\NPn^a\, \nabla \NPl_a 
            - \cconj{\NPm}^a\, \nabla \NPm_a\right)$ & $-\cconj{\NPm}^a\, \nabla \NPn_a$ \\
            \midrule
            $\NPD$ & $\Skappa$ & $\Sepsilon$ & $\Spi$ \\
            $\NPDelta$ & $\Stau$ & $\Sgamma$ & $\Snu$ \\
            $\NPdelta$ & $\Ssigma$ & $\Sbeta$ & $\Smu$ \\
            $\cconj{\NPdelta}$ & $\Srho$ & $\Salpha$ & $\Slambda$ \\
            \bottomrule
        \end{tabular} 
        \nonumber
    \end{align}
    \addtolength{\tabcolsep}{-4pt}
    }%
    where 
    \begin{align}
        \begin{aligned}
            \NPD &= \NPl^a\nabla_a \,, &
            \NPDelta &= \NPn^a \nabla_a \,, 
            \\
            \NPdelta &= \NPm^a \nabla_a \,, &
            \cconj{\NPdelta} &= \cconj{\NPm}^a \nabla_a \,.
        \end{aligned}
    \end{align}
    Using these operators, we can find how tetrad vectors fields change when
    transported. The expressions, called transport equations, are:
    \begin{subequations}
    \begin{align}
            \NPD \NPl^a &= \left( \Sepsilon+\cconj{\Sepsilon} \right) \NPl^a - \cconj{\Skappa}\,\NPm^a - 
            \Skappa\,\cconj{\NPm}^a \,, \\
            \NPDelta \NPl^a &= \left( \Sgamma+\cconj{\Sgamma} \right)\NPl^a-\cconj{\Stau}\,\NPm^a - 
            \Stau\,\cconj{\NPm}^a \,, \\
            \NPdelta \NPl^a &= \left( \cconj{\Salpha}+\Sbeta \right)\NPl^a-\cconj{\Srho}\,\NPm^a - 
            \Ssigma\,\cconj{\NPm}^a \,, \\
            \NPD \NPn^a &= -\left( \Sepsilon+\cconj{\Sepsilon} \right)\NPn^a +\Spi\,\NPm^a 
            +\cconj{\Spi}\,\cconj{\NPm}^a \,,\\
            \NPDelta \NPn^a &= -\left( \Sgamma+\cconj{\Sgamma} \right)\NPn^a +\Snu\,\NPm^a + 
            \cconj{\Snu}\,\cconj{\NPm}^a \,, \\
            \NPdelta \NPn^a &= -(\cconj{\Salpha}+\Sbeta)\NPn^a+\Smu\,\NPm^a + \cconj{\Slambda}\,\cconj{\NPm}^a
            \,, \\
            \NPD \NPm^a &= \cconj{\Spi}\,\NPl^a - \Skappa\,\NPn^a + \left( \Sepsilon-\cconj{\Sepsilon} \right)\NPm^a
            \,, \\
            \NPDelta \NPm^a &= \cconj{\Snu}\,\NPl^a -\Stau\,\NPn^a + \left( \Sgamma-\cconj{\Sgamma} 
            \right)\NPm^a \,, \\
            \NPdelta \NPm^a &= \cconj{\Slambda}\,\NPl^a-\Ssigma\,\NPn^a+\left( \Sbeta-\cconj{\Salpha} 
            \right)\NPm^a \,, \\
            \cconj{\NPdelta}\NPm^a &= \cconj{\Smu}\,\NPl^a -\Srho\,\NPn^a+\left( \Salpha-\cconj{\Sbeta} 
            \right)\NPm^a \,.
    \end{align}
    \label{eq:transport_equations}
    \end{subequations}
    Acting on a scalar, the operators $\NPD$, $\NPDelta$, and $\NPdelta$ obey
    the commutation relations:
    \begin{subequations}
    \begin{align}
        \begin{split}
            \NPD \NPdelta - \NPdelta \NPD &= (\cconj{\Spi} - \cconj{\Salpha} - \Sbeta) \NPD 
            - \Skappa \NPDelta \\
                                &\qquad{}+ (\cconj{\Srho} - \cconj{\Sepsilon} + \Sepsilon)
            \NPdelta + \Ssigma \cconj{\NPdelta} \,, 
        \end{split}
        \\    %
        \begin{split}
            \NPDelta \NPD - \NPD \NPDelta &= (\Sgamma + \cconj{\Sgamma})\NPD + (\Sepsilon +
            \cconj{\Sepsilon}) \NPDelta \\
                                &\qquad{}- (\cconj{\Stau} + \Spi) \NPdelta 
            - (\Stau + \cconj{\Spi}) \cconj{\NPdelta} \,, 
        \end{split}
        \\    %
        \begin{split}
            \NPDelta \NPdelta - \NPdelta \NPDelta &= \cconj{\Snu} \NPD + (\cconj{\Salpha}
            + \Sbeta - \Stau) \NPDelta \\
                                          &\qquad{}+ (\Sgamma - \cconj{\Sgamma} - \Smu) \NPdelta
            - \cconj{\Slambda} \cconj{\NPdelta} \,, 
        \end{split}
        \\    %
        \begin{split}
            \NPdelta \cconj{\NPdelta} - \cconj{\NPdelta} \NPdelta &= (\Smu - 
            \cconj{\Smu}) \NPD + (\Srho - \cconj{\Srho}) \NPDelta \\
                                                          &\qquad{}+ (\cconj{\Salpha}
            - \Sbeta) \cconj{\NPdelta} - (\Salpha - \cconj{\Sbeta}) \NPdelta \,.
        \end{split}
        \label{app:np:eq:commutators}
    \end{align}
    \end{subequations}

    Since all quantities in the NP formalism are projected onto
    the tetrad, except for the spin coefficients, we need also five non-zero
    complex components of the Weyl tensor
    \begin{subequations}
    \begin{align}
            \NPPsi_0 &= C_{abcd}\, \NPl^a\, \NPm^b\, \NPl^c\, \NPm^d \,, \\
            \NPPsi_1 &= C_{abcd}\, \NPl^a\, \NPn^b\, \NPl^c\, \NPm^d \,, \\
            \NPPsi_2 &= C_{abcd}\, \NPl^a\, \NPm^b\, \cconj{\NPm}^c\, \NPn^d \,, \\
            \NPPsi_3 &= C_{abcd}\, \NPl^a\, \NPn^b\, \cconj{\NPm}^c\, \NPn^d \,, \\
            \NPPsi_4 &= C_{abcd}\, \cconj{\NPm}^a\, \NPn^b\, \cconj{\NPm}^c\, \NPn^d \,,
    \end{align}
    \end{subequations}
    and six independent components of the Ricci tensor
    \begingroup
    \allowdisplaybreaks
    \begin{subequations}
        \begin{align}
                \NPPhi_{00} &= - \frac{1}{2} R_{ab}\, \NPl^a\, \NPl^b \,, \\
                \NPPhi_{01} &= - \frac{1}{2} R_{ab}\, \NPl^a\, \NPm^b \,, \\
                \NPPhi_{02} &= - \frac{1}{2} R_{ab}\, \NPm^a\, \NPm^b \,, \\
                \NPPhi_{11} &= - \frac{1}{2} R_{ab} \left(\NPl^a\, \NPn^b + \NPm^a\, \cconj{\NPm}^b\right) , \\
                \NPPhi_{12} &= - \frac{1}{2} R_{ab}\, \NPn^a\, \NPm^b \,, \\
                \NPPhi_{22} &= - \frac{1}{2} R_{ab}\, \NPn^a\, \NPn^b \,, 
        \end{align}
    \end{subequations}
    \endgroup
    where the components with switched indices are complex conjugates,
    $\NPPhi_{ba} = \cconj{\NPPhi}_{ab}$. The trace of the Ricci tensor is given
    by the scalar curvature which in NP formalism is denoted by 
    $\NPLambda = \frac{R}{24}$.

    \subsection{Ricci identities}
        \label{app:np:eq:ricci}
        Since Einstein's equations in NP formalism 
        are just algebraic relations between NP quantities, the true field
        equations are provided by the Ricci and Bianchi identities. By the
        Ricci identities we mean the definition of the Riemann tensor,
        \begin{align}
            2 \nabla_{[c} \nabla_{d]}\, X^a = - R^a{}_{bcd}\, X^b \,.
        \end{align}
        Substituting the vectors of the null tetrad for $X_a$ and projecting them 
        onto the null tetrad, these identities become differential equations
        for the spin coefficients:
        \begin{widetext}
        \begingroup
        \allowdisplaybreaks
        \begin{subequations}
            \begin{align}
                \NPD\Srho - \cconj{\NPdelta} \Skappa &=\Srho ^2 + \left(\Sepsilon
                + \cconj{\Sepsilon}\right) \Srho -\Skappa  \left(3 \Salpha
                +\cconj{\Sbeta }-\Spi \right)-\Stau \cconj{\Skappa }
                +\Ssigma  \cconj{\Ssigma }+\NPPhi_{00} \,,
                \label{app:np:RI:Drho}\\
                \NPD\Ssigma - \NPdelta\Skappa &= (\Srho+\cconj{\Srho}+3\Sepsilon
                -\cconj{\Sepsilon})\Ssigma - (\Stau-\cconj{\Spi}+\cconj{\Salpha}
                +3\Sbeta)\Skappa+\NPPsi_0 \,,
                \label{app:np:RI:Dsigma}\\
                \NPD \Stau - \NPDelta\Skappa &= \Srho(\Stau+\cconj{\Spi})
                +\Ssigma(\cconj{\Stau}+\Spi)+(\Sepsilon-\cconj{\Sepsilon})\Stau
                -(3\Sgamma+\cconj{\Sgamma})\Skappa+\NPPsi_1+\NPPhi_{01} \,,
                \label{app:np:RI:Dtau}\\
                \NPD\Salpha - \cconj{\NPdelta}\Sepsilon &= (\Srho + \cconj{\Sepsilon}
                - 2\Sepsilon)\Salpha + \Sbeta\cconj{\Ssigma}
                - \cconj{\Sbeta}\Sepsilon - \Skappa \Slambda
                - \cconj{\Skappa}\Sgamma + (\Sepsilon+\Srho)\Spi
                + \NPPhi_{10} \,,
                \label{app:np:RI:Dalpha}\\
                \NPD\Sbeta - \NPdelta\Sepsilon &= (\Salpha+\Spi)\Ssigma +
                (\cconj{\Srho}-\cconj{\Sepsilon})\Sbeta - (\Smu+\Sgamma)\Skappa
                - (\cconj{\Salpha}-\cconj{\Spi})\Sepsilon + \NPPsi_1 \,,
                \label{app:np:RI:Dbeta}\\
                \NPD\Sgamma - \NPDelta\Sepsilon &= (\Stau+\cconj{\Spi})\Salpha
                + (\cconj{\Stau}+\Spi)\Sbeta - (\Sepsilon+\cconj{\Sepsilon})\Sgamma
                - (\Sgamma + \cconj{\Sgamma})\Sepsilon + \Stau \Spi - \Snu \Skappa
                + \NPPsi_2 - \NPLambda + \NPPhi_{11} \,,
                \label{app:np:RI:Dgamma}\\
                \NPD\Slambda - \cconj{\NPdelta}\Spi &= (\Srho - 3\Sepsilon
                + \cconj{\Sepsilon})\Slambda + \cconj{\Ssigma}\Smu + (\Spi
                + \Salpha - \cconj{\Sbeta})\Spi - \Snu\cconj{\Skappa}
                + \NPPhi_{20} \,,
                \label{app:np:RI:Dlambda}\\
                \NPD\Smu - \NPdelta\Spi &= (\cconj{\Srho}-\Sepsilon-\cconj{\Sepsilon})\Smu
                + \Ssigma\Slambda + (\cconj{\Spi}-\cconj{\Salpha} + \Sbeta)\Spi
                - \Snu \Skappa + \NPPsi_2 + 2 \NPLambda \,,
                \label{app:np:RI:Dmu}\\
                \NPD\Snu - \NPDelta\Spi &= (\Spi + \cconj{\Stau})\Smu + (\cconj{\Spi}
                + \Stau)\Slambda + (\Sgamma - \cconj{\Sgamma})\Spi - (3\Sepsilon
                + \cconj{\Sepsilon})\Snu + \NPPsi_3 + \NPPhi_{21} \,,
                \label{app:np:RI:Dnu}\\
                \NPDelta\Slambda - \cconj{\NPdelta}\Snu &= -(\Smu + \cconj{\Smu}
                + 3\Sgamma-\cconj{\Sgamma})\Slambda+(3\Salpha+\cconj{\Sbeta}
                + \Spi-\cconj{\Stau})\Snu-\NPPsi_4 \,,
                \label{app:np:RI:Deltalambda}\\
                \NPDelta\Smu - \NPdelta\Snu &= -(\Smu+\Sgamma+\cconj{\Sgamma})\Smu
                - \Slambda\cconj{\Slambda}+\cconj{\Snu}\Spi+(\cconj{\Salpha}
                + 3\Sbeta-\Stau)\Snu-\NPPhi_{22} \,,
                \label{app:np:RI:Deltamu}\\
                \NPDelta\Sbeta-\NPdelta\Sgamma &= (\cconj{\Salpha}+\Sbeta
                -\Stau)\Sgamma - \Smu \Stau + \Ssigma \Snu + \Sepsilon \cconj{\Snu}
                + (\Sgamma-\cconj{\Sgamma}-\Smu)\Sbeta - \Salpha\cconj{\Slambda}
                - \NPPhi_{12} \,,
                \label{app:np:RI:Deltabeta}\\
                \NPDelta\Ssigma-\NPdelta\Stau& = - (\Smu - 3\Sgamma
                + \cconj{\Sgamma})\Ssigma - \cconj{\Slambda}\Srho - (\Stau
                + \Sbeta - \cconj{\Salpha})\Stau + \Skappa \cconj{\Snu}
                - \NPPhi_{02} \,,
                \label{app:np:RI:Deltasigma}\\
                \NPDelta\Srho-\cconj{\NPdelta}\Stau &= (\Sgamma+\cconj{\Sgamma}
                -\cconj{\Smu})\Srho - \Ssigma \Slambda + (\cconj{\Sbeta}-\Salpha
                -\cconj{\Stau})\Stau + \Snu \Skappa - \NPPsi_2 - 2 \NPLambda \,,
                \label{app:np:RI:Deltarho}\\
                \NPDelta\Salpha-\cconj{\NPdelta}\Sgamma &= (\Srho+\Sepsilon)\Snu
                - (\Stau+\Sbeta)\Slambda + (\cconj{\Sgamma}-\cconj{\Smu})\Salpha
                + (\cconj{\Sbeta}-\cconj{\Stau})\Sgamma - \NPPsi_3 \,,
                \label{app:np:RI:Deltaalpha}\\
                \NPdelta\Srho-\cconj{\NPdelta}\Ssigma &= (\cconj{\Salpha}+\Sbeta)\Srho
                - (3\Salpha-\cconj{\Sbeta})\Ssigma+(\Srho-\cconj{\Srho})\Stau
                + (\Smu-\cconj{\Smu})\Skappa -\NPPsi_1 + \NPPhi_{01} \,,
                \label{app:np:RI:deltarho}\\
                \NPdelta\Salpha-\cconj{\NPdelta}\Sbeta &= \Smu\Srho-\Slambda\Ssigma
                + \Salpha\cconj{\Salpha}+\Sbeta\cconj{\Sbeta}-2\Salpha\Sbeta
                + (\Srho-\cconj{\Srho})\Sgamma + (\Smu-\cconj{\Smu})\Sepsilon
                - \NPPsi_2 + \NPLambda + \NPPhi_{11} \,,
                \label{app:np:RI:deltaalpha}\\
                \NPdelta\Slambda-\cconj{\NPdelta}\Smu &= (\Srho-\cconj{\Srho})\Snu
                + (\Smu-\cconj{\Smu})\Spi + (\Salpha+\cconj{\Sbeta})\Smu
                + (\cconj{\Salpha}-3\Sbeta)\Slambda-\NPPsi_3 + \NPPhi_{21} \,.
                \label{app:np:RI:deltalambda}
            \end{align}
            \label{app:np:RI}
        \end{subequations}
        \endgroup

    \subsection{Bianchi identities}
        \label{app:np:eq:bianchi}

        The Ricci identities contain components of the Riemann tensor as 
        unknown functions. Differential equations for them are obtained from
        the Bianchi identities
        \begin{align}
            \nabla_{[e}R_{ab]cd} &=0 \,.
        \end{align}
        Again, projecting the Bianchi identities onto the null tetrad, we get
        the following set of equations in the NP formalism:
        \begingroup
        \allowdisplaybreaks
        \begin{subequations}
            \begin{align}
                \begin{split}
                &{}\NPD\NPPsi_1-\cconj{\NPdelta}\NPPsi_0-\NPD\NPPhi_{01}+\NPdelta\NPPhi_{00} =
                (\Spi - 4 \Salpha) \NPPsi_0 + 2(2\Srho+\Sepsilon)\NPPsi_1
                - 3\Skappa\NPPsi_2 + 2\Skappa\NPPhi_{11} \\
                &\qquad - (\cconj{\Spi} - 2\cconj{\Salpha} - 2\Sbeta) \NPPhi_{00}
                - 2\Ssigma\NPPhi_{10} - 2(\cconj{\Srho}+\Sepsilon)\NPPhi_{01}
                + \cconj{\Skappa}\NPPhi_{02} \,, \label{np:BI:DPsi1} 
                \end{split}
                \\[1.5ex]%
                \begin{split}
                &{}\NPD\NPPsi_2-\cconj{\NPdelta}\NPPsi_1 + \NPDelta\NPPhi_{00}
                - \cconj{\NPdelta}\NPPhi_{01} + 2\NPD\NPLambda = - \Slambda\NPPsi_0
                + 2 (\Spi-\Salpha)\NPPsi_1 + 3\Srho \NPPsi_2 - 2\Skappa\NPPsi_3  \\
                &\qquad + 2\Srho\NPPhi_{11} + \cconj{\Ssigma}\NPPhi_{02}
                + (2\Sgamma + 2\cconj{\Sgamma} - \cconj{\Smu})\NPPhi_{00}
                - 2(\Salpha + \cconj{\Stau})\NPPhi_{01} - 2\Stau\NPPhi_{10} \,,
                \label{np:BI:DPsi2} 
                \end{split}
                \\[1.5ex]%
                \begin{split}
                &{}\NPD\NPPsi_3-\cconj{\NPdelta}\NPPsi_2-\NPD\NPPhi_{21}+\NPdelta\NPPhi_{20}
                -2\cconj{\NPdelta}\NPLambda = -2\Slambda \NPPsi_1+3\Spi\NPPsi_2
                + 2 (\Srho-\Sepsilon)\NPPsi_3 - \Skappa\NPPsi_4  \\
                &\qquad + 2\Smu\NPPhi_{10} - 2\Spi\NPPhi_{11} - (2\Sbeta
                +\cconj{\Spi} - 2\cconj{\Salpha})\NPPhi_{20} - 2(\cconj{\Srho}
                - \Sepsilon)\NPPhi_{21} + \cconj{\Skappa}\NPPhi_{22} \,,
                \label{np:BI:DPsi3} 
                \end{split}
                \\[1.5ex]%
                \begin{split}
                &{}\NPD\NPPsi_4-\cconj{\NPdelta}\NPPsi_3 + \NPDelta\NPPhi_{20}
                - \cconj{\NPdelta}\NPPhi_{21} = - 3\Slambda\NPPsi_2 + 2(\Salpha
                + 2\Spi)\NPPsi_3 + (\Srho - 4\Sepsilon)\NPPsi_4 + 2\Snu\NPPhi_{10} \\
                &\qquad - 2\Slambda\NPPhi_{11} - (2\Sgamma - 2\cconj{\Sgamma}
                + \cconj{\Smu})\NPPhi_{20}- 2(\cconj{\Stau} - \Salpha)\NPPhi_{21}
                + \cconj{\Ssigma}\NPPhi_{22} \,, \label{np:BI:DPsi4} 
                \end{split}
                \\[1.5ex]%
                \begin{split}
                &{}\NPDelta\NPPsi_0 - \NPdelta\NPPsi_1 + \NPD\NPPhi_{02} - \NPdelta\NPPhi_{01}
                = (4\Sgamma-\Smu)\NPPsi_0 -2(2\Stau+\Sbeta)\NPPsi_1+ 3\Ssigma\NPPsi_2 \\
                &\qquad +(\cconj{\Srho} +2\Sepsilon -2\cconj{\Sepsilon})\NPPhi_{02}
                + 2\Ssigma\NPPhi_{11} - 2\Skappa\NPPhi_{12} -\cconj{\Slambda}\NPPhi_{00}
                +2(\cconj{\Spi} -\Sbeta)\NPPhi_{01} \,,
                \label{np:BI:DeltaPsi0} 
                \end{split}
                \\[1.5ex]%
                \begin{split}
                &{}\NPDelta\NPPsi_1 -\NPdelta\NPPsi_2 -\NPDelta\NPPhi_{01}
                +\cconj{\NPdelta}\NPPhi_{02} -2\NPdelta\NPLambda =\Snu\NPPsi_0
                +2(\Sgamma-\Smu)\NPPsi_1 -3\Stau\NPPsi_2 +2\Ssigma\NPPsi_3 \\
                &\qquad -\cconj{\Snu}\NPPhi_{00} + 2(\cconj{\Smu}
                -\Sgamma)\NPPhi_{01} +(2\Salpha+\cconj{\Stau} -2\cconj{\Sbeta})\NPPhi_{02}
                +2\Stau\NPPhi_{11} -2\Srho\NPPhi_{12} \,,
                \label{np:BI:DeltaPsi1} 
                \end{split}
                \\[1.5ex]%
                \begin{split}
                &{}\NPDelta\NPPsi_2 - \NPdelta\NPPsi_3 +\NPD\NPPhi_{22} - \NPdelta\NPPhi_{21}
                + 2\NPDelta\NPLambda = 2\Snu\NPPsi_1-3\Smu\NPPsi_2 + 2(\Sbeta-\Stau)\NPPsi_3
                + \Ssigma\NPPsi_4 \\
                &\qquad - 2\Smu\NPPhi_{11} -\cconj{\Slambda}\NPPhi_{20}
                + 2\Spi\NPPhi_{12}+ 2(\Sbeta+\cconj{\Spi})\NPPhi_{21} + (\cconj{\Srho}
                -2\Sepsilon -2\cconj{\Sepsilon})\NPPhi_{22} \,,
                \label{np:BI:DeltaPsi2} 
                \end{split}
                \\[1.5ex]%
                \begin{split}
                &{}\NPDelta\NPPsi_3-\NPdelta\NPPsi_4\hfill - \NPDelta\NPPhi_{21}
                +\cconj{\NPdelta}\NPPhi_{22} = 3\Snu\NPPsi_2-2(\Sgamma+2\Smu)\NPPsi_3
                +(4\Sbeta-\Stau)\NPPsi_4-2\Snu\NPPhi_{11} \\
                &\qquad - \cconj{\Snu}\NPPhi_{20} + 2\Slambda\NPPhi_{12}
                +2(\Sgamma+\cconj{\Smu})\NPPhi_{21} +(\cconj{\Stau} -2\cconj{\Sbeta}
                -2\Salpha)\NPPhi_{22} \,, \label{np:BI:DeltaPsi3} 
                \end{split}
                \\[1.5ex]%
                \begin{split}
                &{}\NPD\NPPhi_{11}-\NPdelta\NPPhi_{10}+\NPDelta\NPPhi_{00}
                -\cconj{\NPdelta}\NPPhi_{01}+3\NPD\NPLambda = (2\Sgamma
                +2\cconj{\Sgamma}-\Smu-\cconj{\Smu})\NPPhi_{00} \\
                &\qquad + (\Spi-2\Salpha-2\cconj{\Stau})\NPPhi_{01}
                + (\cconj{\Spi}-2\cconj{\Salpha} -2\Stau)\NPPhi_{10}
                +2(\Srho+\cconj{\Srho})\NPPhi_{11}+\cconj{\Ssigma}\NPPhi_{02} \\
                &\qquad + \Ssigma\NPPhi_{20} -\cconj{\Skappa}\NPPhi_{12}
                -\Skappa\NPPhi_{21} \,, \label{np:BI:DPhi11} 
                \end{split}
                \\[1.5ex]%
                \begin{split}
                &{}\NPD\NPPhi_{12}-\NPdelta\NPPhi_{11} +\NPDelta\NPPhi_{01}
                -\cconj{\NPdelta}\NPPhi_{02} +3\NPdelta\NPLambda = (2\Sgamma-\Smu
                -2\cconj{\Smu})\NPPhi_{01} +\cconj{\Snu}\NPPhi_{00}
                -\cconj{\Slambda}\NPPhi_{10} \\
                &\qquad + 2(\cconj{\Spi}-\Stau)\NPPhi_{11} +(\Spi+2\cconj{\Sbeta}
                -2\Salpha-\cconj{\Stau})\NPPhi_{02} +(2\Srho+\cconj{\Srho}
                -2\cconj{\Sepsilon})\NPPhi_{12} \\
                &\qquad + \Ssigma\NPPhi_{21} -\Skappa\NPPhi_{22} \,,
                \label{np:BI:DPhi12} 
                \end{split}
                \\[1.5ex]%
                \begin{split}
                &{}\NPD\NPPhi_{22}-\NPdelta\NPPhi_{21}+\NPDelta\NPPhi_{11}
                -\cconj{\NPdelta}\NPPhi_{12}+3\NPDelta\NPLambda = \Snu \NPPhi_{01}
                +\cconj{\Snu}\NPPhi_{10}-2(\Smu+\cconj{\Smu})\NPPhi_{11}
                -\Slambda\NPPhi_{02} \\
                &\qquad - \cconj{\Slambda}\NPPhi_{20} + (2\Spi-\cconj{\Stau}
                +2\cconj{\Sbeta})\NPPhi_{12} +(2\Sbeta-\Stau
                +2\cconj{\Spi})\NPPhi_{21} \\
                &\qquad + (\Srho+\cconj{\Srho}-2\Sepsilon
                -2\cconj{\Sepsilon})\NPPhi_{22} \,. \label{np:BI:DPhi22}
                \end{split}
            \end{align}\label{eq:bianchi id}
        \end{subequations}
        \endgroup
        $$\\[2.6cm]$$ 
        \end{widetext}

    \subsection{Frame equations}
        In order to close the set of field equations provided by the Ricci
        and Bianchi identities, one needs to supplement them with equations
        for the tetrad itself. These so-called \emph{frame equations}, i.e.\
        equations for the components of the tetrad come from the commutation
        relations (\ref{app:np:eq:commutators}) applied to the coordinates
        $\Kx^\mu = (\Kv,\Kr,\Kx^I)$ introduced in Sec.%
        ~\ref{sec:coordinates_and_tetrad}. The resulting frame equations can be
        divided into two sets. One of them gives us the radial evolution of
        the metric functions
        \begin{align}
            \begin{aligned}
                \NPDelta \TU &= -\Sepsilon - \cconj{\Sepsilon} - \Spi \TOmega
                - \cconj{\Spi} \cconj{\TOmega} \,, \\
                \NPDelta \TX^I &= - \Spi \Txi^I - \cconj{\Spi} \cconj{\Txi}^I \,, \\
                \NPDelta \TOmega &= - \cconj{\Spi} - \Smu \TOmega - \cconj{\Slambda}
                \cconj{\TOmega} \,, \\
                \NPDelta \Txi^I &= - \Smu \Txi^I - \cconj{\Slambda} \cconj{\Txi}^I \,,
            \end{aligned}
            \label{app:np:eq:frame_equations_radial}
        \end{align}
        and the second group describes all other derivatives
        \begin{align}
            \begin{aligned}
                \NPD \TOmega - \NPdelta \TU &= \Skappa + \left(\cconj{\Srho} - \cconj{\Sepsilon}
                + \Sepsilon\right) \TOmega + \Ssigma \cconj{\TOmega} \,, \\
                \NPD \Txi^I - \NPdelta \TX^I &= \left(\cconj{\Srho} - \cconj{\Sepsilon} +
                \Sepsilon\right) \Txi^I + \Ssigma \cconj{\Txi}^I \,, \\
                \cconj{\NPdelta} \TOmega - \NPdelta \cconj{\TOmega} &=
                \Srho - \cconj{\Srho} + \left(\Salpha - \cconj{\Sbeta}\right) \TOmega
                  - \left(\cconj{\Salpha} - \Sbeta\right) \cconj{\TOmega} \,, \\
                \NPdelta \cconj{\Txi}^I - \cconj{\NPdelta} \Txi^I &=
                \left(\cconj{\Salpha} - \Sbeta\right) \cconj{\Txi}^I - \left(\Salpha
                - \cconj{\Sbeta}\right) \Txi^I \,.
            \end{aligned}
        \end{align}

    \subsection{Lorentz transformation}
        \label{app:np:lorentz_transformation}
        In this formalism, the Lorentz transformation is usually decomposed into
        four parts based on their character. There are two transformations with
        a real parameter, the first is the boost with a parameter $A$ which is
        prescribed by
        \begin{align}
            \NPl^a &\mapsto A^2\, \NPl^a\,, &
            \NPn^a &\mapsto A^{-2}\, \NPn^a\,, &
            \NPm^a &\mapsto \NPm^a \,.
        \end{align}
        The spin with a parameter $\chi$ is, on the other hand, given by
        \begin{align}
            \NPl^a &\mapsto \NPl^a \,, &
            \NPn^a &\mapsto \NPn^a \,, &
            \NPm^a &\mapsto \eu^{2\ii \chi} \NPm^a \,.
        \end{align}
        The remaining two transformations are null rotations about the vectors $\NPl^a$
        and $\NPn^a$. The null rotation about $\NPl^a$ with a complex parameter $c$ is
        \begin{align}
            \begin{aligned}
                \NPl^a &\mapsto \NPl^a \,, \\
                \NPn^a &\mapsto \NPn^a + c\, \NPm^a + \cconj{c}\, \cconj{\NPm}^a + \abs{c}^2 \NPl^a \,, \\
                \NPm^a &\mapsto \NPm^a + \cconj{c}\, \NPl^a \,,
            \end{aligned}
            \label{app:np:eq:rotation_l}
        \end{align}
        while the one about $\NPn^a$ with a parameter $d$, which is also complex,
        reads
        \begin{align}
            \begin{aligned}
                \NPl^a &\mapsto \NPl^a + \cconj{d}\, \NPm^a + d\, \cconj{\NPm}^a + \abs{d}^2 \NPn^a \,, \\
                \NPn^a &\mapsto \NPn^a \,, \\
                \NPm^a &\mapsto \NPm^a + d\, \NPn^a \,.
            \end{aligned}
        \end{align}
        Altogether, we get the 6-parameter Lorentz group. Similarly to the tetrad
        itself, all NP scalars transform accordingly.
        Their transformation can be found, e.g., in~\cite{Stewart-1993}.

    \subsection{Killing vectors in the NP formalism}
        \label{app:np:killing_eq}
        Let $K_a$ be a Killing (co)vector of a space-time. We expand it into the
        Newman--Penrose tetrad as
        \begin{equation}
            K_a = K_0\,\NPl_a + K_1\,\NPn_a + K_2\,\NPm_a + \cconj{K}_2\,\cconj{\NPm}_a
            \,.
        \end{equation}
        Then, the independent projections of the Killing equations
        \begin{equation}
            \nabla_aK_b + \nabla_b K_a = 0 
        \end{equation}
        onto the null tetrad read
        \begin{align}
            \begin{aligned}
                \NPD K_1 &=
                \left(\Sepsilon + \cconj{\Sepsilon}\right)K_1
                + \Skappa \, K_2
                + \cconj{\Skappa} \, \cconj{K}_2
                \,,  \\
                \NPD K_0 + \NPDelta K_1 &=
                - \left(\Sepsilon+\cconj{\Sepsilon}\right)K_0
                + \left(\Sgamma+\cconj{\Sgamma}\right)K_1 \\
                &\qquad{}+ \left(\Stau - \cconj{\Spi}\right)K_2
                + \left(\cconj{\Stau} - \Spi\right)\cconj{K}_2
                \,, \\
                \NPD \cconj{K}_2 - \NPdelta K_1 &= \Skappa \, K_0
                - \left(\cconj{\Spi} + \cconj{\Salpha} + \Sbeta\right) K_1 \\
                &\qquad{}- \Ssigma \, K_2
                + \left(\Sepsilon - \cconj{\Sepsilon} - \cconj{\Srho}\right) \cconj{K}_2
                \,, \\
                \NPDelta K_0 &= - \left(\Sgamma+\cconj{\Sgamma}\right)K_0
                - \cconj{\Snu} \,K_2
                - \Snu\,\cconj{K}_2
                \,,  \\
                \NPDelta \cconj{K}_2 - \NPdelta K_0 &=
                \left(\cconj{\Salpha} + \Sbeta + \Stau\right) K_0
                - \cconj{\Snu} \, K_1
                + \cconj{\Slambda} \, K_2 \\
                &\qquad{}+ \left(\Smu + \Sgamma - \cconj{\Sgamma}\right) \cconj{K}_2
                \,, \\
                \NPdelta \cconj{K}_2 &=
                \Ssigma \, K_0
                - \cconj{\Slambda} \, K_1
                - \left(\cconj{\Salpha} - \Sbeta\right) \cconj{K}_2
                \,,  \\
                \NPdelta K_2 + \cconj{\NPdelta}\cconj{K}_2 &=
                \left(\Srho + \cconj{\Srho}\right) K_0
                - \left(\Smu + \cconj{\Smu}\right) K_1 \\
                &\qquad{}+ \left(\cconj{\Salpha} - \Sbeta\right) K_2
                + \left(\Salpha - \cconj{\Sbeta}\right) \cconj{K}_2
                \,.
            \end{aligned}
            \label{app:np:eq:Killing:Killing_equations_projected}
        \end{align}
        As expected, three out of the seven equations are complex, which gives us
        the total of ten components.

            Comparing equations~\eqref{app:np:eq:Killing:Killing_equations_projected} to
            the Ricci identities~\eqref{app:np:eq:ricci}
            or Bianchi identities~\eqref{app:np:eq:bianchi}, it is clear that
            they have a very similar structure and, in fact, they have the same
            relation to the basis coefficients as e.g.\ the Ricci equations have
            to the spin coefficients. Hence, assuming basis coefficients $K_{0,\dots,2}$
            in some region, they allow us to find the Killing vector elsewhere.
            However, neither a general space-time nor a general space-time admitting
            an isolated horizon has a Killing vector. Hence, the Killing equations
            impose additional restrictions on the space-time geometry. In other
            words, in addition to the Killing equations, $K_a$ must satisfy the
            integrability conditions which can be easily derived.

            The second exterior derivative of the 1-form $K_a$ must vanish
            identically, which implies
            \begin{equation}
                \nabla_{[a}\nabla_bK_{c]} = 0 \propto \nabla_a \nabla_b K_c
                + \nabla_c \nabla_a K_b + \nabla_b \nabla_c K_a \,.
            \end{equation}
            Using the antisymmetry of $\nabla_a K_b$ and the definition of the
            Riemann tensor, we arrive at the integrability conditions
            \begin{equation}
                \nabla_c \nabla_a K_b = R_{abcd}\,K^d \,.
                \label{eq:integrability_condition}
            \end{equation}

\section{Series for NP scalars}
    \label{app:scalar_expansions}
    In this section, we present the NP scalars as computed using the expansion
    of the tetrad presented in Sec.~\ref{sec:singular_ode}. The scalars are
    presented up to the first order in $\Kr$ to keep the expressions reasonably
    long.
    Nevertheless, due to the derivatives in the definition of spin coefficients
    this requires the tetrad to be computed to a higher order.

    Although the spin $\Skappa$, and $\NPPsi_0$ are vanishing up to the first
    order, the higher orders are non-zero. On the other hand, spins $\Stau$, 
    $\Sgamma$ and $\Snu$ are identically zero and excluded from the list.
    The behavior of the Weyl scalar $\NPPsi_0$ can be seen in
    Fig.~\ref{fig:Psi0_contour}.

    \begin{widetext}
    \begingroup
    \allowdisplaybreaks
    \begin{subequations}
        \begin{align}
            \begin{split}
                \Salpha &= -\frac{\eu^{\Wnu -\Wlambda } \left(\cot\Ktheta -2 \Wnu_{\pdWtheta}\right)}{4 \sqrt{2} \massBH} + \frac{\Kr\, \eu^{\Wnu -2 \Wlambda }}{8 \sqrt{2} \massBH^2} \Big[\cot\Ktheta +2 \massBH \cot\Ktheta \left(\Wlambda_{\pdWr} - \Wnu_{\pdWr}\right) \\
                        &\qquad{}-2 \Wnu_{\pdWtheta} \left(3\massBH \Wlambda_{\pdWr}-2 \massBH \Wnu_{\pdWr}+1\right) -\massBH \left(\Wlambda_{\pdWrWtheta}-6 \Wnu_{\pdWrWtheta}\right)\Big]+\bigo(\Kr^2) \,,
            \end{split}
            \\[1.5ex]%
            \begin{split}
                \Sbeta &= \frac{\cot\Ktheta\, \eu^{\Wnu -\Wlambda }}{4 \sqrt{2} \massBH}
            -\frac{\Kr\, \eu^{\Wnu -2 \Wlambda }}{8 \sqrt{2} \massBH^2} \Big[\cot\Ktheta +2 \massBH \cot\Ktheta \left(\Wlambda_{\pdWr} - \Wnu_{\pdWr}\right) \\
                       &\qquad{}+2 \Wnu_{\pdWtheta} \left(\massBH \Wlambda_{\pdWr}-2 \massBH \Wnu_{\pdWr}+1\right) + \massBH \left(\Wlambda_{\pdWrWtheta}-2 \Wnu_{\pdWrWtheta}\right)\Big]+\bigo(\Kr^2) \,,
            \end{split}
            \\[1.5ex]%
            \begin{split}
                \Spi &= \frac{\eu^{\Wnu -\Wlambda }\, \Wnu_{\pdWtheta}}{2 \sqrt{2} \massBH}
                -\frac{\Kr\, \eu^{\Wnu -2 \Wlambda }}{4 \sqrt{2} \massBH^2} \Big[2\Wnu_{\pdWtheta} \left(2 \massBH \Wlambda_{\pdWr}-2 \massBH \Wnu_{\pdWr}+1\right)+\massBH \left(\Wlambda_{\pdWrWtheta}-4 \Wnu_{\pdWrWtheta}\right)\Big] +\bigo(\Kr^2) \,,
            \end{split}
            \\[1.5ex]%
            \begin{split}
                \Srho &= -\frac{\Kr\, \eu^{2 \Wnu -2 \Wlambda }}{8 \massBH^2} \left(\Wnu_\pdWthetaWtheta +\massBH \left(\Wlambda_{\pdWr}-2 \Wnu_{\pdWr}\right)+\cot\Ktheta\, \Wnu_{\pdWtheta}-{\Wnu_{\pdWtheta}}^2+1\right)+\bigo(\Kr^2) \,,
            \end{split}
            \\[1.5ex]%
            \begin{split}
                \Ssigma &= -\frac{\Kr\, \eu^{2 \Wnu -2 \Wlambda }}{8 \massBH^2} \left(\Wnu_\pdWthetaWtheta +\massBH \Wlambda_{\pdWr}-\cot\Ktheta\, \Wnu_{\pdWtheta}+{\Wnu_{\pdWtheta}}^2\right)+\bigo(\Kr^2) \,,
            \end{split}
            \\[1.5ex]%
            \begin{split}
                \Smu &= -\frac{\eu^{-\Wlambda } \left(\massBH\left(\Wlambda_{\pdWr}-2 \Wnu_{\pdWr}\right)+1\right)}{2 \massBH}
                +\frac{\Kr\, \eu^{-2 \Wlambda }}{4 \massBH^2} \Big[-2 \massBH^2 \Wlambda_\pdWrWr +4 \massBH^2 \Wnu_\pdWrWr -\massBH \Wlambda_{,\Wr\Wtheta\Wtheta} +2 \massBH \Wnu_{,\Wr\Wtheta\Wtheta} \\
                     &\qquad{}+2 \massBH^2 \Wlambda_{\pdWr}^2+2 \massBH \Wlambda_{\pdWr} \left(1-2 \massBH \Wnu_{\pdWr}\right)-\massBH \cot\Ktheta\, \Wlambda_{\pdWrWtheta}+2 \massBH \cot\Ktheta\, \Wnu_{\pdWrWtheta}+1\Big]+\bigo(\Kr^2) \,,
            \end{split}
            \\[1.5ex]%
            \begin{split}
                \Slambda &= -\frac{\eu^{-\Wlambda }\, \Wlambda_{\pdWr}}{2} 
                +\frac{\Kr\, \eu^{-2 \Wlambda }}{4 \massBH} \Big[-\Wlambda_{,\Wr\Wtheta\Wtheta} -2 \massBH \Wlambda_\pdWrWr +2 \Wnu_{,\Wr\Wtheta\Wtheta} +2 \massBH {\Wlambda_{\pdWr}}^2-2 \cot\Ktheta\, \Wnu_{\pdWrWtheta} \\
                         &\qquad{}+\Wlambda_{\pdWrWtheta} \left(\cot\Ktheta -2 \Wnu_{\pdWtheta}\right)+4 \Wnu_{\pdWrWtheta}\, \Wnu_{\pdWtheta}\Big]+\bigo(\Kr^2) \,, 
            \end{split}
            \\[1.5ex]%
            \begin{split}
                \Sepsilon &= \frac{\eu^{2 \Wnu -\Wlambda }}{8 \massBH}
            -\frac{\Kr\, \eu^{2 \Wnu -2 \Wlambda }}{8 \massBH^2} \left(\massBH \Wlambda_{\pdWr}-4 \massBH \Wnu_{\pdWr}-2 {\Wnu_{\pdWtheta}}^2+1\right)+\bigo(\Kr^2) \,,
            \end{split}
            \\[1.5ex]%
            \begin{split}
                \Skappa &= \bigo(\Kr^2) \,.
            \end{split}
        \end{align}
    \end{subequations}
    %
    \begin{subequations}
        \begin{align}
            \begin{split}
                \NPPsi_0 &= \bigo(\Kr^2) \,,
            \end{split}
            \\[1.5ex]%
            \begin{split}
                \eqc\NPPsi_1 &\eqc= 
                \frac{\Kr\, \eu^{3 \Wnu -3 \Wlambda }}{32 \sqrt{2} \massBH^3} \Big[\massBH \left(-2 \Wlambda_{,\Wr} \cot\Ktheta +\Wlambda_{,\Wr\Wtheta}-6 \Wnu_{,\Wr\Wtheta}\right)-2 \Wnu_{,\Wtheta} \left(-6 \massBH \Wnu_{,\Wr}+2 \Wnu_{,\Wtheta\Wtheta}+3\right)-8 \Wnu_{,\Wtheta}{}^2 \cot \Ktheta +8 \Wnu_{,\Wtheta}{}^3\Big]+\bigo(\Kr^2)
                \,,
            \end{split}
            \\[1.5ex]%
            \begin{split}
                \NPPsi_2 &= -\frac{\eu^{2 \Wnu -2 \Wlambda }}{24 \massBH^2} \Big[2 \massBH \Wlambda_{,\Wr}-6 \massBH \Wnu_{,\Wr}+4 \Wnu_{,\Wtheta} \cot\Ktheta -4 {\Wnu_{,\Wtheta}}^2+2 \Wnu_{,\Wtheta\Wtheta}+3\Big]
                +\frac{\Kr\, \eu^{2 \Wnu -3 \Wlambda }}{48 \massBH^3} \Big[\Wnu_{,\Wtheta} \left(22 \massBH \Wlambda_{,\Wr} \cot\Ktheta \right.\\
                         &\qquad{}-\left.16 \massBH \Wnu_{,\Wr} \cot\Ktheta +3 \massBH \Wlambda_{,\Wr\Wtheta}+22 \massBH \Wnu_{,\Wr\Wtheta}+8 \cot\Ktheta \right)-4 {\Wnu_{,\Wtheta}}^2 \left(7 \massBH \Wlambda_{,\Wr}-4 \massBH \Wnu_{,\Wr}+2\right)\\
                         &\qquad{}+\Wnu_{,\Wtheta\Wtheta} \left(8 \massBH \Wlambda_{,\Wr}-8 \massBH \Wnu_{,\Wr}+4\right)-44 \massBH^2 \Wlambda_{,\Wr} \Wnu_{,\Wr} +8 \massBH^2 {\Wlambda_{,\Wr}}^2+40 \massBH^2 {\Wnu_{,\Wr}}^2+18 \massBH \Wlambda_{,\Wr}-28 \massBH \Wnu_{,\Wr} \\
                         &\qquad{}-6 \massBH^2 \Wlambda_{,\Wr\Wr}+20 \massBH^2 \Wnu_{,\Wr\Wr}-3 \massBH \Wlambda_{,\Wr\Wtheta} \cot\Ktheta -2 \massBH \Wnu_{,\Wr\Wtheta} \cot\Ktheta -\massBH \Wlambda_{,\Wr\Wtheta\Wtheta}-2 \massBH \Wnu_{,\Wr\Wtheta\Wtheta}+9\Big]+\bigo(\Kr^2) \,, 
            \end{split}
            \\[1.5ex]%
            \begin{split}
                \NPPsi_3 &= \frac{\eu^{\Wnu -2 \Wlambda }}{8 \sqrt{2} \massBH} \Big[2 \Wlambda_{,\Wr} \left(\cot\Ktheta -2 \Wnu_{,\Wtheta}\right)-\Wlambda_{,\Wr\Wtheta}+6 \Wnu_{,\Wr\Wtheta}\Big]
                +\frac{\Kr\, \eu^{\Wnu -3 \Wlambda }}{16 \sqrt{2} \massBH^2} \Big[2 \Wlambda_{,\Wr\Wtheta} \left(-11 \massBH \Wnu_{,\Wr}+5 \Wnu_{,\Wtheta} \cot\Ktheta -6 {\Wnu_{,\Wtheta}}^2\right. \\
                         &\qquad{}+\left.2 \Wnu_{,\Wtheta\Wtheta}+5\right) -\Wlambda_{,\Wr} \left(-4 \massBH \Wnu_{,\Wr} \cot\Ktheta -8 \massBH \Wlambda_{,\Wr\Wtheta}+40 \massBH \Wnu_{,\Wr\Wtheta}+2 \cot\Ktheta \right)+52 \massBH \Wnu_{,\Wr} \Wnu_{,\Wr\Wtheta} \\
                         &\qquad{}-8 \massBH {\Wlambda_{,\Wr}}^2 \left(\cot\Ktheta -2 \Wnu_{,\Wtheta}\right)-12 \massBH \Wlambda_{,\Wr\Wr} \Wnu_{,\Wtheta} +4 \massBH \Wlambda_{,\Wr\Wr} \cot\Ktheta -\massBH \Wlambda_{,\Wr\Wr\Wtheta}+10 \massBH \Wnu_{,\Wr\Wr\Wtheta}-20 \Wnu_{,\Wr\Wtheta} \Wnu_{,\Wtheta} \cot\Ktheta \\
                         &\qquad{}+24 \Wnu_{,\Wr\Wtheta} {\Wnu_{,\Wtheta}}^2-8 \Wnu_{,\Wr\Wtheta} \Wnu_{,\Wtheta\Wtheta}-24 \Wnu_{,\Wr\Wtheta} -2 \Wlambda_{,\Wr\Wtheta\Wtheta} \Wnu_{,\Wtheta}+4 \Wnu_{,\Wr\Wtheta\Wtheta} \Wnu_{,\Wtheta}\Big]+\bigo(\Kr^2) \,, 
            \end{split}
            \\[1.5ex]%
            \begin{split}
                \NPPsi_4 &= \frac{\eu^{-2 \Wlambda }}{4 \massBH} \Big[\Wlambda_{,\Wr} \left(4 \massBH \Wnu_{,\Wr}-2\right)-2 \massBH \Wlambda_{,\Wr\Wr}+\Wlambda_{,\Wr\Wtheta} \left(\cot\Ktheta -2 \Wnu_{,\Wtheta}\right)+4 \Wnu_{,\Wr\Wtheta} \Wnu_{,\Wtheta}-2 \Wnu_{,\Wr\Wtheta} \cot\Ktheta -\Wlambda_{,\Wr\Wtheta\Wtheta}+2 \Wnu_{,\Wr\Wtheta\Wtheta}\Big] \\
                         &\qquad{}+\frac{\Kr\, \eu^{-3 \Wlambda }}{8 \massBH^2} \Big[2 \Wlambda_{,\Wr} \left(4 \massBH^2 \Wlambda_{,\Wr\Wr}+4 \massBH^2 \Wnu_{,\Wr\Wr} + \left(8 \massBH \Wnu_{,\Wr\Wtheta} -4 \massBH \Wlambda_{,\Wr\Wtheta}\right) \left(\cot\Ktheta -2 \Wnu_{,\Wtheta}\right)+2 \massBH \Wlambda_{,\Wr\Wtheta\Wtheta}-4 \massBH \Wnu_{,\Wr\Wtheta\Wtheta}+1\right) \\
                         &\qquad{}+8 \massBH^2 \Wnu_{,\Wr} \Wlambda_{,\Wr\Wr}+8 \massBH {\Wlambda_{,\Wr}}^2 \left(1-2 \massBH \Wnu_{,\Wr}\right)-4 \massBH \Wlambda_{,\Wr\Wr}-4 \massBH^2 \Wlambda_{,\Wr\Wr\Wr}-2 \massBH \Wlambda_{,\Wr\Wr\Wtheta}\, \Wnu_{,\Wtheta}+4 \massBH \Wnu_{,\Wr\Wr\Wtheta}\, \Wnu_{,\Wtheta} \\
                         &\qquad{}+\massBH \Wlambda_{,\Wr\Wr\Wtheta} \cot\Ktheta -2 \massBH \Wnu_{,\Wr\Wr\Wtheta} \cot\Ktheta -\massBH \Wlambda_{,\Wr\Wr\Wtheta\Wtheta}+2 \massBH \Wnu_{,\Wr\Wr\Wtheta\Wtheta}-4 \Wnu_{,\Wr\Wtheta} \Wnu_{,\Wtheta}-\Wlambda_{,\Wr\Wtheta} \left(20 \massBH \Wnu_{,\Wr\Wtheta}-2 \Wnu_{,\Wtheta}+\cot\Ktheta \right) \\
                         &\qquad{}+2 \Wnu_{,\Wr\Wtheta} \cot\Ktheta +2 \massBH {\Wlambda_{,\Wr\Wtheta}}^2 +32 \massBH {\Wnu_{,\Wr\Wtheta}}^2+\Wlambda_{,\Wr\Wtheta\Wtheta}-2 \Wnu_{,\Wr\Wtheta\Wtheta}\Big]+\bigo(\Kr^2) \,. 
            \end{split}
        \end{align}
    \end{subequations}
    \endgroup
    \end{widetext}

\section{Numerical solution}
    \label{app:numerical_solution}
    In Sec.~\ref{sec:schwarzschild_deformaiton}, we have found the
    solution for the coordinates, tetrad, and, hence, all subsequent
    variables analytically in the form of a series. This is, of course,
    not the only possible way since we can find the solution to the
    differential equations numerically. We have done so to cross-check
    the solution and to have a tool better suited for figure plotting.
    In this section, we briefly present the numerical problem, and
    methods used to solve it and discuss the precision of the result.

    First, we start from the horizon $\HH$ where we have the vector $\NPn^a$
    at each point ($\Wt_0$, $\Wr_0=2\massBH$, $\Wtheta_0$, $\Wphi_0$).
    These are the initial values for the geodesic equation, and thus we obtain
    a single geodesic, affinely parameterized by $\Kr$, along which $\Wt_0$,
    $\Wtheta_0$, $\Wphi_0$ remain constant. The Bondi-like coordinate $\Ktheta$
    is, on the horizon, identical to $\Wtheta$. We can set up the IH coordinate
    system by varying $\Wt_0$, $\Wtheta_0$, $\Wphi_0$ and evolving a sufficient
    number of geodesics. Second, we want to complete the tetrad by finding the
    parallelly propagated vectors $\NPl^a$ and $\NPm^a$ ($\cconj{\NPm}^a$ is
    given as a complex conjugate). 

    Owing to the symmetries, the only interesting coordinates are 
    $\Kr$ and $\Ktheta$.
    Hence, we have to choose a sample of values of $\Wtheta \horeq \Ktheta$
    at the horizon, find all variables alongside the corresponding
    geodesic, and then interpolate in this angular direction.

    \subsection{Geodesic equation along \texorpdfstring{$\boldsymbol{\NPn^a}$}{n}}
        \label{app:num:differential_equations}
        For the evolution alongside the geodesics, we solve the Hamilton
        equations where components of $\NPn^a$ are the generalized momentums. 
        Although we could integrate for $\NPl^a$ and $\NPm^a$ once the
        coordinates are available, it is more convenient to solve for all these 
        variables at once.

        To find the geodesic motion, we will use the Hamiltonian formulation.
        We define the Hamiltonian as, see e.g.~\cite{Stewart-1993}, 
        \begin{equation}
            H(q,p,\Kr) = - \frac{1}{2} g^{ab}(q)\, p_a p_b \,,
        \end{equation}
        where $q$ are the coordinates, $p$ are the momenta and $\Kr$ is the affine
        parameter of the geodesic. The sign has been chosen so that
        $\frac{\d \Wt}{\d \Kr}$ is negative and hence the congruence is ingoing.
        However, since we only know the value of $F(\Wr,\Wtheta)$ (and most importantly
        its angular derivative) on the horizon, it will be easier to perform
        the geodesic computation in the coordinates of~\eqref{eq:weyl_superimposed}.
        More on this topic in just a moment.
        The Hamiltonian in these coordinates is
        \begin{align}
            \begin{aligned}
                H = \frac{1}{2 \Wr^2} &\left(
                -\frac{\eu^{-2\Wnu} \Wr^2}{\Schf}\, {p_\Wt}^2
                + \eu^{-2\Wlambda + 2\Wnu} \Schf \Wr^2\, {p_\Wr}^2 \right.\\
                    &\qquad{}\left.{}+ \eu^{-2\Wlambda + 2\Wnu} {p_\Wtheta}^2
                + \frac{\eu^{2\Wnu}}{\sin^2\Wtheta}\, {p_\Wphi}^2
                \right) .
            \end{aligned}
        \end{align}

        The Hamilton equations are given by
        \begin{align}
            \begin{aligned}
                \frac{\d q^a}{\d \Kr} &= \frac{\pd H}{\pd p_a} \,, \\
                \frac{\d p_a}{\d \Kr} &= - \frac{\pd H}{\pd q^a} \,.
            \end{aligned}
        \end{align}

        The tetrad on the horizon can be obtained in the same 
        manner as in~\ref{sec:tetrad_on_the_horizon} and reads
        \begin{align}
            \begin{aligned}
                \boldsymbol{\NPl} &\horeq \frac{1}{2}\,\pd_\Wt \,, \\
                \boldsymbol{\NPn} &\horeq \frac{\eu^{-2\Wnu}}{\Schf}\, \pd_\Wt
                    - \eu^{-\Wlambda}\, \pd_\Wr \,, \\
                \boldsymbol{\NPm} &\horeq \frac{\eu^{-\Wnu} \Wnu_\pdWtheta}{\sqrt{2}}\, \pd_\Wt
                    + \frac{\eu^{-\Wlambda + \Wnu}}{2\sqrt{2}\,\massBH}\, \pd_\Wtheta 
                    + \frac{\ii\,\eu^\Wnu}{2\sqrt{2}\,\massBH\sin\Wtheta}\, \pd_\Wphi \,.
            \end{aligned}
        \end{align}
        Besides the tetrad, we are interested in the value of the covector
        $\NPn_a$ which is
        \begin{equation}
            \NPn_\mu \d x^\mu \horeq \d\Wt + \frac{\eu^{\Wlambda - 2\Wnu}}{\Schf}\, \d\Wr \,.
        \end{equation}

        Since the component $p_\Wphi \horeq 0$ and $\frac{\d p_\Wphi}{\d \Kr} = 0$
        we have $p_\Wphi = 0$. Similarly, $p_\Wt = 1$.
        However, the momentum component $p_\Wr$ is singular on the horizon
        because $\Schf \horeq 0$. Therefore, the equations for $\frac{\d p_\Wr}{\d \Kr}$
        and $\frac{\d p_\Wtheta}{\d \Kr}$ are also singular. This should not
        surprise us, since~\eqref{eq:weyl_superimposed} already diverges
        on the horizon.  Should we have
        stayed with the coordinates given by~\eqref{eq:star_transformation}?
        Although we could, we would have to prescribe both $\Wnu$ and $\Wlambda$
        everywhere to be able to compute $F(\Wr,\Wtheta)$. Using the
        coordinates~\eqref{eq:weyl_superimposed} we can make do with less
        knowledge. We shall still need $\Wnu$, but for $\Wlambda$ its value
        on the horizon will be enough. This is useful since $\Wlambda$ is
        generally not available analytically in the literature. This is possible
        because the functions $\Wnu$ and $\Wlambda$ have to satisfy the Einstein
        equations, and we will use them to integrate the value of $\Wlambda$.
        On the other hand, we need to remedy the divergence on the horizon.

        Given that the function $\Schf$ is singular, we can define new
        variables by multiplying with a power of $\Schf$ as
        \begin{align}
                \hat{p}_\Wr &= \Schf^k p_\Wr \,, &
                \hat{p}_\Wtheta &= \Schf^k p_\Wtheta \,.
        \end{align}
        While simple inspection of the equation for $\frac{\d p_\Wr}{\d \Kr}$
        seems to indicate that $k$ must be at least $\geq 2$ it is
        sufficient to use $k=1$ for both $p_\Wr$ and $p_\Wtheta$.

        Because the horizon is singular in the Weyl coordinates, we need 
        separate formulas for \emph{on} and \emph{off} horizon computation.
        \begin{alignat}{2}
            \frac{\d\Wt    }{\d \Kr} &= \frac{\eu^{-2\left(\Wlambda+\Wnu\right)}\!\left(-\eu^{2\Wlambda} \Wr - \eu^{4\Wnu} \!\left(\Schf - 1\right) \hat{p}_\Wr \Wt\right)}{\Wr} 
                                   & &\horeq \frac{\eu^{-\Wlambda} \Wt}{2\massBH} - \eu^{-2\Wnu} \,,\nonumber\\
            \frac{\d\Wr    }{\d \Kr} &= \eu^{-2\Wlambda + 2\Wnu} \hat{p}_\Wr 
                                   & &\horeq \eu^{-\Wlambda} \,,\nonumber\\
            \frac{\d\Wtheta}{\d \Kr} &= \frac{\eu^{-2\Wlambda + 2\Wnu}}{\Schf \Wr^2} \hat{p}_\Wtheta 
                                   & &\horeq 0 \,,\nonumber\\
            \frac{\d\Wphi  }{\d \Kr} &= \frac{\eu^{2\Wnu}}{\Wr^2 \sin^2\Wtheta} p_\Wphi 
                                   & &\horeq 0 \,,
        \end{alignat}

        \begin{widetext}
        \begin{alignat}{2}
            \frac{\d p_\Wt}{\d \Kr} &= 0
                                   & &\horeq 0 \,, \nonumber\\
            \frac{\d \hat{p}_\Wr}{\d \Kr} &= 
            \eu^{-2\Wlambda+2\Wnu} \left(\frac{2\hat{p}_\Wtheta{}^2
                + \left(1 - \Schf\right) \hat{p}_\Wr{}^2 \,\Wr^2}{2 \Schf \Wr^3}
                + \left(\hat{p}_\Wr{}^2 + \frac{\hat{p}_\Wtheta{}^2}{\Schf \Wr^2}\right)
                \!\Big(\Wlambda_\pdWr - \Wnu_\pdWr\Big)\!\right)
                - \frac{\eu^{-2 \Wnu} \left(1 - \Schf + 2 \Schf \Wr\, \Wnu_\pdWr\right)}
                {2 \Schf \Wr}
                                        & &\horeq \eu^{-\Wnu} \left(\Wlambda_\pdWr - 2\Wnu_\pdWr\right) ,\nonumber\\
            \frac{\d \hat{p}_\Wtheta}{\d \Kr} &=
            \eu^{-2\Wlambda+2\Wnu} \left(\frac{\left(1 - \Schf\right) \hat{p}_\Wr 
                \hat{p}_\Wtheta}{\Schf \Wr}
                + \left(\hat{p}_\Wr{}^2 + \frac{\hat{p}_\Wtheta{}^2}{\Schf \Wr^2}\right)
                \!\Big(\Wlambda_\pdWtheta - \Wnu_\pdWtheta\Big)\!\right)
                - \eu^{-2 \Wnu} \Wnu_\pdWtheta
                                   & &\horeq 0 \,,\nonumber\\
            \frac{\d p_\Wphi}{\d \Kr} &= 0
                                   & &\horeq 0 \,,
        \end{alignat}
        \end{widetext}

        The vectors $\NPl^a$ and $\NPm^a$ are parallelly transported:
        \begin{align}
            \begin{aligned}
                \frac{\d \NPl^a}{\d \Kr} &= \ChGamma^a{}_{bc}\,\NPn^b \NPl^c \,, \\
                \frac{\d \NPm^a}{\d \Kr} &= \ChGamma^a{}_{bc}\,\NPn^b \NPm^c \,.
            \end{aligned}
        \end{align}
        We know the Weyl metric, and the corresponding Christoffel
        symbols are easy to find. However, because of the singularity on the
        horizon, it is essential to reduce all terms before evaluating them. We
        shall not write the final expressions down as they are lengthy and
        simple to obtain.
            
        We need to complete the differential equations presented in the
        previous section by the initial data given on the horizon. All the
        values can be found in the same way as the series expansion of the
        tetrad. Let us summarize the results.

        The coordinates and momentums are
        \begin{align}
            &\begin{aligned}
                \Wt &\horeq 0 \,,\\
                \Wr &\horeq 2\massBH \,,\\
                \Wtheta &\horeq \Ktheta \,,\\
                \Wphi &\horeq 0 \,,
            \end{aligned}
            &
            &\begin{aligned}
                p_\Wt &\horeq 1 \,,\\
                \hat{p}_\Wr &\horeq \eu^{\Wlambda - 2\Wnu} \,,\\
                \hat{p}_\Wtheta &\horeq 0 \,,\\
                p_\Wphi &\horeq 0 \,.
            \end{aligned}
        \end{align}
        where we have chosen $\Wt$ and $\Wphi$ arbitrarily thanks to the
        symmetries.

        For the remaining vectors we have:
        \begin{align}
            &\begin{aligned}
                \NPl^\Wt &\horeq \frac{1}{2} \,,\\
                \NPl^\Wr &\horeq 0\vphantom{\frac{1}{1}} \,,\\
                \NPl^\Wtheta &\horeq 0\vphantom{\frac{1}{1}} \,,\\
                \NPl^\Wphi &\horeq 0\vphantom{\frac{1}{1}} \,.
            \end{aligned}
            &
            &\begin{aligned}
                \NPm^\Wt &\horeq \frac{\eu^{-\Wnu} \Wnu_\pdWtheta}{\sqrt{2}} \,,\\
                \NPm^\Wr &\horeq 0\vphantom{\frac{1}{1}} \,,\\
                \NPm^\Wtheta &\horeq \frac{\eu^{-\Wlambda + \Wnu}}{\sqrt{2}\,\Wr} \,,\\
                \NPm^\Wphi &\horeq \frac{\eu^{\Wnu}}{\sqrt{2}\,r\,\sin{\Wtheta}} \,.
            \end{aligned}
        \end{align}

    \begin{figure}
        \centering
        \figinput{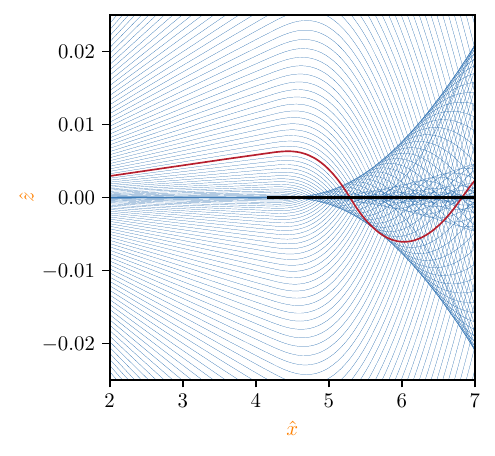}
        \caption{Region around the equator with unclipped geodesics
            illustrating the oscillating behavior of the solution
            when extended beyond the touching point at the disk.
            One arbitrary geodesic has been highlighted. Note that
            the plot does not have equal coordinate scales and shows
            a region farther away from the black hole than other plots.
            Moreover, a denser set of geodesics has been used for this
            plot and the geodesics were mirrored with respect to the 
            $\WEx$ axis to fill the region with $\WEz < 0$.}
        \label{fig:geodesics_at_disk}
    \end{figure}
    \begin{figure}
        \centering
        \figinput{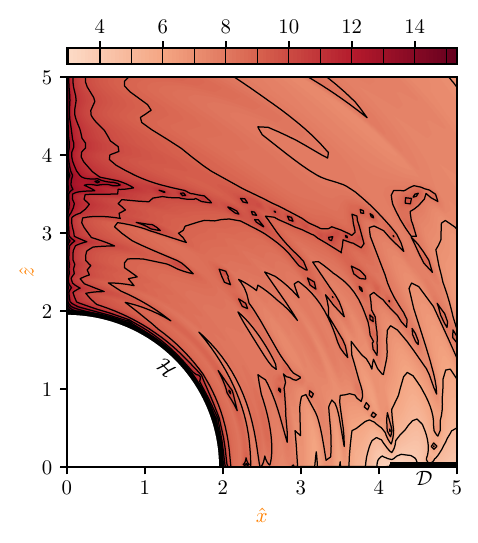}
        \caption{Precision of the numerical solution as demonstrated by
        $\log \abs{\Spi / \left[\Spi - \left(\Salpha + \cconj{\Sbeta}\right)\right]}$.
        The value corresponds to the number of correct digits. The data for
        the figure
        has been made by evaluating the 2-dimensional interpolating
        functions representing the spin coefficients on a grid of 
        $\approx10^2$ points in each dimension, this grid was then
        used by matplotlib to construct the contours.}
        \label{fig:error_estimate}
    \end{figure}
    \begin{figure}
        \centering
        \figinput{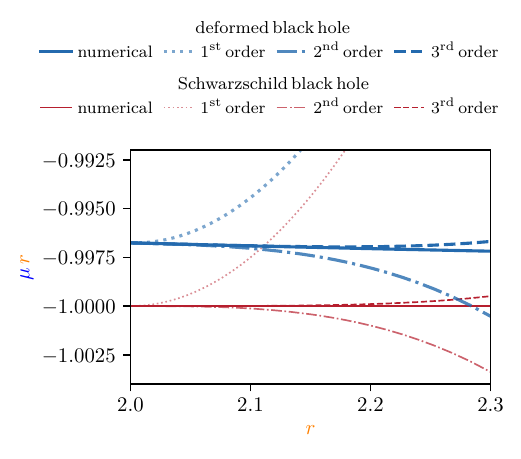}
        \caption{The value of $\Smu$ computed by the numerical solution is
            compared to its analytic expansion for both the example deformation
            by a thin disk (thicker blue lines) and the pure Schwarzschild term
            (red thinner lines). The expansion, demonstrated up to the 
            first, second and third
            orders, is displayed by dashed
            and dotted lines, while the numerical solution is drawn by a solid
            line. The values are multiplied by $\Wr$ to highlight the
            differences.}
        \label{fig:expansion_vs_numerics}
    \end{figure}

    \subsection{Numerical methods}
        While the disk~\cite{Kofron-2023} used as an example has analytic
        expressions for both metric functions $\Wlambda$ and $\Wnu$ available
        in the whole region outside the black hole, usually the function
        $\Wlambda$ is known only at the horizon and symmetry axis. Therefore,
        while we could have used directly the numerical value of $\Wlambda$ and
        $\Wnu$, we had used only their value on the horizon paired with the
        derivatives of $\Wnu$ outside to numerically integrate the function
        using the Einstein equations~\eqref{eq:lambda_nu_einstein_eq:lambda_r}
        and~\eqref{eq:lambda_nu_einstein_eq:lambda_theta}.

        To solve the set of differential equations and initial data
        from Sec.~\ref{app:num:differential_equations}
        given at their singular point (the horizon), we used
        Mathematica's sixth-order \name{ExplicitRungeKutta} method with the
        \name{DoubleStep} option, which provided the numerical solution with
        precision $\approx 10^{-9}$.  Because the implemented step-size control
        led to a very small step size near the horizon, we had to use multiple
        precision arithmetic available in Mathematica.

        For interpolation, we wanted to use the discrete cosine 
        transform (DCT). However, we encountered issues caused by the Gibbs
        effect. Hence, we decided to use the same built-in Mathematica's
        interpolation as was used by the radial solution, but the number of the
        used geodesics had to be increased.
        While in Fig.~\ref{fig:geodesics} only a few geodesics were shown
        to illustrate the problem, for the numerical solution, 200 geodesics were 
        computed, the starting angles for the geodesics were chosen to be at the
        Chebyshev nodes of the first kind scaled to the interval 
        $(0, \frac{\Cpi}{2})$, \cite{Boyd-2001}.

        One notable difficulty is that the Hamilton equations may become
        invalid when the geodesic touches or comes very close to the disk.
        This issue is illustrated in Fig.~\ref{fig:geodesics_at_disk}.
        Each geodesic was clipped at the point where it touches the disk
        to prevent overlapping coordinates, which could cause problems. The
        crossing of geodesics at the disk is expected and has been
        discussed in~\cite{Krishnan-2012}.

    \subsection{Numerical errors}
        To check the validity and precision of the numerical solution, we
        verified selected identities that should be satisfied by the computed
        quantities.
        One particular example is shown in Fig.~\ref{fig:error_estimate}.

        Moreover, we also checked that the numerical solution and the series
        yield the same results near the horizon where the series corresponds
        to the full solution well. In Fig.~\ref{fig:expansion_vs_numerics},
        we illustrated the difference for one particular NP scalar.

\section{Model Weyl metric: black hole and disk}
    \label{app:disk}
    The Morgan--Morgan class of disks is comprised of disks with mass $\massDisk$
    and Newtonian density profile
    \begin{equation}
        \begin{aligned}
            \sigma^{(n)}_\text{MM} (\Wrho \leq \diskR)& = \frac{(2n + 1) 
            \massDisk}{2\Cpi \diskR^2} \left( 1 - \frac{\Wrho^2}{\diskR^2} \right)^{n-\frac{1}{2}}\,,\\
            \sigma^{(n)}_\text{MM} (\Wrho \geq \diskR)& = 0 \,.
        \end{aligned}
    \end{equation}  
    The Newtonian potential $\Wnu$ was found in \cite{Morgan-1969}. 

    By performing the Kelvin transformation, we get inverted Morgan--Morgan
    disks---holey disks with an inner rim at $\Wrho=\diskR$ and density
    distributions as 
    \begin{align}
        \begin{aligned}
            \sigma_\text{iMM}^{(n)} (\Wrho \leq \diskR) &= 0\,, \\
            \sigma_\text{iMM}^{(n)} (\Wrho \geq \diskR) &= 
                \frac{2^{2n}(n!)^2 \massDisk\, \diskR}{(2n)!\, \Cpi^2 \Wrho^3} 
                \left( 1 - \frac{\diskR^2}{\Wrho^2} \right)^{n-\frac{1}{2}} \,. 
        \end{aligned}
    \end{align}
    Such a disk can be, in Weyl coordinates, easily superposed with the 
    Schwarzschild black hole.

    The exact analytical solution for both of the metric functions has been 
    recently provided in \cite{Kofron-2023}. For the sake of simplicity, we use 
    the simplest case $n=2$ (for $n=1$ the density has a sharp onset) as a 
    particular example. We thus used the following expressions
    given in the oblate spheroidal coordinates for our numerical
    calculations:
    \begin{widetext}
        \begingroup
        \allowdisplaybreaks
        \begin{subequations}
        \begin{align}
            \begin{split}
                \Wnu_\text{d} &= -\frac{\massDisk}{4 \diskR\Cpi \distDisk^{\frac{9}{2}}} \,\mathcal{P}^\nu_1(\OSxi,\OSzeta)
                \arccot\!\left( \frac{|\OSxi|}{\sqrt{\distDisk}} \right)
                +\frac{\massDisk^2}{12 \diskR\Cpi \distDisk^4} \mathcal{P}^\nu_0(\OSxi,\OSzeta) \,,
            \end{split}
            \\[1.5ex]
            \begin{split}
            \Wlambda_\text{d} &= 
            \frac{\massDisk^2}{1440 \diskR^2\Cpi^2\distDisk^9} \mathcal{P}^d_0(\OSxi,\OSzeta)
            -\frac{\massDisk^2}{48 \diskR^2\Cpi^2\distDisk^{\frac{19}{2}}} \mathcal{P}^d_1(\OSxi,\OSzeta)
            \arccot\!\left( \frac{|\OSxi|}{\sqrt{\distDisk}} \right)
            +\frac{\massDisk^2}{32 \diskR^2\Cpi^2\distDisk^{10}} \mathcal{P}^d_2(\OSxi,\OSzeta)
            \arccot^2\!\left( \frac{|\OSxi|}{\sqrt{\distDisk}} \right) ,
            \end{split}
            \\[1.5ex]
            \begin{split}
                \Wlambda_\text{i} &=
                \frac{2\left(\diskR^2+\massBH^2\right)^2\massDisk}{\massBH^5\Cpi}\,\Bigl[
                    \arctan\!\left(-\massBH\OSzeta + \diskR\OSxi\,, l_\text{p}(\OSxi,\OSzeta)\right)
                +\arctan\!\left(\massBH\OSzeta+\diskR\OSxi\,, l_\text{m}(\OSxi,\OSzeta)\right) \Bigr] \\
                &\qquad+\frac{\massDisk}{24 \diskR \massBH^5\distDisk^{\frac{9}{2}}}\Biggl[
                \left(-2\massBH \distDisk\, \mathcal{P}^i_1(\OSxi,\OSzeta)
                + \Bigl(\mathcal{P}^i_2(\OSxi,\OSzeta)+\mathcal{P}^i_3(\OSxi,\OSzeta)\Bigr)
                \,\arccot\!\left(\frac{|\OSxi|}{\distDisk}\right)\!\right)l_\text{p}(\OSxi,\OSzeta) \\
                &\phantom{\qquad+\frac{\massDisk}{24 \diskR \massBH^5\distDisk^{\frac{9}{2}}}\Bigl[}
                +\left(2\massBH \distDisk\, \mathcal{P}^i_1(-\OSxi,\OSzeta)
                + \Bigl(\mathcal{P}^i_2(\OSxi,\OSzeta)-\mathcal{P}^i_3(\OSxi,\OSzeta)\Bigr)
            \,\arccot\!\left(\frac{|\OSxi|}{\distDisk}\right)\!\right)l_\text{m}(\OSxi,\OSzeta)\Biggr]\,,
            \end{split}
        \end{align}
        \label{app:eq:disk_potentials}
        \end{subequations}
        \endgroup
    \end{widetext}
    where $\mathcal{P}^{\bullet}_\bullet(\OSxi,\OSzeta)$ are polynomials in their 
    arguments whose particular form is postponed to Sec.~\ref{app:disk_polynomials}. 
    The function $\arctan(x,y)$ with two arguments gives the proper angle $\phi$ 
    taking into account the quadrants in which a point $(x,y)$ lies; i.e.\ 
    such that $x=\cos\phi$ and $y=\sin\phi$. The distance $\distDisk=1+\OSzeta^2-\OSxi^2=(\Wrho^2+\Wz^2)/\diskR^2$ and the functions $l_\text{p}(\OSxi,\OSzeta)$ and $l_\text{m}(\OSxi,\OSzeta)$
    are identical to $l_\text{p}$ and $l_\text{m}$ from~\eqref{eq:lplm} but
    expressed in the oblate spheroidal coordinates:
    \begin{align}
        \begin{aligned}
            l_\text{p}&=\sqrt{\left(\distDisk\,\OSxi\,\OSzeta+\massBH\right)^2
                +\distDisk^2\left(1-\OSxi^2\right)\left(1+\OSzeta^2\right)}\,, \\
            l_\text{m}&=\sqrt{\left(\distDisk\,\OSxi\,\OSzeta-\massBH\right)^2
                +\distDisk^2\left(1-\OSxi^2\right)\left(1+\OSzeta^2\right)}\,.
        \end{aligned}
    \end{align}

    \pagebreak
    The oblate spheroidal coordinates $\OSzeta\in[0,\infty)$ and 
    $\OSxi\in[-1,+1]$, in which the expressions related to the disk take the
    simplest form, are defined by
    \begin{align}
        \Wrho^2 &= \diskR^2 \left(1 + \OSzeta^2\right)\!\left(1 - \OSxi^2\right) , &
        \Wz &= \diskR\, \OSzeta\, \OSxi \,.
    \end{align}
    The inverse relations read
    \begin{align}
        \OSzeta &= \frac{\sqrt{2} \left|\Wz\right|}{\sqrt{\sqrt{u^2 + 4\diskR^2 \Wz^2} - u}} \,,&
        \OSxi &= \frac{\Wz}{\diskR \OSzeta} \,,
    \end{align}
    where we set $u = \Wrho^2 - \diskR^2 + \Wz^2$.

    The final solution of the Einstein field equations (EFEs) is given by
    \begin{align}
        \Wnu &= \Wnu_\text{S} + \Wnu_\text{d}\,,&
        \Wlambda &= \Wlambda_\text{S} + \Wlambda_\text{i}+\Wlambda_\text{d}\,,
    \end{align}
    where $\Wnu_\text{S}$ is the Schwarzschild Newtonian potential and 
    $\Wnu_\text{d}$ is the Newtonian potential of the disk itself. Owing to the 
    non-linearity of EFEs, the $\Wlambda$ has three different contributions: 
    from the Schwarzschild black hole solely $\Wlambda_\text{S}$, from the disk solely 
    $\Wlambda_\text{d}$ and the interaction term $\Wlambda_\text{i}$.

    If one adheres to the counter-rotating dust streams interpretation of a 
    static disk, one has to ensure that the speed of individual particles is 
    subluminal (i.e.\ $0<v^2<1$). The individual dust particles should follow a
    time-like geodesics of constant radius in the equatorial plane. Then their speed 
    $v$ can be calculated as
    \begin{align}
        v^2 &= \frac{\Wrho\, \Wnu_{,\Wrho}}{1-\Wrho\, \Wnu_{,\Wrho}}\,,
    \end{align}
    in Weyl coordinates.
    
    This gives us a constrain on the admissible parameter space $\massBH$, $\massDisk$
    and $\diskR$. For our particular choice of disks, we have
    \begin{equation}
        0 < \frac{8 \Wrho ^5 \sqrt{\massBH^2+\Wrho ^2}}
        {8 \Wrho^5 \left( \sqrt{\massBH^2+\Wrho ^2} - \massBH \right)
            -p\massDisk \sqrt{\massBH^2+\Wrho ^2}}-1 < 1 \,,
    \end{equation}
    where
    \begin{equation}
        p = 15\diskR^4 - 24\diskR^2 \Wrho^2 + 8\Wrho^4\,,
    \end{equation}
    which has to hold for all $\Wrho > \diskR$.

    In other words, the inner rim of the disc must be positioned above the 
    equatorial photon orbit (whose radius is $3\massBH$ for Schwarzschild). The radius 
    of this orbit is influenced by the presence of the gravitating disk itself.

    In principle, since the orbits between the photon orbit and the innermost
    stable circular orbit (ISCO), \cite{Misner-2017}, are unstable, in an
    astrophysically relevant system, the inner rim of the disk should be set
    above the ISCO. As we want to make the gravitational influence of the disk as
    prominent as possible, we push the parameters to the extreme. A detailed
    discussion of the stability of these discs can be found in \cite{Semerak-2003}.

    \subsection{The disk polynomials}
        \label{app:disk_polynomials}
        The explicit form of the polynomials appearing in the disk solution from%
        ~\cite{Kofron-2023} 
        is as follows
        \begin{widetext}
            \begingroup
                \allowdisplaybreaks
                \begin{align}
                    \begin{split}
                        \mathcal{P}^\nu_0 (\OSxi,\OSzeta)&=\left[40 \OSzeta ^6+8 \OSzeta ^4 \left(\OSxi ^2+2\right)+\OSzeta ^2 \left(39 \OSxi ^4-6 \OSxi ^2-33\right)+9 \left(\OSxi
                        ^2-1\right)^2 \left(2 \OSxi ^2-1\right)\right] \left|\OSxi\right|,
                    \end{split}
                    \\[1.5ex]%
                    \begin{split}
                        \mathcal{P}^\nu_1(\OSxi,\OSzeta)&= -8 \left(\OSzeta ^2+3\right) \OSxi ^6+\left(35 \OSzeta ^4+54 \OSzeta ^2+27\right) \OSxi ^4-2 \left(\OSzeta ^2+1\right) \left(4 \OSzeta ^4+23 \OSzeta ^2+7\right) \OSxi ^2\\
                        &\qquad +\left(\OSzeta
                       ^2+1\right)^2 \left(8\OSzeta^4 + 8\OSzeta^2 + 3\right)+8 \OSxi ^8 .
                    \end{split}
                    \\[3ex]
                    \begin{split}
                        \mathcal{P}^d_0(\OSxi,\OSzeta) &= \left(\OSxi
                            ^2-1\right) \Bigl[320 \OSzeta ^{14} \left(201 \OSxi ^2-32\right)+320 \OSzeta ^{12} \left(235 \OSxi ^4+484 \OSxi ^2-160\right)\\
                                        &\qquad{}+240 \OSzeta ^{10} \left(1287 \OSxi ^6-698 \OSxi ^4+679 \OSxi
                        ^2-448\right)\\
                        &\qquad+4 \OSzeta ^8 \left(40449 \OSxi ^8+100729 \OSxi ^6-153381 \OSxi ^4+61179 \OSxi ^2-30976\right)\\
                        &\eqc\qquad+\OSzeta ^6 \left(\OSxi^2 -1\right) \left(228521 \OSxi ^8-9179 \OSxi ^6+317611 \OSxi ^4-262969
                        \OSxi ^2+86016\right)\\
                        &\qquad+3 \OSzeta ^4 \left(\OSxi ^2-1\right)^2 \left(12192 \OSxi ^8+84557 \OSxi ^6-12478 \OSxi ^4+54017 \OSxi ^2-12288\right)\\
                        &\qquad+9 \OSzeta ^2 \left(\OSxi ^2-1\right)^3 \left(2004 \OSxi
                        ^8+2364 \OSxi ^6+8059 \OSxi ^4-4451 \OSxi ^2+1024\right)\\
                        &\qquad-\left(\OSxi ^2-1\right)^4 \left(304 \OSxi ^8-6436 \OSxi ^6+7584 \OSxi ^4-4501 \OSxi ^2+1024\right)\Bigr] \,,
                    \end{split}
                    \\[1.5ex]%
                    \begin{split}
                        \mathcal{P}^d_1(\OSxi,\OSzeta) &= \left(\OSxi ^2-1\right) 
                        \Bigl[832 \OSzeta ^{16}+64 \OSzeta ^{14} \left(73 \OSxi ^2+33\right)+16 \OSzeta ^{12} \left(649 \OSxi ^4+358 \OSxi ^2-15\right)\\
                        &\qquad +4 \OSzeta
                        ^{10} \left(4313 \OSxi ^6+1917 \OSxi ^4-1157 \OSxi ^2-1393\right)+5 \OSzeta ^8 \left(2797 \OSxi ^8+2632 \OSxi ^6-3046 \OSxi ^4-64 \OSxi ^2-1359\right)\\
                        &\eqc\qquad +2 \OSzeta ^6 \left(\OSxi^2 -1\right) \left(4789 \OSxi
                        ^8+6849 \OSxi ^6+3889 \OSxi ^4-5321 \OSxi ^2+1794\right)\\
                        &\qquad +2 \OSzeta ^4 \left(\OSxi ^2-1\right)^2 \left(1200 \OSxi ^8+5395 \OSxi ^6+2331 \OSxi ^4+4203 \OSxi ^2-529\right)\\
                        &\qquad+6 \OSzeta ^2 \left(\OSxi
                        ^2-1\right)^3 \left(72 \OSxi ^8+328 \OSxi ^6+653 \OSxi ^4-193 \OSxi ^2+40\right)\\
                        &\qquad +3 \left(\OSxi ^2-1\right)^4 \left(80 \OSxi ^6-64 \OSxi ^4+38 \OSxi ^2-9\right)\Bigr] \left|\OSxi\right| ,
                    \end{split}
                    \\[1.5ex]%
                    \begin{split}
                        \mathcal{P}^d_2(\OSxi,\OSzeta) &= \left(\OSzeta ^2+1\right) \left(\OSxi ^2-1\right) 
                            \Bigl[64 \OSzeta ^{16}+64 \OSzeta ^{14} \left(7 \OSxi ^2+5\right)+16 \OSzeta ^{12} \left(157 \OSxi ^4-18 \OSxi ^2+45\right)\\
                            &\qquad +4 \OSzeta
                        ^{10} \left(949 \OSxi ^6+785 \OSxi ^4-1025 \OSxi ^2+251\right)\\
                            &\qquad +5 \OSzeta ^8 \left(1241 \OSxi ^8-560 \OSxi ^6+730 \OSxi ^4-1288 \OSxi ^2+197\right)\\
                            &\eqc\qquad +4 \OSzeta ^6 \left(\OSxi^2 -1\right) \left(949 \OSxi
                        ^8+1649 \OSxi ^6-1356 \OSxi ^4+929 \OSxi ^2-171\right)\\
                            &\qquad +2 \OSzeta ^4 \left(\OSxi ^2-1\right)^2 \left(1256 \OSxi ^8+942 \OSxi ^6+2453 \OSxi ^4-606 \OSxi ^2+155\right)\\
                            &\eqc\qquad +4 \OSzeta ^2 \left(\OSxi ^2-1\right)^3 \left(2 \OSxi -1\right) \left(2 \OSxi +1\right)
                        \left(\OSxi ^2+4\right) \left(28 \OSxi ^4-3 \OSxi ^2+5\right)\\
                            &\qquad +\left(\OSxi ^2-1\right)^4 \left(64 \OSxi ^8-64 \OSxi ^6+80 \OSxi ^4-44 \OSxi
                        ^2+9\right)\Bigr] \,,
                    \end{split}
                    \\[3ex]
                    \begin{split}
                        \mathcal{P}^i_1(\OSxi,\OSzeta)&=24 \diskR^3 \OSzeta  \left(\OSzeta^2-\OSxi ^2+1\right)^3-12 \diskR^2 \massBH \OSxi  \left(\OSzeta ^2-\OSxi ^2+1\right)^2 \left(2 \OSzeta ^2+\OSxi ^2-1\right)\\
                        &\qquad+4 \diskR \OSzeta  \massBH^2 \left[10 \OSzeta^6+\OSzeta ^4 \left(26-20 \OSxi ^2\right)
                        +\OSzeta ^2 \left(25 \OSxi ^4-47 \OSxi ^2+22\right)
                        -3 \left(\OSxi ^2-1\right)^2 \left(5 \OSxi^2-2\right)\right]\\
                        &\qquad-\massBH^3 \OSxi  \left[40 \OSzeta ^6+8 \OSzeta^4
                        \left(\OSxi ^2+2\right)+\OSzeta ^2 \left(39 \OSxi^4-6 \OSxi^2
                        -33\right)+9 \left(\OSxi ^2-1\right)^2 \left(2 \OSxi ^2-1\right)\right],
                    \end{split}
                    \\[1.5ex]%
                    \begin{split}
                        \mathcal{P}^i_2(\OSxi,\OSzeta)&=-6\, \Bigl[8 \diskR^4 \left(\OSzeta ^2-\OSxi
                        ^2+1\right)^4+4 \diskR^2 \massBH^2 \left(\OSzeta ^2-\OSxi ^2+1\right)^2 \left(4 \OSzeta ^4+\OSzeta ^2 \left(7-5 \OSxi ^2\right)+4 \OSxi ^4-7 \OSxi ^2+3\right)\\
                        &\qquad +\massBH^4 \Bigl(8 \OSzeta ^8-8 \OSzeta ^6 \left(\OSxi
                        ^2-3\right)+\OSzeta ^4 \left(35 \OSxi ^4-54 \OSxi ^2+27\right)+\OSzeta ^2 \left(-8 \OSxi ^6+54 \OSxi ^4-60 \OSxi ^2+14\right)\\
                        &\qquad+\left(\OSxi ^2-1\right)^2 \left(8 \OSxi ^4-8 \OSxi
                        ^2+3\right)\Bigr)\Bigr] \,,
                    \end{split}
                    \\[1.5ex]%
                    \begin{split}
                        \mathcal{P}^i_3(\OSxi,\OSzeta)&=24 \diskR \OSzeta \massBH \OSxi \left(\OSzeta ^2-\OSxi ^2+1\right) \left[2 \diskR^2 \left(\OSzeta ^2-\OSxi ^2+1\right)^2+\massBH^2 \left(4 \OSzeta ^4+\OSzeta ^2 \left(5-3 \OSxi
                        ^2\right)+4 \OSxi ^4-5 \OSxi ^2+1\right)\right] .
                    \end{split}
                \end{align}
            \endgroup
        \end{widetext}

        \usetikzlibrary{calc}
        \begin{tikzpicture}[remember picture, overlay]
            \node (A) at (current page.south west){};
            \node (B) at (current page.south east){};
            \node (C) at (current page.north west){};
            \node (D) at (current page.north east){};
            \node (E) at ($ (A)!0.51!(B) $){};
            \node (F) at ($ (C)!0.51!(D) $){};
            \node (G) at ($ (E)!0.26!(F) $){};
            \node (H) at ($ (B)!0.29!(D) $){};
            \filldraw[color=white, opacity=1.0] (G) rectangle (H);
        \end{tikzpicture}


\input{paper.bbl}

\end{document}

%% file: paper.bbl
%